\documentclass[12pt, twoside]{mitthesis}

\usepackage{cmap}
\usepackage[T1]{fontenc}
\usepackage{graphicx}
\usepackage{epstopdf}
\usepackage{tikz-cd}
\usepackage{lgrind}
\usepackage{amsmath}
\usepackage{amssymb}
\usepackage{color}
\usepackage[colorlinks,citecolor=blue,linkcolor=blue,urlcolor=blue]{hyperref}
\usepackage{braket}
\usepackage{nicefrac}
\usepackage{slashed}
\usepackage{textcomp}
\usepackage{amsmath,amsfonts,amssymb,latexsym}
\usepackage{hhline}
\usepackage{graphicx}
\usepackage[arrow,matrix]{xy}
\usepackage{tikz}
\usepackage{tikz-cd}
\usetikzlibrary{positioning,arrows}
\usetikzlibrary{decorations.pathmorphing}
\usetikzlibrary{decorations.markings}
\usepackage{braket}
\usepackage{csquotes}

\usepackage{bm}
\usepackage{verbatim}
\usepackage{mathrsfs}
\def\all{all}
\ifx\files\all  \else 

\renewcommand{\c}{\hat c}
\newcommand{\co}{\emph{(Color Online)}}
\newcommand{\psibar}{ \psi^\dagger}
\renewcommand{\H}{\mathcal{H}}
\newcommand{\Z}{\mathbb{Z}}
\newcommand{\R}{\mathbb{R}}
\newcommand{\cN}{\mathcal{N}}
\newcommand{\cM}{\mathcal{M}}
\newcommand{\cB}{\mathcal{B}}
\newcommand{\ka}{\kappa}

\newcommand{\cV}{\mathcal{V}}

\newcommand{\RZ}{{\mathbb{R}/\mathbb{Z}}}

\newcommand\hcup[1]{\underset{{\scriptscriptstyle #1}}{\smile}}
\newcommand\toZ[1]{\lfloor #1 \rceil}
\newcommand\E{\mathcal{E}}
\newcommand\T{\mathcal{T}}

\newcommand\dd{\text{d}}

\newcommand{\beq}{\begin{equation}}
\newcommand{\eeq}{\end{equation}}

\newcommand{\eq}[1]{\begin{align} #1 \end{align}}

\newcommand{\Va}{V_{\alpha}}
\newcommand{\Vb}{V_{\beta}}

\newcommand{\Vab}{V^{1}_{\alpha\beta}}

\newcommand{\Vabgs}{V^{2}_{\alpha\beta\gamma\sigma}}

\renewcommand{\aa}{a^{\alpha}}
\newcommand{\ab}{a^{\beta}}

\newcommand{\pab}{\phi^{\alpha\beta}}

\newcommand{\ta}{\theta^{\alpha}}
\newcommand{\tb}{\theta^{\beta}}
\newcommand{\mab}{m^{\alpha\beta}}

\renewcommand{\L}{\mathcal{L}}

\newcommand{\Hol}{\text{Hol}}

\begin{document}

\title{Chiral Phases on the Lattice}

\author{Michael Austin DeMarco}
\prevdegrees{B.S., Physics, Rice University (2014) \\
B.A., Mathematics, Rice University (2014)\\
M.A.S.t., Theoretical Physics, University of Cambridge, (2015)}
\department{Department of Physics}

\degree{Doctor of Philosophy in Physics}

\degreemonth{May}
\degreeyear{2022}
\thesisdate{February 11, 2022}


\supervisor{Xiao-Gang Wen}{Cecil and Ida Green Professor of Physics}

\chairman{Deepto Chakrabarty}{Professor of Physics, Associate Department Head}

\maketitle



\cleardoublepage
\setcounter{savepage}{\thepage}
\begin{abstractpage}
%
%
%

While chiral quantum field theories (QFTs) describe a wide range of physical systems, from the standard model to topological quantum matter, the realization of chiral QFTs on a lattice has proved to be difficult due to the Nielsen-Ninomiya theorem and the possible presence of quantum anomalies. 
In this thesis, we use the connection between chiral phases of matter and chiral quantum field theories (QFTs) to define chiral QFTs on a lattice and allow a huge class of exotic field theories to be simulated numerically. Our work builds on the `mirror fermion' approach to the problem of defining chiral theories on a lattice, which defines chiral field theories as the edge modes of chiral phases. We begin by reviewing the deep connections between chiral phases of matter, chiral field theories, and anomalies. We then develop numerical treatments of an $SU(2)$ chiral field theory, and provide a semiclassically solvable definition of Abelian $2+1$ chiral topological orders. This leads to an exactly solvable definition of chiral $U(1)$ SPT phases with zero correlation length, which we use to extract the edge chiral field theories exactly. These zero-correlation length models are vastly more simple than previous approaches to defining chiral field theories on the lattice.

\end{abstractpage}


\cleardoublepage

\section*{Acknowledgments}

No PhD is accomplished alone. Even so, I feel that I have been particularly fortunate to be surrounded by so many wonderful people over these past years. 

I have had the most extraordinary advisor in Xiao-Gang. Writing about Sir Isaac Newton, the economist Keynes said: ``Newton was not the first of the age of reason. He was the last of the magicians...'' There is something of that magic in Xiao-Gang. I have never met anyone like Xiao-Gang before, and I doubt I ever will again. His mind really does pierce the mysteries of the universe to reveal the secrets of quantum mechanics. He dreams up solutions (that even he cannot explain how he found) which show us entirely new aspects of this strange world. So great is his contribution that there are two historical periods characterizing our understanding of the phases of matter: before and after Xiao-Gang. I would like the reader to understand something even more extraordinary about him: with his great mind and all of his achievements, never once did he chastise me for basic questions, even at the beginning of my PhD. Xiao-Gang was always encouraging, kind, and genius. I have been incredibly fortunate to be his student.

I have also been lucky to have had excellent teachers and collaborators throughout my PhD. Xiao-Gang's other students Hamed Pakatchi and Wenjie Ji graciously mentored me as I joined the group. Ethan Lake and Jing-Yuan Chen taught me much of what I know about Chern-Simons Theories on the lattice, and I collaborated with Ethan on much of the work in this thesis. Cyprian Lewandowski and Michal Papaj were great friends and patient teachers of solid state physics; in addition to being a great friend (and hosting a surprise party after my defense!), Kushal Seetharam has given me a wonderful introduction to quantum devices. Michael Pretko has been a friend and mentor, as has Alex Dear. Josue Lopez has been a friend, lab mate, and teacher for many years, and I was thrilled to see him marry another friend, Jessica Ruiz. Shane Alpert has been a consistent companion through this Physics journey. I am also grateful to my committee, Senthil Todadri and Will Detmold, for their supervision.

I am especially grateful to the many friends and family who made this possible. My first thanks go to my parents, who raised me, encouraged (and later tolerated) my interest in science for all these years, always stood by me and supported me, and graciously housed my girlfriend and I as we fled Covid in the early days. I could never have done this without their love and examples. Thanks also go to my mother for her extensive corrections to this thesis. I am deeply grateful to my girlfriend, Rachel Grasfield, whose strength I have borrowed and counsel I have relied on many times over the past four years. Sharing the various highs and lows of graduate school with her has been a joy in itself. 

Nancy and Steve Salmon have always been there for me, and I am deeply grateful for their love. All of my family have been wonderful throughout this journey: Janice and Rick, Lynda and Carlos, Bob and Tracy, Bill, Katie and Nico, Jon and Christie, Connor and Katie, Ben and Melina, Alicia, and Lauren. To my grandfather Carlo, my grandmother Jane, my grandfather Bill, and my grandmother Lee: I wish you could have seen this. 

There are so many others. My High School teachers, Mr. Chuang and Mr. Barrows, took a wayward teenager and led him to the light of physics. Tom and Jennifer Pekar have been dear friends. Cathy Modica has been a deeply wise friend, mentor, and guide to whom I will always be grateful; her advice was a light in many confusing moments during my PhD. Sydney Miller has been essential in the thesis process. I deeply enjoyed serving on the Federal Affairs Board with Ben Lane, Seamus Lombardo, Jimmy McCrae, Jordan Harrod, and all my friends there; I look forward to hearing about their work to continually strengthen our country. To Lisa Vo and Matt Zinman, who took a plunge with me to found a technology start-up: I will always be grateful for the faith you placed in me. And to our advisors, Nicco Mele, Julia Lestage, and Steven Liss: thank you for your wise teaching throughout the journey. To Rahul Srivathsa, Sameer Abraham, Kamna Kathuria, Vinay Ramprasad, Faiyad and Sheza Ahmad, Purva Jain, Teresa Modigel, Peeya Tak, David Chen, Alex Kendall, Joe Fisher, Francesca Bastianello, Michael Schmid, Michael Bryden, Christi Economy, Arthur Kouyoumidjan, Mariel Pettee, Christian Henry, and Nick Ryder: thank you for your friendship, kindness, and antics. Jerry and Susu Meyer have been kind friends and wise mentors; I often hear Jerry's advice ringing in my ears and guiding me towards the wiser path. Pastor Jon Lee guided me through one of the toughest times of graduate school. To Dominic Valenti, my brother and friend, and his family, Lupita, Peepaw, and others: I am so deeply grateful for your presence in my life. 

You---the family, the friends, the teachers---are what made this possible, and I will be grateful for the rest of my life. I have absolutely loved my time here at MIT. As much as I could not be more thrilled for what comes next, it is a bittersweet moment. In the words of Winnie the Pooh, ``How lucky am I to have something that makes saying goodbye so hard.''


\pagestyle{plain}
\tableofcontents
\newpage
\listoffigures
\newpage
\listoftables

\chapter{Introduction}\label{chap:Introduction}

Quantum Field Theories are wild things. It has been nearly a century since Born, Heisenberg, Jordan, and Dirac discovered Quantum Field Theories (QFTs) \cite{shifman_2012, doi:10.1063/1.2914365}, and yet these theories remain mysterious. They are among our most precise theories of physics, delivering predictions accurate to one part in a trillion in some cases. However, as mathematical objects they often make no sense, requiring careful regulation to tame their divergences. QFTs are extremely versatile: they govern the behavior of all known quantum systems, ranging from fireballs at trillions of Kelvin inside the Large Hadron Collider to the cold condensates at at fractions of a billionth of a Kelvin, just one building over from where this thesis was written. Much of theoretical physics, from Schwinger and Feynman to Witten and Wen, is devoted to their study, simulation, and calculation. 

QFTs are so powerful and versatile because they capture the universal, long-distance behavior of quantum systems of many particles. QFTs were discovered in a context close to modern high-energy physics, where not only do the energies involved create many particles but the physics itself becomes sensitive to the immense number of virtual particles winking in and out of existence. However, QFTs find a natural application to the theory of condensed matter systems, where we may consider ${\sim}10^{23}$ particles interacting on a vast lattice. From our condensed matter perspective, the `long-distance behavior' of particles on a lattice describes the characteristic properties of a phase of matter near a gapless point, where correlation lengths grow and the dynamics happen at long distances on the lattice scale. There the QFT can reveal the fundamental nature of the phase: What are the low-energy excitations? How does the system respond to a probe? QFTs capture the essence of a system in the vicinity of a gapless point; in turn, the physics of many particles on a lattice can be used to study and regulate quantum field theories.

It is a testament to the wonder of physics that studying QFTs more closely makes them more interesting and more mysterious. In the earliest days, the precise definition of field theories was a pedagogical complication that could be avoided with perturbative renormalization. QFTs were allowed to be mathematically ill-defined because their predictions were so incredibly accurate, while their formalization was left to mathematicians and mathematical physicists. However, as our understanding of them grew, we found that many of their most important properties---anomalies, strange conservation laws, and beyond---were revealed when QFTs were defined precisely. This relationship is at the core of the relationship between condensed matter and high energy theorists that has recently ignited a golden age in the study of topological phases and physics. In turn, a better definition of QFTs allows us to develop a deeper understanding of the laws that govern the universe and to use those laws to create new science and technology. 

This thesis is anchored around the application of condensed matter theory to the simulation of chiral phase and QFTs. Chiral phases and field theories, where left and right-handed excitations behave differently, are particularly challenging to define owing to the appearance of the `quantum anomalies' that will be defined in Chapter \ref{chap:CMQFT}. In particular, we set out to define and regulate chiral QFTs in the lattice, in a way which either cancels out their anomalies or renders those anomalies explicitly calculable. 
We will begin with a numerical simulation of a previously proposed approach to defining chiral QFTs, and end with an exactly solvable approach that more efficiently creates $U(1)$ chiral lattice field theories and reveals their subtle properties.Our results also lead to new phases of chiral matter in condensed matter systems. 

Chapters \ref{chap:CMQFT} and \ref{chap:CFTQM} form the background of this thesis. In Chapter \ref{chap:CMQFT}, we will set out a few of the properties of QFTs that we wish to study and demonstrate them in the context of simple lattice models. The most important of these are quantum anomalies. We establish a simple picture of the `chiral' or `axial' anomaly in $1+1$d with a lattice model, and we discuss the general relevance of quantum anomalies to actually defining QFTs on the lattice. In Chapter \ref{chap:CFTQM}, we will lay out the connections between phases of matter and quantum field theories that we defined previously. We will introduce the concepts of `topological' order and `SPT' order and establish the relationships between the bulk quantum order and the anomalies of the edge theories. We will also elaborate specific quantum field theories that we wish to simulate in the following chapters, and comment on their classification.

Our new results are presented in Chapters \ref{chap:SU2}-\ref{chap:Chiral}.  In Chapter \ref{chap:SU2}, we follow a previously proposed approach to create a new regularization of a $1+1$d $SU(2)$ chiral field theory. This approach depends on creating a non-chiral field theory, then giving a subset of fields a large gap so that only the chiral theory is left at low energies, which we are able to successfully demonstrate. Nonetheless, this approach faces technical obstacles to generalization. In Chapter \ref{chap:KMatrix}, we approach a related problem, the simulation of gapped Abelian topologically ordered theories in $2+1$d. We are able to create rotor models which can be reliably solved semiclassically, a significant achievement for these theories. Following this, in Chapter \ref{chap:U1SPT}, we `ungauge' the theories of Chapter \ref{chap:KMatrix} to create exactly solvable models of SPT ordered phases. These exactly solvable theories are also a significant achievement, and we use them to create commuting projector models previously thought to be impossible. In an extraordinary turn, these SPT models immediately lead to exact chiral field theory models as a corollary, and we explore these chiral field theories and their anomalies in Chapter \ref{chap:Chiral}.

This thesis details an extended, and successful, search \cite{Borges} for fruitful lattice models of chiral QFTs. We begin by defining a chiral field theory in a $1+1$d $SU(2)$ model and, while successful, find obstacles to future work. We then take a meandering path through higher-dimensional phases, eventually uncovering an SPT formalism which immediately yields a chiral field theory of the sort we had originally sought. We hope the reader will find the journey enlightening and useful.

\chapter{Condensed Matter Quantum Field Theories}\label{chap:CMQFT}

In this chapter, we lay out some basics of quantum field theories, establishing our notions of chirality and symmetry, for the perspective of a reader broadly familiar with physics and calculus. The expert should refer directly to Chapter \ref{chap:CFTQM}. 

As mentioned in the previous chapter, QFTs capture the long-distance, universal behavior of a system of many particles. However, with renormalization group (RG) theory, we can go further. RG leads to equivalence classes of QFTs which have the same long-distance behavior by erasing or `integrating out' short-distance degrees of freedom. Under this process, QFTs morph into one another, until eventually only infinite-range behavior is captured. These infinite correlation length `fixed-point' QFTs are in fact representatives of phases of matter. To study any QFT, fixed point or otherwise, is to study a phase of matter and gain insight into the collective behavior of a system. 

Something extraordinary arises in QFTs: universality. Because there are relatively few classes of fixed-point QFTs, there are relatively few patterns of behavior in the physical world. This allows relationships between various systems and their phase transitions to be derived, and is why the pattern of cracks in metals and rocks mirror the image of lightening (eg \cite{Cracks}), while a similar argument applies to the shifts of sand and snow in piles (eg \cite{MANNA1991249}). The Superfluid-Mott-Insulator transition we study in Chapter \ref{chap:U1SPT} is in fact in the same universality class as the condensation of a superfluid and the ferromagnetic ordering of two-dimensional `XY' magnets, which have both been studied extensively. This universality is a crowning triumph of modern QFTs and is a principle behind the ordering of all matter.

Throughout the decades of QFT, significant progress has involved both the condensed matter and high-energy communities, from the theory of symmetry multiplets and particles to the theory of phases and Wilsonian renormalization group (RG). We are currently in a `golden age' \cite{GoldenAge} arising from just such a collaboration, where the theory of quantum anomalies, higher symmetries, and topological phases have been unified and the two communities are working closer than ever, though differences of ideas and language remain. This thesis, and the research herein, has its ideas firmly rooted in the condensed matter side, but seeks to build tools that both the condensed matter and high-energy communities will find useful. In particular, we set our sights on the Chiral Fermion Problem, which has plagued simulations of both condensed matter and high-energy theory for decades. We will demonstrate a solution to it in Chapter \ref{chap:SU2}, before developing the theory to create a far more efficient solution in Chapter \ref{chap:Chiral}.   

\section{Practical Lessons in QFTs}

A huge amount of research has gone into defining just what a QFT is: as sums of all functions, renormalizable path integrals, quantizations of phase space, operator-valued distributions, and more. Even more complex is the relationship between traditional QFTs defined in continuous spacetime and lattice QFTs defined in discrete spacetime.

We will sidestep all of this and instead take a simple-minded approach. In this thesis, we always define QFTs as lattice QFTs. Any continuum quantities we write down will be only useful abstractions that help us reason about the dynamics of an underlying lattice model. This approach will render many of the more surprising and subtle properties of QFTs extremely clear, as on the lattice it is difficult for any complication to hide.

Let us now see this in practice. We will consider a simple $1+1$d lattice model and use it to see some of the major features of QFTs: dynamics, symmetry, chirality,  anomalies, and the relationship between the lattice and the continuum as well as the Hamiltonian and Lagrangian formalisms.

\begin{figure}
    \centering
    \includegraphics[height = 3.5 cm]{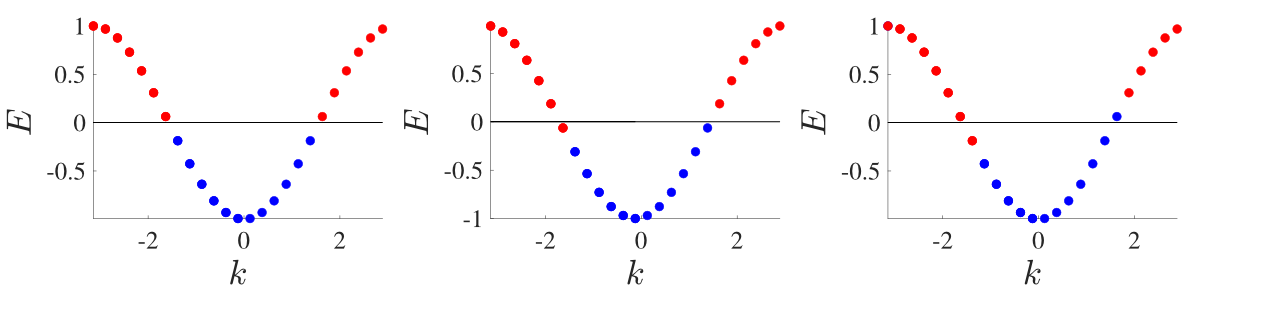}
    \caption[Axial Anomaly in a Hamiltonian Model]{\co. \emph{(left)} Eigenenergies of the Hamilonian \eqref{ch2:eq:Energies} with $\mu=0$. Filled states, below the chemical potential, are shown in blue, while unfilled states are shown in red. \emph{(middle)} As we adiabatically change the background field $A$, the energies evolve as $-\cos(k - A)$ \emph{(right)} After adiabatically evolving the background gauge field from $A=0$ to $A= \frac{2\pi}{N}$, we transfer charge from the left-moving mode to the right-moving mode. This is the axial anomaly.}
    \label{ch2:fig:cosk}
\end{figure}

Consider fermions on a lattice in one spatial dimension. We label the $N$ lattice sites by $i=1, ..., N$ with periodic boundary conditions, and introduce creation and annihilation operators on each site which satisfy $\{\hat c_{i}, \hat c_{j}^\dagger\} = \delta_{ij}$, $\{\c_i, \c_j\} = \{\c_i^\dagger, \c_j^\dagger\} = 0$. The Hamiltonian for our lattice model is given by:
\begin{equation}
    \hat H = -\frac{t}{2} \sum_{i}\left(\c_i^\dagger \c_j + \c_j^\dagger\c_i\right) - \mu\sum_i \c_i^\dagger \c_i \label{ch2:eq:Ham}
\end{equation}
This free theory describes particles hopping left or right, with a chemical potential $\mu$. It can be diagonalized in momentum space by setting $\hat c_k^\dagger = \frac{1}{\sqrt{N}}\sum_j e^{i k j} \c_j^\dagger$, where $k = \frac{2\pi m}{N}$, $m=1, ..., N$. Doing so, we obtain the Hamiltonian:
\begin{equation}
    \hat H_k = -\sum_k(t\cos (k) + \mu) \c^\dagger_k \c_k \label{ch2:eq:Energies}
\end{equation}
(We will always assume unit lattice spacing). The eigenspectrum of this Hamiltonian with $t=1, \mu = 0$ is shown in Figure \ref{ch2:fig:cosk}. At zero temperature, the states with energy $E< \mu$ will be filled by fermions and are depicted in blue, while states with $E>\mu$ will be unfilled. The filled states are referred to as the `Fermi Sea', while the states right at $\mu = 0$ are the `Fermi Surface' (It is indeed a surface in higher dimensions) and have Fermi momenta $\pm k_F$.

In the spirit of understanding the long distance physics, it is important to understand what are the low-energy excitations of this system. With some energy $\epsilon \ll t$, the only possible charge-conserving excitations involve moving a fermion from a filled state just below the Fermi surface to an empty state just above the Fermi surface. Crucially, each point on the Fermi surface comes with an associated group velocity:
\begin{equation}
    v_F = \frac{dE}{dk} = \pm t\sin(k_F)
\end{equation}
The disturbance we may create by exciting a particle near one of these Fermi points will move to the left or right along our 1d spatial lattice with velocities given by $v_F$.

To understand the phase of matter, we need only understand the physics near those two points. We do so by writing down a Lagrangian that captures two fermionic modes: one right-moving and one left moving. The correct model is:
\begin{equation}
    S = \int dx dt \left[
    \psibar_L (\partial_t + \partial_x )\psi_L
    +
    \psibar_R (\partial_t - \partial_x)\psi_R
    \right]\label{ch2:eq:Lag}
\end{equation}
Here $\psibar_{L, R}, \psi_{L,R}$ are anticommuting Grassman fields and we have rescaled space so that $v_F = 1$. Note that varying with respect to $\psibar_{L, R}$ reproduces the equations of motion:
\begin{align}
    (\partial_t + \partial_x)\psi_L = 0 \\
    (\partial_t - \partial_x)\psi_R = 0
\end{align}
These are the wave equations in $1+1$d for an excitation moving to the left or right, respectively. 

In just a few short lines, we have written down both a Hamiltonian and Lagrangian description of particles hopping in one dimension. Surprisingly, quite a lot can be derived from these models, including their symmetries and a specifically quantum phenomenon called a quantum anomaly. 

The Hamiltonian models \eqref{ch2:eq:Ham}, \eqref{ch2:eq:Energies} have an important $U(1)$ global symmetry given by:
\begin{align}
    \c_i\to e^{i\theta}\c_i \hspace{1 cm} \c_i^\dagger \to e^{-i\theta}\c_i^\dagger \\
    \c_k\to e^{i\theta}\c_k \hspace{1 cm} \c_k^\dagger \to e^{-i\theta}\c_k^\dagger \\
\end{align}
where $\theta$ is a constant. Crucially, this implies that $\sum_i \c_i^\dagger \c_i$ commutes with $\hat H$, $[\sum_i \c_i^\dagger \c_i, H] = 0$, and so we have a conserved quantity, namely conservation of charge.

We can see the same behavior in the Lagrangian formalism by sending:
\begin{equation}
    \psi_{L, R} \to e^{i\theta} \psi_{L, R} \hspace{1cm}
    \psibar_{L, R} \to e^{-i\theta} \psibar_{L, R}\label{ch2:eq:vector}
\end{equation}
which leaves \eqref{ch2:eq:Lag} invariant. Using Noether's theorem, we see that the charge:
\begin{equation}
    Q= \int dx (\psibar_L\psi_L + \psibar_R \psi_R)\label{ch2:eq:U1}
\end{equation}
is a conserved quantity. 

Examining the Lagrangian \eqref{ch2:eq:Lag}, it would appear that we could rotate the phase independently on each of the left-moving and right-moving fields. Formally, we would implement this by, in addition to the `vector' symmetry \eqref{ch2:eq:vector} employing an `axial' symmetry:
\begin{align}
    \psi_{L} \to e^{i\theta} \psi_{L} \hspace{1cm}
    \psibar_{L} \to e^{-i\theta} \psibar_{L} \\
    \psi_{R} \to e^{-i\theta} \psi_{R} \hspace{1cm}
    \psibar_{R} \to e^{i\theta} \psibar_{R}
    \label{ch2:eq:axias}
\end{align}
which rotates the phases on left and right-moving fields oppositely. Between the vector and axial symmetries, we are able to independently rotate the phase on the left and right moving modes. Instead of the the single conserved quantity \eqref{ch2:eq:U1}, we would obtain two:
\begin{align}
    Q_L = \int dx~\psibar_L\psi_L\label{ch2:eq:U1_L} \\
    Q_R = \int dx~\psibar_R\psi_R\label{ch2:eq:U1_R}
\end{align}
These two charges would have the interpretation of the number of excitations at the left and right Fermi points, respectively.

Turning back to the Hamiltonian model, we see that there is no local way to interpret these conserved charges in terms of the $\c_i, \c_i^\dagger$ operators, as there is no way in the Hamiltonian formulation to address the two Fermi points separately. This is our first clue that something might be wrong with the axial `symmetry.' 

More directly, we can consider coupling the model to a background gauge field, whereby we modify the Hamiltonian to be:
\begin{equation}
    \hat H = -\frac{t}{2} \sum_{i}\left(\c_i^\dagger e^{i A} \c_j + \c_j^\dagger e^{-i A}\c_i\right) - \mu\sum_i \c_i^\dagger \c_i \label{ch2:eq:Ham_gauged}
\end{equation}
where we may take $A$ to be constant in space, but not time. In momentum space, this Hamiltonian is:
\begin{equation}
    \hat H_k = - t \sum_k \cos(k - A) \c_k^\dagger \c_k
\end{equation}
where we have set $\mu = 0$. So we see the effect of this background gauge field is to shift the energies. Recalling that $k = \frac{2\pi m}{N}$ for $m=1, ..., N$, consider the effect of slowly changing $A$ from $A_i=0$ to $A_f = \frac{2\pi}{N}$ as shown in Figure \ref{ch2:fig:cosk}. The spectrum is identical for both $A=A_i$ and $A= A_f$. However, adiabatically evolving the system would transfer a filled state above the Fermi sea, effectively transferring charge from the left-moving mode to the right-moving mode. Thus we see that the left and right moving modes \emph{do not independently conserve charge} in the presence of a gauge field. Only the sum of left and right moving is conserved. The axial `symmetry' is not a symmetry it all!

It is remarkable that this breaking of the axial symmetry is in fact universal. Translating the argument above into an equation, we have seen that the change of the axial charge is given by:
\begin{equation}
    \partial_t Q_A = \frac{2 e}{2\pi} \partial_t A
\end{equation}
In a more general covariant theory with $e=1$, this can be rewritten as:
\begin{equation}
    \partial_\mu j_A^\mu = \frac{1}{2\pi} \epsilon^{\mu\nu}F_{\mu\nu}
\end{equation}
where $j_A^\mu$ is the axial conserved current which is predicted to be conserved by Noether's theorem. This is the exact expression that can also be derived by expanding the action for $A$ in terms of Feynman diagrams \cite{Peskin:257493}. Moreover, this result is general: any lattice model that has charge-$n$ right and left moving modes will contain a similar anomaly term, with the right-hand side multiplied by\footnote{One factor of $n$ arises because the mode is $n$ times as sensitive to the gauge field, and so we would lose $n$ times as many particles. The second factor arises because each particle carries a charge of $n$.} $n^2$. The result is independent of the details of the model and so is \emph{universal}.

\begin{figure}
    \centering
    \includegraphics[height = 2cm]{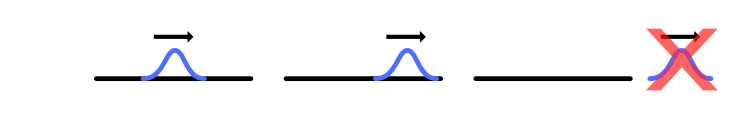}
    \caption[Axial Fermion Anomaly from the Equation of Motion]{\co. A right-moving mode is created in the middle of an open wire. If right and left moving charge are independently conserved, then the right moving mode would continue right off the wire, which is impossible. Hence the axial symmetry must break at a boundary.}
    \label{ch2:fig:Anomaly_Edge}
\end{figure}

We can also see this failure from a very physical argument. Suppose that, instead of closed boundary conditions, we consider the theory on an open line segment as shown in Figure \ref{ch2:fig:Anomaly_Edge}. We create a right-moving excitation in the center of the system. If the axial symmetry held, then the right-moving excitation could only move to the right, and could not scatter into left-moving modes. It would continue to the right, right off the end of the system, thereby violating the axial (and vector) symmetry. This must not happen, and so it must be true that the boundary also breaks the axial symmetry. In fact, the symmetry breaking on a boundary should be considered as a consequence of the breaking with a gauge field, as an electric potential could be used to create a boundary. However, in practice the boundary formulation is often more convenient.

Whenever a theory would seem to have a symmetry, but that symmetry is broken in the presence of a gauge field or a boundary, we say that the theory has an \emph{anomalous symmetry} or that the symmetry has a \emph{quantum anomaly}. In this case it is a `mixed anomaly' between the global $U(1)$ and the axial $U(1)$ which breaks axial symmetry. This anomaly is our first glimpse of a `chiral' anomaly. The next section will dive much deeper into these anomalies and their relationship to a decades-old problem in lattice gauge theory.

\section{The Chiral Fermion Problem}

One feature we noted of the axial `symmetry' is that it is not present in the original Hamiltonian lattice model, nor is it clear how to write down a lattice model in one dimension that would have a axial symmetry affecting the left and right-moving modes differently. In many ways, this difficulty of the lattice model is a saving grace: it is telling us that the theory is anomalous. If we had some local lattice model with an axial symmetry, we could couple it to a background gauge field and obtain a well-defined, non-anomalous field theory. But the anomaly is universal and so this is impossible---and that is what the difficulty in defining an axial lattice symmetry is hinting at. 

The anomaly discussed in the previous section has the unique property that it is \emph{chiral}, meaning that it treats left and right differently, and we now adopt the term \emph{chiral anomaly} for the behavior we have seen, instead of axial anomaly which is more common in high-energy physics. More generally, a theory, symmetry, or field is chiral if it is not invariant under the inversion of one spatial dimension. In general, we will be concerned with chiral anomalies and their relations to chiral phases of matter.

In the last section, we wrote down a Lagrangian \eqref{ch2:eq:Lag} describing left and right moving modes. It is very tempting to attempt to take only half of that Lagrangian, say:
\begin{equation}
    \int dx dt \psibar_R (\partial_t - \partial_x) \psi_R
\end{equation}
as a model for a \emph{chiral field theory}. This theory would effectively live under a ``vector $+$ axial symmetry'', with an anomaly:
\begin{equation}
    \partial_\mu j^\mu = \frac{e}{4\pi} \epsilon^{\mu\nu}F_{\mu\nu} = \frac{e}{2\pi} F_{01}
\end{equation}
This anomaly is even more severe than the axial one, as it is not `mixed' but direct: flux in the gauge field directly breaks the $U(1)$ symmetry to which the gauge field is coupled. Just as the axial symmetry lacked a lattice definition, there is a severe obstruction to defining this theory on the lattice, and for the same reason: because the anomaly is universal, it must also appear on the lattice. On the other hand, it is not immediately clear how such an anomaly could possibly appear in a lattice model.

The difficulty in defining a chiral field theory on a lattice is the subject of the decades-old \emph{chiral fermion problem} (CFP). Specifically, the CFP refers to the difficulty to define a fermion theory in odd spatial dimensions which satisfies:
\begin{itemize}
    \item The Hilbert space of the theory factorizes as a product of local Hilbert spaces.
    \item The theory can be coupled to a background gauge field.
    \item The theory has Hamiltonian or Lagrangian formulation involving only local terms.
    \item The theory has symmetry $U(\theta)$ which factorizes as a product of operators on each site $U(\theta) = \otimes_i U_i(\theta)$.
    \item The symmetry $U$ is not inversion symmetric, i.e. it treats left and right-handed modes differently (note that this may include a case where there are different numbers of left and right handed modes).
\end{itemize}
In this thesis, we will solve this problem in several ways. For simplicity, we will often relax the condition that the theory be fermionic. In the most successful way, we will have a solvable model, but will have to allow a non-on-site symmetry $U(\theta)$. In that case, we will in fact have a lattice model that is local, captures the quantum anomaly, and satisfies all the other above conditions. However, we have many chapters before we encounter that theory. 

\begin{figure}
    \centering
    \includegraphics[height = 5cm]{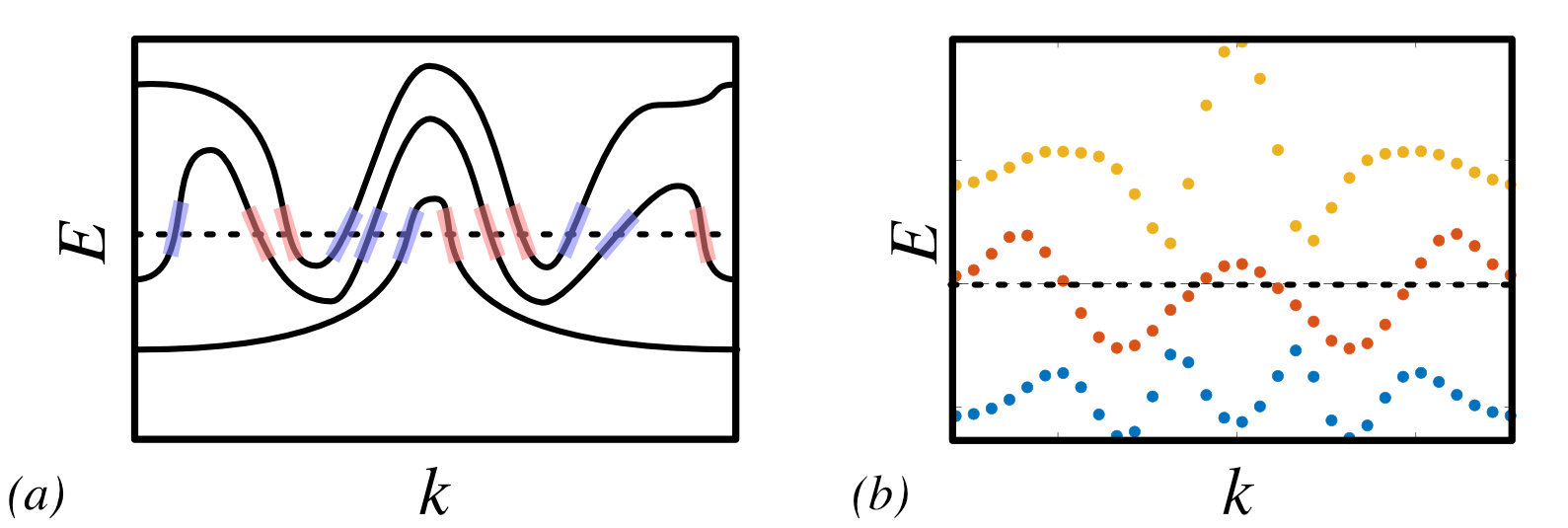}
    \caption[Band Structure and the Nielsen-Ninomiya Theorem]{\co. \emph{(a)} Schematic depiction of a periodic band structure in one spatial dimension. The chemical potential is denoted by the dotted line. Right moving modes are highlighted in blue and left moving modes are highlighted in red. \emph{(b)} Actual band structure of a random translation-symmetric Hamiltonian with third-nearest-neighbor hopping.}
    \label{ch2:fig:NNT}
\end{figure}

Nielson and Ninomiya proved that that the CFP is impossible to solve without interactions \cite{NIELSEN198120, NIELSEN1981173}. The basic argument is already visible in the results of the previous section. There we saw that the eigenvalues of the Hamiltonian $E_j(k)$ are a periodic function of $k$. We assume that we fill up eigenstates up to some Fermi energy, and we assume that $\frac{dE_j}{dk}\neq 0$ at the Fermi energy, so that the $E_j(k)$ crosses the Fermi energy linearly (see Figure \ref{ch2:fig:NNT}). Because the functions are periodic, we are assured that the number of crossings with $\frac{dE_j}{dk} >0$ (right-movers) is equal to the number crossings with $\frac{dE_j}{dk} <0$. Hence there are equal numbers of right-and left movers. Moverover, just as the two modes we examined share a chiral anomaly between them, in the general case modes pair up so as to share anomalies, and this will imply that that the charges of the right-moving modes under the non-chiral $U(1)$ symmetry will mirror exactly the charges of the left-moving modes, and the total theory is not chiral. The generalization to higher dimensions involves more mathematics, but the basic idea is the same.

The difficulty in defining a chiral theory on a lattice has been a serious frustration for the lattice gauge theory (LGT) community because of the critical role chiral phenomena play in the standard model. For one, all observed neutrinos are left-handed, while all observed anti-neutrinos are right-handed \cite{griffiths}. Neutrinos couple to the dynamical gauge field of the weak sector and simulating this interaction would be useful to the lattice gauge theory community. Beyond particle physics, simulating a chiral field theory with its attendant anomalies would have considerable use for condensed matter theory, which regularly sees chiral field theories in edge theories, a fact which we will devote almost the entire next chapter to.

Considerable work to evade the Nielsen-Ninomiya result and solve the CFP has been performed over the intervening decades. The first thing one might ask is why can we not start with the momentum-space Hamiltonian \eqref{ch2:eq:Energies} and restrict to one point near the Fermi level? Doing so, one effectively creates a discontinuous Hamiltonian in momentum space, so when it is Fourier transformed back to real space the Hamiltonian is infinitely long ranged. Far more elegant versions of this idea exist \cite{Neuberger:1997fp, Neuberger:1998wv, Hernandez:1998et}, and all suffer from the same non-locality. In some cases, these models can still be coupled to a weak background gauge field, but the non-locality violates the conditions of the CFP that we set out above. Two other approaches soon came out of the lattice field theory community. In the overlap formalism \cite{NARAYANAN199362,
NARAYANAN1997360, Luscher:1998du, Luscher:2000hn}, the chiral theory is defined as the overlap of successive ground states. However, the physical interpretation of this formalism can be difficult, and it is not clear if the Hilbert space factorizes as a product of local Hilbert spaces. In a somewhat related model, one can realize the chiral theory as fermions living on a domain wall \cite{KAPLAN1992342, SHAMIR199390} inside a higher-dimensional space. However, the gauge field which couples to the fermions propagates in that higher-dimensional space.

A vein of approaches to the CFP involve similar ideas to what we will explore in the next chapter, realizing the chiral theory as the boundary of a higher-dimensional system \cite{Wen:2013ppa,Wang:2013yta, PhysRevB.91.125147, Grabowska:2015qpk, Mirror1, Montvay:1992eg, Giedt:2007qg,
PhysRevD.94.114504}. This will be explored in detail in the next chapter, in conjunction with the development of ideas around topological order and SPT phases that we will explore in that chapter.

We have taken a short stroll through quantum field theory, with a condensed matter lens. The great advantage of this approach, blending between continuum arguments and lattice models, is that it can unite the best of both worlds: the continuum arguments build intuition for the physical picture, while the lattice model nails down technical details and clarifies subtleties. This is particularly true in the case of anomalies, where the use of lattice models allows us to demonstrate the $U(1)$ anomaly in just a few lines. We then parlayed this anomaly picture to illustrate the Chiral Fermion Problem, and the problems with chiral field theories in general. In the next section, we will explore in detail how chiral theories can appear naturally as boundary theories. 

\chapter{Chiral Field Theories and Quantum Matter}\label{chap:CFTQM}

We have discussed the relationship between field theories and anomalies in the context of a simple lattice model. Now we will elaborate on the relationship both of these have with quantum phases of matter, before laying out the general properties and classification of some of those phases that we will require in subsequent chapters.

\section{Chiral Field Theories, Edge Theories, and the Mirror Fermion Approach}

In the previous chapter, we discussed several chiral anomalies and found that they indicated the failure of conservation of charge in a system. We discussed the non-conservation of charge in a system consisting only of a right-moving mode in one spatial dimension, and indicated that its anomaly would make the theory difficult to define on a lattice. A theory with a single right-moving mode in which charge is not conserved would indeed be hard to make sense of. Particles would be appearing and vanishing whenever an electric field is applied, and at a boundary the particles would have to continue out of the system and into the vacuum. 

\begin{figure}
    \centering
    \includegraphics[height = 4cm]{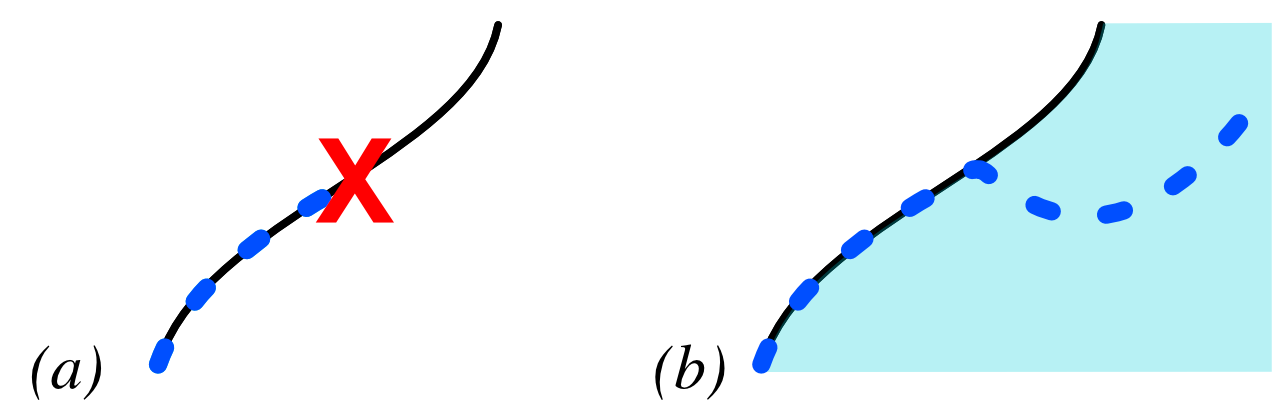}
    \caption[Anomalous Theories as Edge Theories]{\co. \emph{(a)} In a chiral theory in one spatial dimension, charge is not conserved, and particles (blue dotted line) may travel along before disappearing (red X). \emph{(b)} We now conceive of the one-dimensional system as the gapless edge theory of the gapped two-dimensional bulk. Particles do not disappear, but instead travel into the bulk.}
    \label{ch3:fig:Anomaly_Edge}
\end{figure}

However, there is a very physical way to avoid all of this and define the theory of a chiral mode on a lattice. We can solve the seeming contradiction of particles vanishing from the one-dimensional spacetime by understanding the one-dimensional system as the edge of a two dimensional bulk, as shown in Figure \ref{ch3:fig:Anomaly_Edge}. We consider the edge to be gapless, and the bulk gapped. Most of the time, any particle is confined to traveling along the edge. However, when an electric field is applied, it can tunnel into the bulk. From the perspective of the edge, the particle has disappeared, while from the perspective of the bulk a particle has suddenly appeared. Accounting for both theories, particle number is actually conserved. Mathematically, this will mean that the edge theory and the bulk theory have equal and opposite anomalies which cancel only when the two theories are considered together. This approach also nullifies the anomaly which happened at the boundary of the one-dimensional system: the one-dimensional system cannot have a boundary because it is itself a boundary of the two-dimensional system, and boundaries cannot have boundaries \cite{Hatcher}. 

At first glance, this may seem somewhat contrived. However, exactly such a theory describes the Quantum Hall states. In the next subsection, we describe a detailed lattice theory for these models and extract the relevant behavior. 

\subsection{Integer Quantum Hall on a Lattice}

The Integer Quantum Hall (IQH) effect \cite{2016arXiv160606687T} has been extremely well studied. Here, we concoct a simple lattice model that reproduces its properties and demonstrates the chiral edge mode.

In the IQH, we take a system with two spatial dimensions and apply a transverse magnetic field. The IQH occurs when there are an integer number of particles per flux quantum of the magnetic field, i.e. $\nu \in \mathbb Z$, where $\nu$ is the filling fraction. To implement this on a lattice, consider a lattice with nearest neighbor and next-nearest-neighbor hopping. Labeling the points by $x, y \in \{1, ..., L_x\}\times \{1, ..., L_y\}$ We apply a magnetic field $A_x= 2\pi \frac{y}{2} $, $A_y = 0$, which leads to a field strength $F = \partial_x A_y - \partial_y A_x = -\pi $ per plaquette. With a half flux quantum per plaquette, at half filling, the system should reproduce the IQH state. 

\begin{figure}
    \centering
    \includegraphics[width = .9\columnwidth]{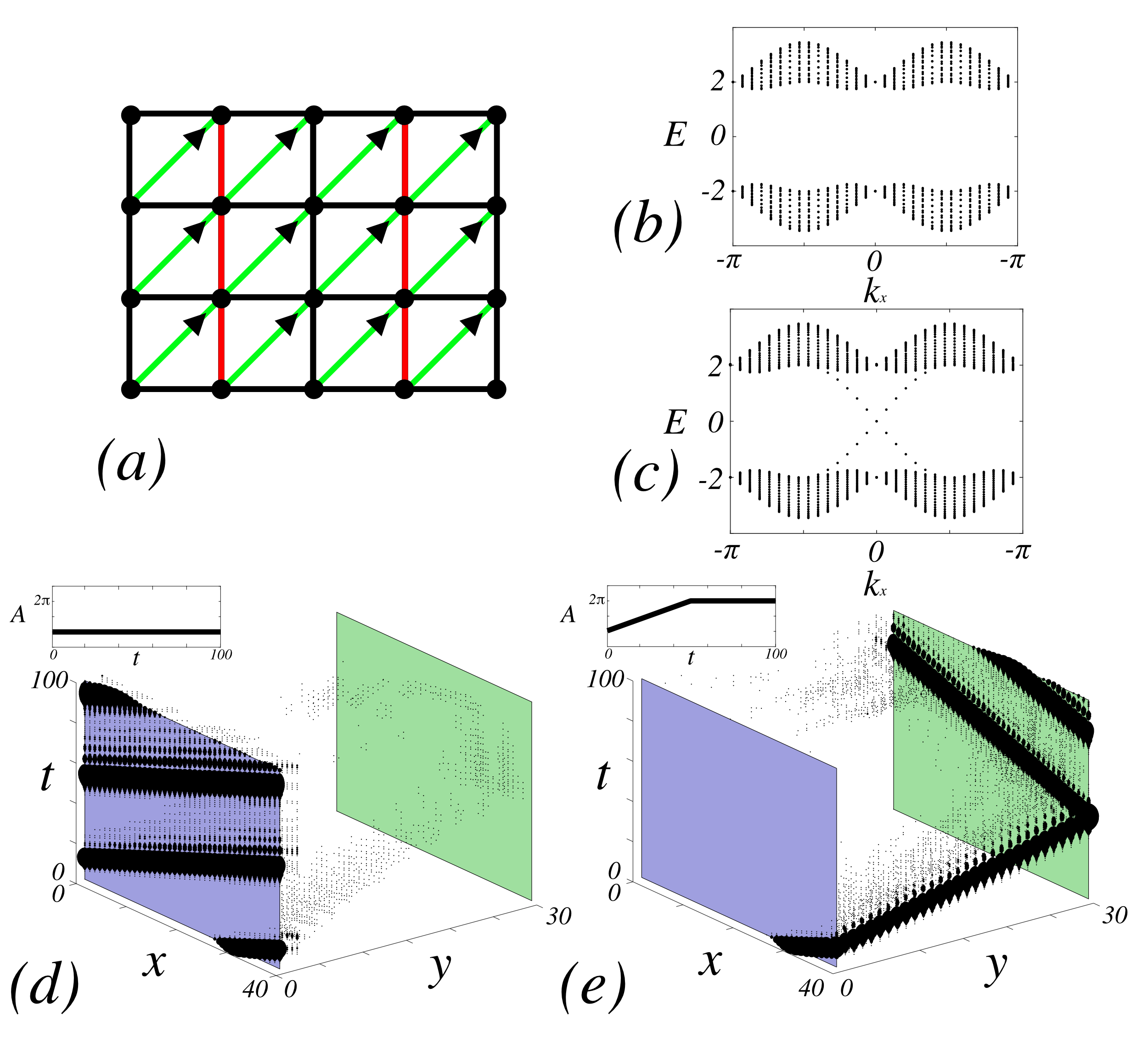}
    \caption[Lattice Models for the Integer Quantum Hall Effect]{\co. \emph{(a)} Spatial Lattice for the Integer Quantum Hall Effect. Black links indicate a hopping of $+1$, red a hopping of $-1$, and green links a hopping of $+i$ in the direction of the arrow. The flux through any plaquette is $\pi$, so at half filling this realizes a $\nu = 1$ IQH state. \emph{(b, c)} Fourier transforming in the $x$ direction, we plot the band structure as a function of $k_x$ for \emph{(b)} closed boundary conditions, where we see a gapped bulk, and \emph{(c)} open boundary conditions, where we see the two gapless modes. \emph{(d, e)} Spacetime propagation of a particle wavefunction in an IQH system. We consider the IQH lattice model with open boundary conditions in the y direction and periodic boundary conditions in the $x$ direction. The two edges are shown in purple and green, respectively. We inject a particle wavepacket with $k\approx 0$ at time $t=0$. The amplitude for the particle to propagate to point $(x, y, t)$ is indicated by the size of the black markers at $(x, y, t)$, and the strength of the background gauge field is shown in the inset.}
    \label{ch3:fig:IQH_Lattice}
\end{figure}

The lattice model is shown in Figure \ref{ch3:fig:IQH_Lattice}a. Writing it out explicitly, we have:
\begin{equation}
    \hat H = \sum_{x} \Big[\c^\dagger_{x+1, y}\c_{x, y} + (-1)^{x}\c^\dagger_{x, y+1}\c_{x, y} + i \c^{\dagger}_{x+1, y+1}\c_{x, y} + h.c.\Big]\label{ch3:eq:IQH_Lattice}
\end{equation}
where we have set the hopping to be unit strength. One may check that hopping counter-clockwise around any triangle leads to a factor of $+i$, and therefore the flux through each square plaquette is $\pi$, as desired. 

To understand the edge theory, we Fourier transform in the $x$ direction. The resulting band theory is shown in Figure \ref{ch3:fig:Anomaly_Edge}b for closed boundary conditions and Figure \ref{ch3:fig:Anomaly_Edge}c for open boundary conditions. For closed boundary conditions, we see a gapped bulk theory across the entire one-dimensional Brillouin zone. The various bands result because we have not taken the Fourier transform in the $y$ direction. Note that if we place the chemical potential in the range $-1.5\leq \mu \leq 1.5$, then we fill up half the bands, leading to the desired half filling. 

For open boundary conditions, we see two modes appearing inside the bulk gap. One is clearly right-moving ($dE/dk>0$) and the other left-moving ($dE/dk <0$). In this case, however, the two modes are in fact spatially separated---each of them on their own edge.

We can see the spatial separation of edge modes quite clearly by calculating the motion of a particle added to the system within the gap. To do so, we calculate the propagator of a spacetime lattice which uses \eqref{ch3:eq:IQH_Lattice} as its spatial model. The results are presented in Figures \ref{ch3:fig:IQH_Lattice}d and \ref{ch3:fig:IQH_Lattice}e. There we see a spacetime representation of the IQH system. The two $1+1$d edges are shown in purple and green, respectively. We inject a particle with a wavepacket with low momentum ($k\approx 0 $) modes at $t=0$. The magnitude of the amplitude for particle propagation is shown in black. In \ref{ch3:fig:IQH_Lattice}d, the particle propagates smoothly as a slightly dispersing, right moving mode. It remains confined to the edge and simply moves to the right. Next, we can effect a time-varying gauge potential by rotating the boundary conditions\footnote{This is a slightly different approach from how we treated the Hamiltonian in the previous section but is computationally easier. Note that in this case the wavepacket stays localized as it tunnels through the bulk---that is an artifact of how we have chosen the gauge variation and is not universally true.} from $0$ to $2\pi$. The boundary conditions are shown in the inset. In \ref{ch3:fig:IQH_Lattice}e, we change the boundary conditions, and this tunnels the resulting electron through the bulk and to the other edge, where it then becomes a left-mover. We thus have three theories each with their own anomalies: two edges and one bulk, and taken together they must in total conserve charge, i.e. their anomalies must cancel. 

So we see that, even though the gapless modes are spatially separated (and they may be macroscopically so), they are still connected in the same way that the left and right moving modes of the ``$\cos(k)$'' model \eqref{ch2:eq:Ham_gauged} were. One may similarly couple in a spatially constant background gauge field and, by varying it, transfer charge from the left-moving mode to the right-moving mode. In this case, one also ends up transferring charge across the system. This is the Integer Quantum Hall Effect! We have essentially reproduced Laughlin's argument \cite{PhysRevB.23.5632}, and we see that the Hall conductance is quantized because the charge of the particles is quantized.

Thus there is a model which satisfies the rather stringent conditions that we laid out. This model has also pointed the way to beginning to unify these lattice field theory considerations with deep principles from condensed matter, and we will explore those connections in the following. We return to defining chiral field theories on the lattice in Section \ref{chap3:sec:Mirror}.

\section{Topological Order, SPT Phases, and Local Unitary Quantum Gates}

In the last section, we saw that the two counter-propagating gapless modes in an IQH state may actually be spread across a sample, with one gapless mode appearing one one edge and the other mode on the other edge. This was a significant feat, as it allowed us to recognize that a chiral gapless mode can indeed appear alone. However, it is not true that the modes are isolated from one another, as they are linked by the anomaly. Exposing one edge to an electric field actually transfers charge across the system to the other. 

The source of the link between these two gapless edge modes, which may be macroscopically separated, is the same as the source of all `spooky actions at a distance' in quantum mechanics: it is entanglement. In the quantum Hall system, degrees of freedom in the system arbitrarily far away are entangled. Moreover, this entanglement cannot be unwound.

We can formalize this into a working definition on how to classify quantum phases of matter \cite{RevModPhys.89.041004, 2015arXiv150802595Z}. Suppose we have states $\ket\psi, \ket{\psi'}$ in a Hilbert space $\H$ which factorizes as a tensor product over local Hilbert spaces $\H = \otimes_i \H_i$. We wish to know under what conditions $\ket{\psi}$ and $\ket{\psi'}$ can be considered to belong to the same phase of matter. Ultimately, this will cover macroscopic observables like Hall conductance and anomalies, as well as mappings of the low-energy Hilbert spaces of the systems. We define a local unitary operator on a site $i$ as a unitary operator $\hat U_{i}$ which acts as the identity on all the local Hilbert spaces except those in a finite radius $\xi$ of site $i$. We define a local unitary operation as a product of local unitary operators. We say that two states $\ket \psi$ and $\ket{\psi'}$ are in the same phase if they can be deformed into one another via a finite number (\emph{i.e.} independent of system size) of local unitary operations.

From the condensed matter perspective, these local unitary operations can be roughly thought of as finite-time evolution under gapped Hamiltonians. The locality demands the presence of a gap, since otherwise information and entanglement could propagate at luminal velocities. Hence we are simply considering the finite-time evolution of our system under perturbations that are local, and gapped. This also has an appealing interpretation in terms of quantum gates. Local unitary operations are simply (local) quantum gates, and two states belong to the same phase if they can be deformed into one another by a finite depth\footnote{Formally, we should say that two states belong to the same phase if they require a depth that grows less than linearly (e.g. logarithmically) with the system size.} circuit. 

The local unitary operations, or local quantum gates, allow us to partition the set of states in a given Hilbert space. If we augment local unitary transformations by allowing the addition of unentangled `ancilla' qubits that enlarge the physical system, we now have a proper definition of phases of matter. 

\begin{figure}
\centering
\includegraphics[width = .75\textwidth]{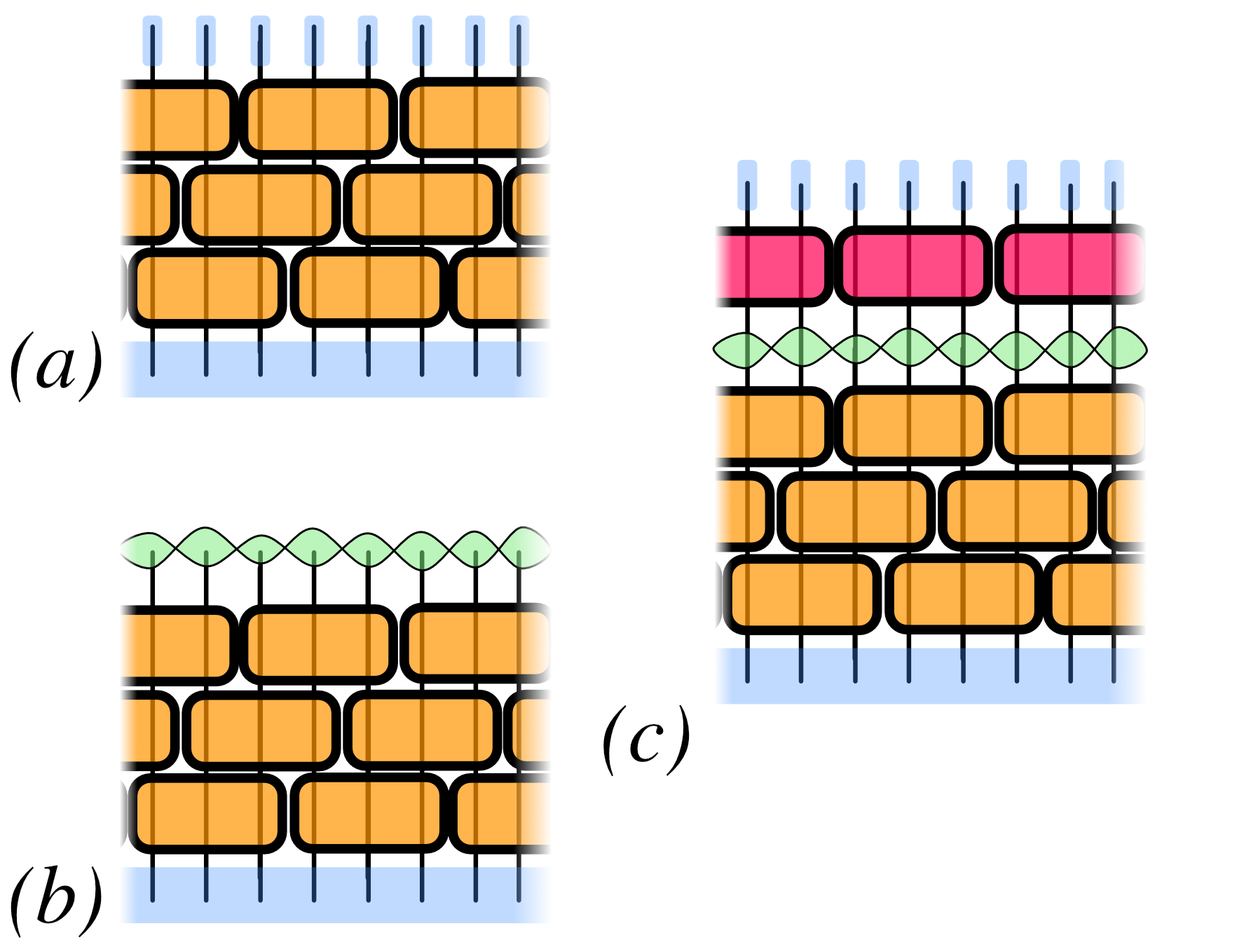}
\caption[Local Unitary Circuits and the Classification of Quantum Phases]{\co. \emph{(a)} A trivial state (blue at bottom) may be unwound by a finite-depth local unitary circuit (orange blocks) into a product state (blue at top). \emph{(b)} A topologically ordered state exhibits long-range entanglement which cannot be unwound by a finite-depth circuit. Even after application of the circuit, the entanglement remains (green). \emph{(c)} A SPT state contains entanglement that can be unwound into a product state by a finite-depth circuit only if we apply gates which break the symmetry (red blocks).}\label{ch3:fig:LUs}
\end{figure}

Within this definition, there is one class of states that is `trivial,' namely the product states. Any state which factorizes as $\ket\psi = \otimes_i \ket{\psi_i}$ with $\ket{\psi_i}\in \H_i$ is trivially unentangled (Figure \ref{ch3:fig:LUs}a), and the equivalence class of states that it generates is known as the `trivial phase.' 

In addition to the trivial phase, there are many non-trivial phases which cannot be deformed into the trivial phase or into each other. These are the topologically ordered states. The basic ingredient leading to non-trivial phases is the aforementioned long-range entanglement \cite{Kitaev:1997wr}. Any short-range entanglement can be unwound by a finite-depth circuit. However, a long-range pattern of entanglement cannot be unwound (Figure \ref{ch3:fig:LUs}b) and it is this which characterizes the nontrivial topologically ordered states. On the other hand, these states are physically characterized by their thermal Hall conductance, i.e. their chiral central charge, the braiding of their excitations, and their topologically protected ground state degeneracy. Just as the Hall conductance was connected to a $U(1)$ anomaly, the thermal Hall conductance appears in field theory as a gravitational anomaly. Formally, each topological order is defined by a tensor category \cite{2020JHEP...09..093K}, though for our purposes we will not need the most general definition. In Section \ref{chap3:sec:KMatrix}, we will review the field theory description and classification of $2+1$ Abelian topological orders in terms of $U(1)$ $K$-matrix Chern-Simons theory, as in Chapter \ref{chap:KMatrix} we will create a lattice definition of these states.

Including a symmetry adds another layer to this classification. In doing so, we should restrict to local unitary gates that commute with the symmetry. This restriction leads to new phases. There will be phases that were formerly trivial which are now non-trivial. These phases feature short range-entanglement, but their entanglement patterns cannot be unwound by any finite-depth circuit which respects the symmetry. However, they can be unwound by a finite-depth circuit which breaks the symmetry (Figure \ref{ch3:fig:LUs}c). Known variously as Symmetry-Protected Trivial or Symmetry-Protected Topological (SPT) states, these states have trivial ground state degeneracy, but often host nontrivial, anomalous edges that we describe below. In Section \ref{chap:U1SPT}, we will create an exactly solvable lattice path integral and Hamiltonian model for a large class of $U(1)$ SPT states.

\begin{figure}
\centering
\includegraphics[width = .75\textwidth]{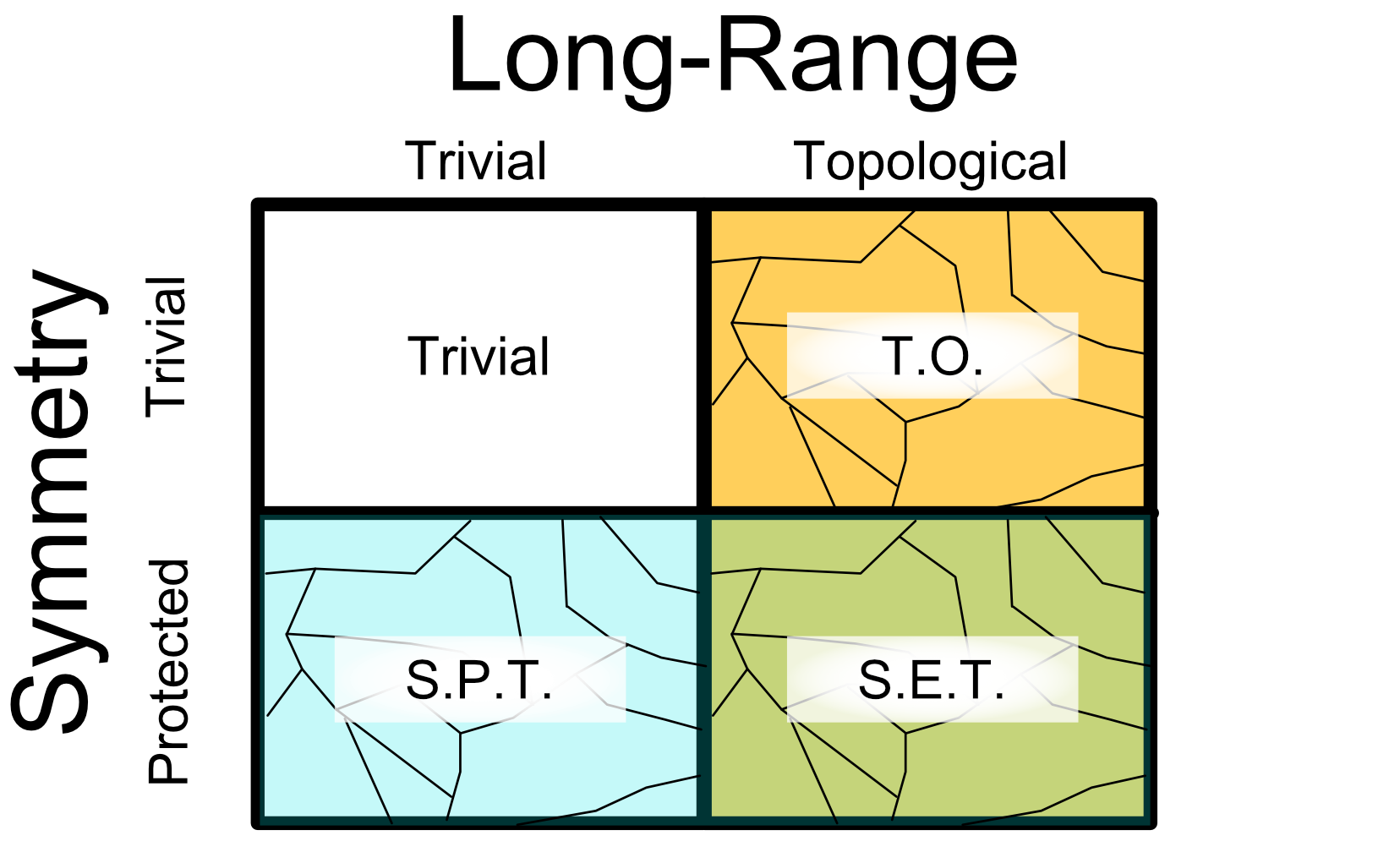}
\caption[Four Types of Quantum Phases]{\co. Phases of matter may exhibit short or long-range entanglement as well as a symmetry that may or may not protect the state. These two properties lead to a four-fold classification of the types of symmetric phases of matter: trivial states, topologically ordered states, symmetry-protected trivial states, and symmetry-enhanced topological states.}\label{ch3:fig:fourfold}
\end{figure}

These two considerations: long-range entanglement and symmetry protection, lead to a four-fold classification of symmetric phases shown in Figure \ref{ch3:fig:fourfold}. There we see the trivial phase, as well as the topologically ordered phases, the SPT phases, and the symmetry enhanced topological SET phases, which are both topologically ordered and enjoy symmetry protection. Our focus in this thesis will be on the topologically ordered and SPT phases\footnote{Note that this classification is for symmetric phases. Spontaneous symmetry breaking, of the sort we will see in the superfluid phase in Chapter \ref{chap:U1SPT}, effectively adds a third dimension to this classification that is not drawn in Figure \ref{ch3:fig:fourfold}.}.

We have already seen one of the the simplest examples of topological order: the IQH state. There, the nontriviality of the phase is revealed not only by the Hall conductance, which requires $U(1)$ symmetry to reveal, but by the thermal Hall conductance and its associated gravitational anomaly. Just as it is impossible to trivialize the bulk due to the long-range entanglement, it is in fact impossible to gap out the edge due to the gravitational anomaly.

\begin{figure}
\centering
\includegraphics[width = .75\textwidth]{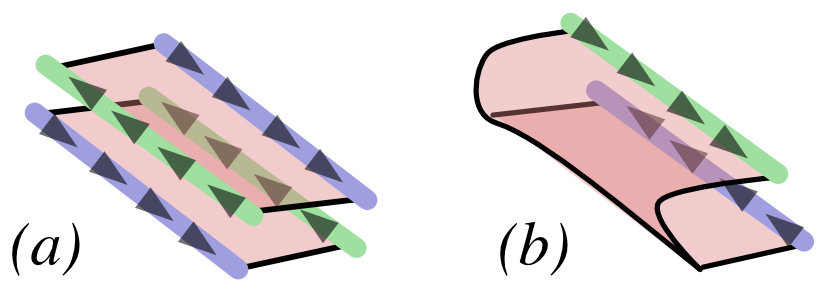}
\caption[Gapping an SPT]{\co. \emph{(a)} To create an SPT system, we consider a two layer system composed of an IQH state and its time-reversal conjugate, with the particles in the two layers having opposite charges. \emph{(b)} We may gap out the edge of the system by inserting lattice links to turn the model into an $\text{IQH}_{+1}$ system folded over on itself, but only at the cost of breaking the $U(1)$ symmetry defined in the text.}\label{ch3:fig:SPTgapping}
\end{figure}

We have not directly seen an SPT phase, but it is easy to create a model using the ingredients we already have. Consider the two-layer system shown in Figure \ref{ch3:fig:SPTgapping}a composed of the IQH state we have already examined, which we denote by $\text{IQH}_{+1}$ and its time-reversal conjugate, which we denote by $\text{IQH}_{-1}$. We define a $U(1)$ symmetry with the particles in the $\text{IQH}_{+1}$ state having charge $+1$ and the particles in the $-1$ state having charge $-1$. 

If we do not consider the $U(1)$ symmetry, then the edge is trivial. As mentioned above, both layers carry gravitational anomalies, but these cancel because gravity couples to all particles with the same sign and the two layers are time-reversal conjugates. The edge can be gapped out by scattering the respective left and right-moving modes into one another. Indeed, one can simply augment the lattice model so that it becomes the $\text{IQH}_{+1}$ state folded over on itself (Figure \ref{ch3:fig:SPTgapping}b). However, this explicitly breaks $U(1)$ symmetry, as it scatters positive charges from the $\text{IQH}_{+1}$ into the $\text{IQH}_{-1}$ layer, where they become negative charges. In fact, if we consider the $U(1)$ anomaly, we see that, because the particles have opposite charges and the states are time-reversal conjugates of each other, the gauge anomalies add instead of canceling. Hence the edge and the bulk are nontrivial when we require $U(1)$ symmetry, and we have an example of an SPT state. The model we describe in Chapter \ref{chap:U1SPT} has a similar edge theory to this toy model.

\begin{table}[]
    \centering
    \begin{tabular}{|c|c|c|}\hline
         & \textbf{Topological Order} & \textbf{SPT Order} \\ \hline
        \textbf{Entanglement} & Long-Ranged & Short-Ranged but Symmetry-Protected \\ \hline
        \textbf{Bulk} & Thermal Hall Conductance & Charge Hall Conductance \\ \hline
        \textbf{Boundary} & Gravitational Anomaly & Gauge Anomaly\\ \hline
    \end{tabular}
    \caption{Properties of Chiral Topological Orders and Chiral SPT Orders.}
    \label{ch3:tab:TOvsSPT}
\end{table}

These two examples---the IQH state and the $U(1)$ charged SPT state composed of $\text{IQH}_{+1}$ and $\text{IQH}_{-1}$ states---illustrate some of the fundamental relationships that we have been exploring, which are summarized in Table \ref{ch3:tab:TOvsSPT}. Topologically ordered states host long-ranged entanglement in the bulk, while the entanglement pattern for SPT order is short-ranged by symmetry-protected. In turn, \emph{chiral} topologically ordered phases suffer gravitational anomalies that correspond to bulk thermal Hall conductance, while the \emph{chiral} SPT phases exhibit gauge anomalies which reflect bulk charge Hall conductance (often referred to simply as Hall conductance)\footnote{There is a state, the $E_8$ state, which hosts chiral edge modes and thermal Hall conductance \cite{Lu:2012dt} but has trivial bulk statistics, and there is some discussion as to whether $E_8$ is topologically ordered or an SPT phase.}. 

In Chapter \ref{chap:SU2}, we will use many of these ideas to construct a Chiral fermion model in $1+1$d. Following that, in Chapter \ref{chap:KMatrix}, will focus on developing a lattice description of $K$-matrix Chern Simons theory, which encompasses both topologically ordered and SPT states in $2+1$ dimensions. We review this $K$-matrix theory in the following section. In Chapter \ref{chap:U1SPT}, we will `ungauge' the $K$-matrix model to yield an SPT model, and we review the classification of SPT phases in section \ref{chap3:sec:GroupCohomology}. In Chapter \ref{chap:Chiral}, we use the ungauged model of Chapter \ref{chap:U1SPT} to develop a $1+1$d chiral lattice field theory that far surpasses the model of Chapter \ref{chap:SU2}. 

\section{Abelian Chern-Simons Theories}\label{chap3:sec:KMatrix}

Much of our preliminary discussion has focused on the connection between chiral field theories in $1+1$d and topological phases in $2+1$d. However, it we be good to have a field theory description of those $2+1$d topological phases. Here we review the $U(1)$ Chern-Simons theories which describe $2+1$d Abelian topological phases \cite{2004qftm.book.....W, 2010cmft.book.....A}. 

What the topological phases we have described have in common is a conserved current. Starting from a phenomenological approach in terms of this conserved current, we can build a field theoretic description of the IQH states and $U(1)$ SPT states we have examined so far. In fact, the general $K$-matrix formalism we will describe goes far beyond that, capturing all $2+1$d Abelian topological orders \cite{Lu:2012dt}.

The usual way of describing a conserved current, in terms of derivatives of matter fields that are conserved at the level of the equation of motion, will not be strong enough for what we want to do. We will not be concerned with the dynamics of the theories we consider and so we do not wish to rely on an equation of motion to conserve a current. Instead, we describe a phenomenological conserved current in terms of a $U(1)$ gauge field $a_\mu$ by writing:
\begin{equation}
    j^\mu = \frac{1}{2\pi}\epsilon ^{\mu\nu\lambda}\partial_\nu a_\lambda \hspace{1cm} \text{or} \hspace{1cm} j = \frac{1}{2\pi}\star da
\end{equation}
where we have included a factor of $2\pi$ because while flux is quantized to $2\pi\Z$ particle number should be quantized to $\Z$. On the right-hand side, we have introduced the differential form notation \cite{Bott_Tu} which we will make use of liberally in this section. The conserved current is gauge invariant under $a\to a+d\theta$ and, moreover, is automatically conserved, as:
\begin{equation}
    \partial_\mu j^{\mu } = \frac{1}{2\pi}\epsilon^{\mu\nu\lambda}\partial_\mu\partial_\nu a_\lambda = 0 \hspace{1cm} \text{or} \hspace{1cm} d^\dagger j = (\star d \star)\left(\frac{1}{2\pi}\star da\right) = \frac{1}{2\pi}\star d^2a = 0
\end{equation}
We do not need to worry about the origins of the conserved current, but instead should take it as emergent. Much of the power of the formalism we describe here consists in abstracting away from the microscopics of the system. 

With our current $j = \star da$ in hand, we should consider what sort of actions can be built. One obvious choice is to give the current energy, via:
\begin{equation}
    S = -\frac{1}{g}j^2 = - \frac{1}{(2\pi)^2g}(f\star f) 
\end{equation}
where $f=da$. This is the usual Maxwell term familiar from electrodynamics. Accordingly, it gives rise to dynamic suppression of current. We will generically assume either no or a small Maxwell term to be present, but the topological properties we are after will have to go beyond dynamics.

However, owing to the fact that we are working in $2+1$d, there is a unique term which we can write down, namely, the Chern-Simons action:
\begin{equation}
    S = \frac{m}{4\pi}\int ada = \frac{m}{4\pi}\int d^3x \epsilon^{\mu\nu\lambda}a_\mu \partial_\nu a_\lambda 
\end{equation}
At first glance, this action would appear to fail to be gauge invariant. However, it is indeed gauge invariant under small (continuously connected to the identity) gauge transformations on a closed manifold. Under large gauge transformations, one can show that it is gauge invariant if $m \in \Z$. (If $m$ is odd, then the theory is fermionic and will require a spin structure.) To couple the conserved current to a background gauge field $A$, we write $\star A j = Ada$ and integrate by parts so that the action is always invariant under gauge transformations of the background field. We thus add to the term an action:
\begin{equation}
    \frac{q}{2\pi} \int adA
\end{equation}
where the coefficient has been chosen so that when $q\in\Z$ the action is invariant under large gauge transformations. 

Putting these together, we now have the action:
\begin{equation}
    S = \frac{m}{4\pi} \int ada + \frac{q}{2\pi} \int adA
\end{equation}
This Chern-Simons action is a gauge invariant action specified by the integers $m$ and $q$ describing a quantum system in two spatial dimensions. $q$ should be thought of as the charge of the quasiparticle, while $m$ is known as the \emph{level} of the Chern-Simons theory. 

What then are we to make of this action, and what phase should it describe? We can match this action to real systems by understanding its Hall conductance. Since the action is quadratic, we can work classically. Varying $a$, we see that the equation of motion of the action is:
\begin{equation}
    m da = q dA \leftrightarrow j = \frac{q}{m} \star F
\end{equation}
where $F = dA$. Hence we see that the action `attaches' quasiparticle current to the flux of the gauge field. In the case of magnetic flux, so that $F_{xy}\neq 0$, the Chern-Simons equation of motion implies that $j^0\neq 0 $, and so the system localizes quasiparticles proportional to the flux. On the other hand, suppose that couple in an electric field in the $x$ direction, so that, say $F_{xt} \neq 0, F_{xy} = F_{yt} = 0$. This equation then says that:
\begin{equation}
    j^{y} = \frac{q}{m} F_{xt}
\end{equation}
Let us set $m=1$. Then we see that applying an electric field results in current in the $y$ direction. Noting that the electric current is related to the quasiparticle current by $j^{e} = q j$, we see that this is the Quantum Hall Effect, with Hall conductance $\sigma_{xy} = \frac{1}{2\pi} \frac{q^2}{m}$ ($\hbar = e = 1$). Specifically, for $m=1$, this is the basic $IQH_{+1}$ state, while for higher $m$ this leads to the basic Laughlin series fractional quantum Hall effects. We can obtain higher integer Hall states and more varied fractional quantum Hall effects by including multiple quasiparticle currents $a^1, a^2, ...$ and allowing them to couple to both each other and the background gauge field. The resulting action is the so-called $K$-matrix Chern-Simons theory:
\begin{equation}
    S = \frac{1}{4\pi}\sum_{I, J} K_{IJ} \int a^I da^J + \frac{1}{2\pi} \sum_{I} q_I\int a^I dA
\end{equation}
Here $K_{IJ}$ is a symmetric integer-valued matrix which describes the quasiparticle currents, while $q_I$ is an integer-valued vector that describes their charges. 

The $K$-matrix formulation captures the essence of Abelian topological orders in $2+1$d. To understand its properties, we begin with the Hall conductance. Running a similar analysis as above, one can see that the Hall conductance is given by $\sigma_{xy} = \frac{1}{2\pi} q \cdot K^{-1} \cdot q$. On the other hand, the number of right (left) moving chiral edge modes is given by the number of positive (negative) eigenvalues of $K$, and the thermal Hall conductance is determined by the signature of $K$. If any of the diagonal entries is odd, then it is a theory of fundamental fermions, and one must specify a spin structure. On a torus, the system will have a ground state degeneracy of $|\det K|$.

Not all theories with different $K$-matrices represent different phases. The formulation is redundant under changing the $N\times N$ $K$-matrix by any matrix $M \in GL(N, \Z)$ via $K\to M^{-1} K M$; this leads to the $K$-matrix classification of Abelian topological orders, and one can augment this description to allow for the inclusion of internal symmetries.

One crucial aspect of these theories that will play a role in our SPT work is the notion of braiding of excitations. Suppose we create two point-like excitations in our two-dimensional system. If we `braid' one particle around the other, the many-body wavefunction can transit through the many-body Hilbert space. In Abelian topological orders, we assume that when the particles return to their original positions, the many-body wavefunction returns to its original state (In non-Abelian topological orders there may be a matrix acting on the excitation state manifold). While the state may remain the same, the many-body wavefunction may pick up a $U(1)$ geometric phase resulting from the transit of the wavefunction through the Hilbert space. 

\begin{figure}
\centering
\includegraphics[width = .5\columnwidth]{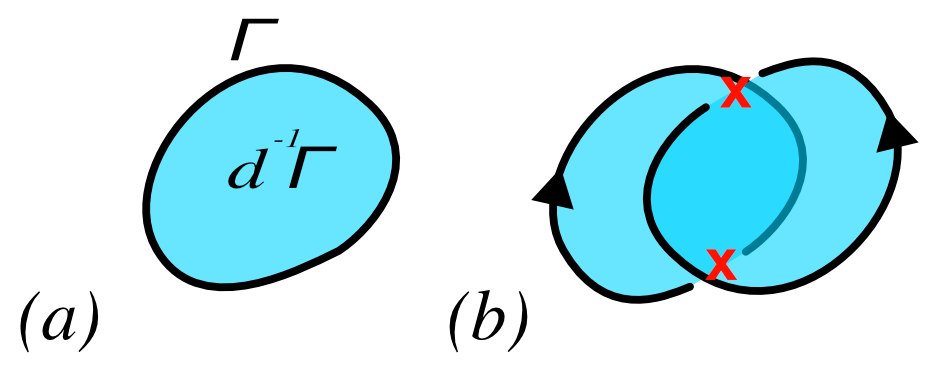}
\caption[Braiding Statistics for Chern-Simons Theories]{\co. \emph{(a)} We define $d^{-1}\Gamma$ as a one-form dual to a surface bounded by $\gamma$, which is dual to the two-form $\Gamma$. \emph{(b)} Calculating $\int \Gamma d^{-1} \Gamma'$ leads to the linking number $L(\gamma, \gamma')$.}\label{chap3:fig:Braiding}
\end{figure}

We can actually see these particle statistics quite simply in the $K$-matrix theory.  Excitations in this theory are generated by Wilson lines $e^{iw^I\oint a^I}$, where the integral vector $w^I$ describes the composite quasiparticle character of the excitation. Including $N$ Wilson lines with vectors $w_{n}^N$ along closed paths $\gamma_n$, the $K$-matrix action becomes:
\begin{align}
    S = \sum_{n = 1}^{N} \sum_{I} \int_{\gamma_n} w^I a^I+
    \frac{1}{4\pi}\sum_{I, J} & K_{IJ} \int a^I da^J
    \\
    =\sum_{n = 1}^{N} \sum_{I} &\int w_n^I a^I\Gamma_n +
    \frac{1}{4\pi}\sum_{I, J} K_{IJ} \int a^I da^J \nonumber
\end{align}
where we have set the background gauge field $A$ to be zero. Here $\Gamma_n$ is the two-form which is Poincar\'{e} dual to $\gamma_n$. We assume a contractible manifold so that we can choose one-forms that we denote by $d^{-1}\Gamma_n$ such that $d(d^{-1}\Gamma_n )= \Gamma_n$ since $\gamma_n$ is closed and so $d\Gamma_n = 0$. As illustrated in Figure \ref{chap3:fig:Braiding}a, one may think of $d^{-1} \Gamma_n$ as being one on a surface (blue) bounded by $\Gamma$ (black) and zero otherwise. Shifting the integration variables $a^{I} \to a^{I} - 2\pi \sum_n K_{IL}^{-1} w_n^L d^{-1}\Gamma_n$, the action becames:
\begin{equation}
    S = \frac{1}{4\pi}\sum_{I, J} K_{IJ} \int a^I da^J  - \pi \sum_{n, n'} (w_n \cdot K^{-1} \cdot w_n') \int \Gamma_n d^{-1}\Gamma_{n'}
\end{equation}
We have successfully eliminated the Wilson line variable and exchanged it for the second term, which contributes an overall phase to the path integral. We can simplify it by noting that, as illustrated in Figure \ref{chap3:fig:Braiding}b, $\int \Gamma_n d^{-1}\Gamma_{n'}$ is the linking number $L(\gamma_n, \gamma_{n'}$ between the curves $\gamma_n, \gamma_{n'}$. Reducing to the case of just two curves $\gamma_1, \gamma_2$, this phase becomes
\begin{equation}
    2\pi i (w_{1}\cdot K^{-1}\cdot w_2)L(\gamma_1, \gamma_2)
\end{equation}
This is the expression for the braiding statistics. Repeating a similar analysis, one encounters the ``self-statistics'' for a particle turning about itself:
\begin{equation}
    \pi i (w\cdot K^{-1}\cdot w) L(\gamma, \gamma)
\end{equation}
where $L(\gamma, \gamma)$ may be thought of as the number of times a particle turns. (In general, care must be taken to define $L(\gamma, \gamma)$, usually by using the framing in spacetime \cite{cmp/1104178138}).

These results on mutual and self statistics are a critical aspect of Chern-Simons theory. In particular, the statistics characterize topologically ordered states. Even more strangely, the self-statistics make clear that the excitations themselves may be fermions ($ (w\cdot K^{-1}\cdot w)$ odd), bosons ($ (w\cdot K^{-1}\cdot w)$ even), or neither (general $ (w\cdot K^{-1}\cdot w)$). Note that these are all excitations described by a nominally bosonic field $a^I$; instead it is the topological term which gives them statistics. This will play a crucial role in Chapter \ref{chap:U1SPT}, where we will encounter a topological term which turns a bosonic lattice model into a model of emergent fermions.

Returning to the case of a single Chern-Simons gauge field,
\begin{equation}
    \frac{m}{4\pi}\int ada
\end{equation}
there is a subtlety in this definition that we must uncover and that will play a crucial role in Chapter \ref{chap:KMatrix}. For any $U(1)$ connection, we may not be able to define $a$ everywhere in spacetime. For example, we know that the flux over any closed surface, say a sphere, is quantized to be an integer. However, if we have a globally defined gauge field $a$, then trivially $\int_{S^2}da = 0$, as $S^2$ has no boundary. Instead, one typically defines $a$ in contractible patches that cover spacetime, allowing the the fields on overlapping patches to differ by a gauge transformation.

Because the Chern-Simons action is gauge invariant only after integration by parts, if one takes a gauge field defined in patches, the action is no longer well defined. One can add terms to the action that depend on the connections in the various patches and the gauge transformation they differ by; this leads to the Cech-Deligne-Beilenson formulation \cite{Alvarez:1984es, Thuillier2015}. However, we will take another route, which instead seeks to reformulate the Chern-Simons action in terms of the field strength $f=da$ which is gauge invariant and insensitive to the problems of patches. We have two powers of $a$ in the Chern-Simons action that we wish to rewrite in terms of $f$. However, $f$ is a two form, and so we would naturally find ourselves facing a four-dimensional integral. The trick here is to actually define the Chern-Simons term in $2+1$d as the boundary of a $3+1$d action. Namely, let $\cN^4$ be a four-manifold, with boundary $\cM^3 = \partial \cN^4$. Then:
\begin{equation}
    \int_{\cN^4} f\wedge f = \int_{\cN^4}d(a\wedge da) = \int_{\cM^3} a\wedge da
\end{equation}
So to evaluate $\int_{\cM^3}ada$, we extend $a$ into a fourth dimension and evaluate $\int_{\cN^4}f\wedge f$. Because the four-dimensional theory depends only on $f$, and not on $a$, it is insensitive to the concerns over patches which the three-dimensional ``$ada$'' suffers. This leads to the Witten's `$\eta$ invariant' formulation of Chern-Simons theory, and we will use the $f\wedge f$ term to define a topological field theory on a lattice in Chapter \ref{chap:KMatrix}.

There is one remaining phenomena to address in Chern-Simons theories. We have asserted that Chern-Simons theories may host anomalous edge theories. We have also stated that they are gauge invariant when on a closed manifold, but we have not discussed the case for an open manifold. In fact, the Chern-Simons action is not gauge invariant on an open manifold, but rather suffers an anomaly there. We can exactly match the anomaly suffered at the edge by the Chern-Simons with an anomaly of an edge theory, and in fact this tells us that an edge theory must appear at the boundary of the Chern-Simons theory. 

One can directly match the anomaly of the edge to the boundary theory, ensuring that they cancel. This is the field theory statement of the behavior we encountered in Chapter \ref{chap:CMQFT}, where charge was transferred from an edge into the bulk and back. The resulting edge theory is:
\begin{equation}
    S_\text{edge} = \frac{1}{4\pi} \sum_{I, J} K_{IJ} \int dx dt~ \partial_t \phi^I \partial_x \phi^J + \frac{1}{2\pi} \sum_I q_I \int dx dt~ \phi^I dA
\end{equation}
This is the standard `bosonized' edge theory for chiral fermion theories, the Luttinger liquid, and more, depending on the value of $K_{IJ}$. For a bosonic theory with $\det K=1$, one can use an $GL(2, \Z)$ transformation to transform any $2\times 2$ $K$-matrix into:
\begin{equation}
    K = \left(\begin{array}{cc}
        0 & 1 \\
        1 & 0
    \end{array}\right)
\end{equation}
which we can rewrite as:
\begin{equation}
    S_\text{edge} = \frac{1}{2\pi} \partial_t \phi \partial_x \theta
\end{equation}
in the absence of a background field. In Chapters \ref{chap:U1SPT} and \ref{chap:Chiral}, we will discuss a lattice field theory that has a similar continuum limit. 

\section{Lattice Field Theories and the Group Cohomology Picture of SPTs}\label{chap3:sec:GroupCohomology}

In the previous section, we elaborated on the continuum field theory description of Abelian topological and SPT phases in $2+1$ dimensions. Here we will lay out a parallel area of progress, namely the classification of bosonic SPT phases in all dimensions using Group Cohomology (GC) \cite{GroupCohomology}. 

In a remarkable paper \cite{Chen:2011pg}, Chen, et al. argued that SPT phases are in one-to-one with the cohomology classes of maps from the group which protects the SPT to $U(1)$. The GC result is extraordinary: it is as if one discovered the periodic table \emph{before} being able to predict the physical properties of the elements, much less actually being able to isolate any of them. In Chapter \ref{chap:U1SPT}, we will actually construct models for a large class of these phases and understand their physical properties. 

Our starting point for the group cohomology classification is the formalism of lattice models on simplicial complexes \cite{Hatcher}. Let $G$ be a group, and suppose that $\cM^{d+1}$ is a $d+1$ dimensional spacetime lattice with sites labeled by $i, j, k,...$. We consider a lattice path integral in terms of a field $g_i$ which assigns a $G$-valued variable to each lattice site $i$. An action in these models is a function $\nu: G^{d+2} \to U(1)$ which assigns a $U(1)$ phase to each $d+1$-dimensional simplex, each of which has $d+2$ lattice sites:
\begin{equation}
    \nu(g_0, g_1, ..., g_{d+1}) \equiv \nu(\{g_i\})
\end{equation}
We require that each action be symmetric under the action of $G$, namely that:
\begin{equation}
    \nu(\{\bar g g_i\}) = \nu(\{g_i\})
\end{equation}
for any $\bar g\in G$. 

Given any function $\nu:G^{n} \to U(1) $, we can contruct a function $(d\nu):G^{n+1}\to U(1)$ by defining:
\begin{equation}
    (d\nu)(g_0, ..., g_n) = \prod_{\ell = 0}^{n} \nu^{(-1)^{\ell}}(\{g_i \setminus g_\ell\})
\end{equation}
where $\{g_i \setminus g_\ell\}$ means $g_0, ..., g_n$ excluding $g_\ell$. 
Because SPT states are trivial and have no topologically protected ground state degeneracy, the action should vanish on any closed manifold. In other words, the action should be only a surface term. In \cite{Chen:2011pg}, Chen et al argue that this means that the action should satisfy the cocycle condition
\begin{equation}
d\nu = 0 \label{ch3:eq:cocycle}
\end{equation}
At the same time, they argue that phases whose actions differ by an exact cochain, i.e. $\nu, \nu'$ such that:
\begin{equation}
\nu' = \nu + d\mu\label{ch3:eq:cocycle_mod}
\end{equation}
are equivalent. One way to understand this is to consider the ground state wavefunction on a particular lattice. Changing the action by $d\mu$ corresponds to multiplying the ground state:
\begin{equation}
\ket{\psi} \to \ket{\psi'} = \prod_{S} \mu(\hat g_i \in S) \ket{\psi}
\end{equation}
where the product is over all simplices on a spatial lattice. In particular, this is clearly a symmetric local unitary transformation as discussed previously, and hence $\ket{\psi}$ and $\ket{\psi'}$ belong to the same phase.

The classifications of cocycles which satisfy eqn. \eqref{ch3:eq:cocycle} modulo symmetric local unitary transformations \eqref{ch3:eq:cocycle_mod} is known as group cohomology. Specifically, the results of Chen et al argue that the classification of bosonic SPTs protected by a symmetry $G$ in $d+1$ spatial dimensions is given by:
\begin{equation}
H^{d+1}(G, U(1))
\end{equation}
Moreover, the group cocycles carry a natural abelian `addition' operation which corresponds to stacking of SPT phases. 

Applying the group cohomology classification to bosonic SPT phases protected by $U(1)$ in $d$ spatial dimensions, we obtain the following classification:
\begin{equation}
H^{d+1}(U(1), U(1)) = \begin{cases}
\Z & d = 2 \\
\Z_1 & d = 1, 3
\end{cases}
\end{equation}
The nontrivial SPT classification in $2$ spatial dimensions corresponds to states which feature a quantized Hall conductance of $2k \frac{e^2}{h}$. These are the states which we will construct in Chapter \ref{chap:U1SPT}.

Ultimately, while the group cohomology theory provides an exceptionally powerful classification of SPT phases, it does not provide explicit field theories for these cases. In the typical way that mathematics can be strange, it is much easier to enumerate all possible cohomology classes than it is to write down representatives of those classes. In turn, this is why the work of Chapter \ref{chap:U1SPT} is so critical: it provides the first examples of cocycles theories for continuous groups. These allow for a detailed physical picture as charged vortex condensates of the $U(1)$ phases in $2$ spatial dimensions to emerge, and the models we discuss there exhaust all the $U(1)$ bosonic SPTs in two dimensions.

We have now examined the deep connections between anomalies and quantum phases of matter. In the next chapter, we use these connections to develop a solution to the Chiral Fermion Problem.

\chapter{A Proposal for a Non-Perturbative Lattice Regularization of an Anomaly-Free $1+1$d $SU(2)$ Chiral Fermion Theory}\label{chap:SU2}

Here we develop an early-stage numerical treatment of a novel non-perturbative lattice
regularization of a $1+1$D $SU(2)$ Chiral Gauge Theory. Our approach follows
recent proposals that exploit the connection between anomalies
and topological (or entangled) states to create a lattice
regularization of any anomaly-free chiral gauge theory, and mirrors much of the discussion of `mass without mass terms' \cite{BenTov:2014eea,PhysRevD.93.081701,Ayyar2016,BenTov:2015gra,
Catterall:2015zua,Catterall:2016dzf, Xue:1996da,Xue:2000du, Xue:1999xa}. In comparison to other
methods, our regularization enjoys (ultra) local on-site fermions and gauge action, as well
as a physically sensible on-site fermion Hilbert space. 

Before proceeding further, let us further specify the problem that we mean to solve and the conditions under which we do so. We mean to find a  path integral on a space-time lattice which produces
the desired chiral fermion theory as the low energy effective theory, subject to the following conditions:
\begin{enumerate}
    \item The path integral and the resulting chiral fermion theory have the same space-time dimension.
    \item There are finitely many degrees of freedom on each lattice point.
    \item The lattice path integral is local, {\it i.e.} all terms in the Lagrangian have a finite range (this is sometimes termed ultra-local in the lattice literature).
    \item The lattice theory has the symmetry $G$ of the chiral fermion theory. This symmetry should act on-site, so it can be gauged \cite{W1313}, and is not spontaneously broken.
    \item The emergent chiral fermion theory may or may not have Lorentz symmetry (which would require different species of fermions to have the same velocity), but must break parity and time-reversal symmetry. In particular, the right and left-moving modes carry different representations of the gauge group. (Ensuring Lorentz symmetry will be the subject of future work).
\end{enumerate}
The condition of on-site symmetry discussed above merits further discussion. The on-site condition means that the quantum operator representing any `internal' symmetry factorizes as a tensor product of operators that act only on a single site. That is, $U(g) = \otimes_{i}U_{i}(g)$, with each $U_{i}(g)$ acting only on fermions at site $i$. The conventional solution to the Ginsparg-Wilson equation \cite{PhysRevD.25.2649} is not on-site, and this is precisely why chiral symmetry breaking is anomalous. In contrast, we are interested in models without anomalies and with on-site symmetry.  

We achieve a solution to this problem by using the mirror fermion approach described in the next section. We first create a lattice regularization of both
the chiral theory and its mirror conjugate and then introduce interactions
induced by a Higgs field that gap out only the mirror theory.  In particular,
we show that a space-time random Higgs field (which preserves $SU(2)$ symmetry on average) can gap the fermion spectrum. We then use topological arguments to argue that the resulting effective Higgs action can be gapped as well. 

The lattice theory can be
easily gauged, and the gauged theory is guaranteed to be $G$-gauge invariant
({\it i.e.} there is no gauge anomaly), due to the on-site $G$-symmetry and the locality. For weak gauge coupling, the gauged lattice path integral
produces a low energy effective chiral fermion gauge theory and the mirror sector will remain gapped, though we do not study the gauged theory (dynamically or otherwise). Furthermore, since there are finitely many degrees of freedom per site, the path integral is well
defined for any finite space-time volume and the chiral fermion
theory is fully regulated. In this case, the chiral fermion theory is defined
non-perturbatively.

\section{The Mirror Fermion Approach and the Bulk-Boundary Correspondence}\label{chap3:sec:Mirror}

\begin{figure}
    \centering
    \includegraphics[width = .75\textwidth]{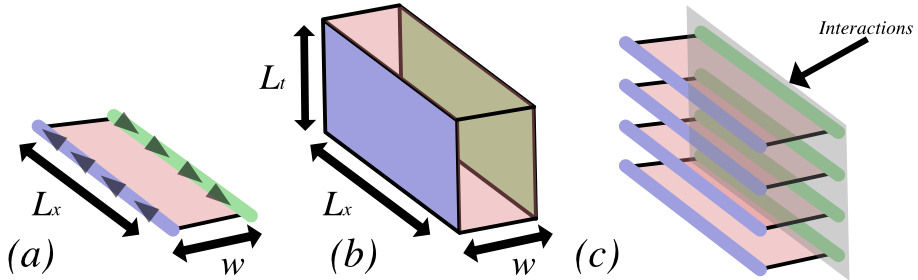}
    \caption[The Mirror Approach]{\co. We define a chiral lattice theory as either \emph{(a)} a spatial model or \emph{(b)} a spacetime model by considering a thin slab of material. A chiral theory appears on one edge (purple), while the mirror of the chiral theory appears on the other (green), and they are separated by a bulk (pink). \emph{(c)} We then try to gap out the mirror sector using interactions, leaving only the chiral theory.}
    \label{ch3:fig:slabs}
\end{figure}

Now we can take the ideas behind topological order and SPT phases and build an approach towards defining a chiral theory on a lattice. We have seen that the major obstacle to defining a successful chiral field theory is the anomaly, and that one way to resolve this is to define the theory as an edge theory. Let us formalize this slightly: we consider spatial model consisting a thin slab of system of $L_x \times w$, as shown in Figure \ref{ch3:fig:slabs}. We take periodic boundary conditions in the $x$ direction and open boundary conditions in the other, so that $w$ is the distance between the two edges and $L_x$ is the length of the periodic edges. As before, a chiral field theory (e.g. a left-moving mode) appears on one edge, while its mirror conjugate theory (e.g. a right-moving mode) appears on the other edge. 

When taking a thermodynamic limit, we scale the lengths $L_x$ and, in a spacetime lattice, $L_t$. However, we keep $w$ fixed. This ensures that the system is truly $1+1$d in the thermodynamic limit.

Now it would seem that we have successfully defined a lattice chiral theory. However, while it is true that we have a chiral theory on one edge, this model also contains the mirror conjugate of the chiral theory on the opposite edge, and the total low-energy theory remains non-chiral. Moreover, the chiral theory and its mirror are still coupled, as in the presence of a background gauge field charge may tunnel from one side to another. The key then is to choose an anomaly-free chiral theory, so that the edges decouple.

However, even if the edges are decoupled, the low-energy sector still contains the chiral theory and its mirror conjugate and so is non-chiral. This is exactly where the results of the previous chapter are crucial: it has been conjectured \cite{Wen:2013oza}, and in some cases proven \cite{Wang:2013yta}, that being free of anomalies is equivalent to having trivial bulk ordering. In turn, this implies that the mirror edge can be gapped. Hence we can introduce interactions to the mirror edge to gap it out, and all that should remain is the chiral theory. This is the mirror fermion approach \cite{Wen:2013ppa, PhysRevB.91.125147, Grabowska:2015qpk, Montvay:1992eg, Giedt:2007qg,
PhysRevD.94.114504} we mentioned previously. It is related to the Eichten-Preskill approach \cite{Mirror1}, which seeks to use composite fermions to gap out a mirror edge, but the sufficient conditions we propose are much more strict (it it has been shown \cite{Wen:2013ppa} that the Eichten-Preskill conditions cannot be strict enough, as they propose to gap out an anomalous theory). This chapter will use this approach to define an anomaly-free $SU(2)$ gauge theory on the lattice.

\section{Anomaly Cancellation}

We require that our edge theory be free of all anomalies. As discussed in Chapter \ref{chap:CFTQM}, this implies that the bulk theory is trivially ordered. Confirming that a theory is free of all anomalies is not generally an easy task. The cancellation of all Adler-Bell-Jackiw (ABJ) \cite{Bell1969,PhysRev.177.2426} type anomalies can be ensured using the usual anomaly cancellation conditions \cite{Peskin:257493, Wang:2013yta}, which examine the Lie Algebra of the gauge group $G$ to provide powerful constraints (see Section \ref{ch4:sec:CM_gen}). However, anomalies beyond those detectable from the Lie Algebra can still occur (e.g. \cite{WITTEN1982324}) which result from the non-trivial homotopic structure of the gauge group. For our $1+1d$ system defined on the edge of a $2+1d$ bulk, we ought to na\"{i}vely require that $\pi_{n}(G)=0$, $n\leq3$. 

In practice, our sufficient condition is slightly more forgiving. Consider a $d+1$ dimensional theory with a chiral theory on one edge, the mirror conjugate theory on the other, and an otherwise gapped bulk. In order to gap out the mirror edge, we couple the fermion fields there to a Higgs field $\phi$ transforming in the fundamental representation of $G$. Suppose that, in the symmetry-breaking ($\phi=\text{const.}$) case, the Higgs field gaps out the mirror edge and breaks the symmetry group $G$ symmetry down to $G_\text{grnd}$. If $\pi_n(G/G_\text{grnd})=0$ for $n\leq d+2$, then the connection with entangled states conjectures that we should be able to restore the symmetry by demanding that $\phi$ be in a disordered phase. In this paper, we provide numerical support for this conjecture for a specific $1+1$d theory. 

We take the Lie Group $G=SU(2)$. To ensure ABJ anomaly cancellation, we consider the $U(1) < SU(2)$ $S^z$ subgroup of $SU(2)$ and ensure that it cancels. This requires:
\begin{equation}
\sum_{L}\frac{4}{3} s_i(s_i+1)(2s_i + 1) = \sum_{R}\frac{4}{3} s_i(s_i+1)(2s_i + 1)
\end{equation} 
We also require the cancellation of all gravitational anomalies, which is equivalent to:
\begin{equation}
\sum_L (2s_i+1) = \sum_R (2s_i+1)
\end{equation}
We do not consider any other symmetries for now. 

The simplest chiral $SU(2)$ representation that satisfies the anomaly cancellation conditions is $1_{R}\oplus(0_{R})^{5}\oplus(\nicefrac{1}{2}_{L})^{4}$, where subscripts indicate a collection of left or right-movers. Topologically, $SU(2)\simeq S^{3}$, and so while $\pi_{1}(SU(2))=\pi_{2}(SU(2))=0$, $\pi_{3}(SU(2))=\mathbb{Z}$. Fortunately, this simply reflects the possibility of a Wess-Zumino-Witten (WZW) \cite{WESS197195} term, and corresponds to the perturbative ABJ anomalies which are absent in our model by design. Hence we can replace the requirement that $\pi_{d+2}(G/G_\text{grnd})=0$ by the weaker condition $\pi_{d+2}(G/G_\text{grnd})=\mathbb{Z}$, given that the theory is free of all ABJ anomalies. (In this paper, $d=1$, $G=SU(2)$, and $G_\text{grnd}=1$.)  Although there is no $SU(2)$ symmetric mass term, we will show that the $SU(2)$ chiral fermion theory can be fully gapped by strong interactions without breaking the $SU(2)$ symmetry.

Note that this sufficient condition is far from necessary. A theory with topological defects in the Higgs field could form a gas of defects and (possibly) still gap out the mirror sector. But this restrictive, sufficient condition keeps the theory simple, and it is more than enough for an interesting system. In fact this condition can regularize far more complicated and topical theories: a similar proposal uses the same condition to suggest a regularization for an $SO(10)$ gauge theory in $3+1$ dimensions \cite{Wen:2013ppa}.

\section{Lattice Model}\label{ch4:sec:Lattice}

We consider a lattice path integral in imaginary time, which has a form
\begin{align}
\label{ch4:eq:model}
 Z &= \int [\prod_i \phi_i] 
\text{e}^{-S_\phi (\phi)}
\int \prod_i [\text{d} \psi_{ia} \text{d} \psi_{ia}^\dag]
\text{e}^{-\psi_{ia}^\dag  M^{1+1D}_{ia,jb}(\phi) \psi_{jb}}
\nonumber\\
&= \int D[\phi(x)] 
\text{e}^{-S_\phi (\phi)} \text{Det}[M^{1+1D}(\phi)]
\end{align}
where $\psi_{ia}$ is the fermion field, $\phi_{i}$ is the Higgs field on the
site-$i$ of 1+1D space-time lattice, and index $a$ labels different components
of the fermion field.  Here $\phi_{i} \in SU(2)$ and $\psi_{a}$ is a
representation of $SU(2)$. Also, the $\phi$-action $S_\phi (\phi)$ and the
$\phi$ dependent matrix $M^{1+1D}_{ia,jb}(\phi)$ have a $SU(2)$ symmetry.  We
claim that if we choose $S_\phi (\phi)$ and $M^{1+1D}_{ia,jb}(\phi)$ properly,
the above model is described by the following $SU(2)$ chiral fermion effective
theory 
\begin{align}
\label{ch4:eq:cferm}
 {\cal L}=
\Psi_{R \alpha}^\dag ( \partial_\tau - v^R_\alpha \text{i}\partial_x) \Psi_{R \alpha}
+
\Psi_{L \alpha}^\dag ( \partial_\tau + v^L_\beta \text{i}\partial_x) \Psi_{L \alpha}
\end{align}
at low energies, where $\alpha=1,\cdots,8$.  $\Psi_{R \alpha}$ and $\Psi_{L
\alpha}$ transform as two different $SU(2)$ reducible representations:
$\Psi_{R \alpha}$ is formed by one $SU(2)$ triplet and five $SU(2)$ singlets,
while $\Psi_{L \alpha}$ is formed by four $SU(2)$ doublets. The velocity parameters in eq. \ref{ch4:eq:cferm} are all positive, but may different in magnitude. Our model breaks parity, which is already prohibited by the Nielsen-Ninomiya theorem, while promoting this to full Lorentz invariance will be the subject of a future work. 

Our lattice model consists of a hopping part and a Yukawa coupling to a (quenched) Higgs field. We label the 1+1D space-time lattice site by a pair $i=(i_\tau,i_x)$, where
$i_x=1,\cdots, L_x$ and $i_\tau=1,\cdots, L_\tau$. We label the fermion species
by three indices $a=(o,i_w,\alpha)$, where $o=R,L$, $i_w=1,\cdots,L_w$ and
$\alpha=1,\cdots,8$.  We choose the fermion lattice action to have a form
\begin{align}
\label{ch4:eq:L3D}
&\ \ \ \ \psi_{ia}^\dag  M^{1+1D}_{ia,jb}(\phi) \psi_{jb}
\nonumber\\
&=
\psi_{R,i_\tau i_x i_w,\alpha}^\dag  
M^{1+2D,R}_{i_\tau i_x i_w;j_\tau j_x j_w}
\psi_{R,j_\tau j_x j_w,\alpha} 
\nonumber\\
&
+
\psi_{L,i_\tau i_x i_w,\alpha}^\dag  
M^{1+2D,L}_{i_\tau i_x i_w;j_\tau j_x j_w}
\psi_{L,j_\tau j_x j_w,\alpha}
\nonumber\\
&
+[
\psi_{R,i_\tau i_x i_w,\alpha}^\dag  
M^{1+2D,RL}_{i_\tau i_x i_w;\alpha\beta}(\phi)
\psi_{L,i_\tau i_x i_w,\beta} + h.c.].
\end{align}
Here $\psi_{R,\alpha}$ is formed by one $SU(2)$ triplet and five $SU(2)$
singlets.  $\psi_{L,\alpha}$ is formed by four $SU(2)$ doublets.  This
fermion action can be viewed as a fermion action on 2+1D space-time, where the
$w$-direction has a finite thickness $L_w$ measured in lattice spacing.

Viewing the above as a 2+1D system and following the mirror fermion
approach, \cite{Mirror1, Montvay:1992eg, Giedt:2007qg, PhysRevD.94.114504} we
choose $M^{1+2D,R}_{i_\tau i_x i_w;j_\tau j_x j_w}$ ($M^{1+2D,L}_{i_\tau i_x i_w;j_\tau j_x j_w}$) such that the fermions $\psi_{R,\alpha}$ are gapped in the 2+1D bulk, and have 8 massless right-moving (left-moving)
modes on the $i_w=1$ boundary and 8 massless left-moving (right-moving) modes on the $i_w=L_w$
boundary. Note that $M^{1+2D,R}_{i_\tau
i_x i_w;j_\tau j_x j_w}$ and $M^{1+2D,L}_{i_\tau i_x i_w;j_\tau j_x j_w}$ are
independent of the $SU(2)$ Scalar field $\phi$. Their detailed expression
will be given in Section \ref{ch4:sec:Lattice}. Using condensed matter terminology,
$M^{1+2D,R}_{i_\tau i_x i_w;j_\tau j_x j_w}$ and $M^{1+2D,L}_{i_\tau i_x
i_w;j_\tau j_x j_w}$ both describe a single filled Landau level but with opposite
magnetic fields. 

When viewed as a 2+1D system, the  $SU(2)$ scalar field $\phi_{i_\tau,i_x}$
only lives on the $i_w=L_w$ boundary and only couples to the fermions on the
$i_w=L_w$ boundary.  We choose the coupling form in $M^{1+2D,RL}_{i_\tau i_x i_w;\alpha\beta}(\phi)$
such that a constant $SU(2)$ scalar field $\phi$ can give all the right-moving and
left-moving chiral fermions on the $i_w=L_w$ boundary a finite mass $M_\phi$ and then consider configurations where $\phi$ varies smoothly over the $1+1$d spacetime. 

\subsection{Hopping Terms}
\begin{figure}
\centering
\includegraphics[width=.7\textwidth]{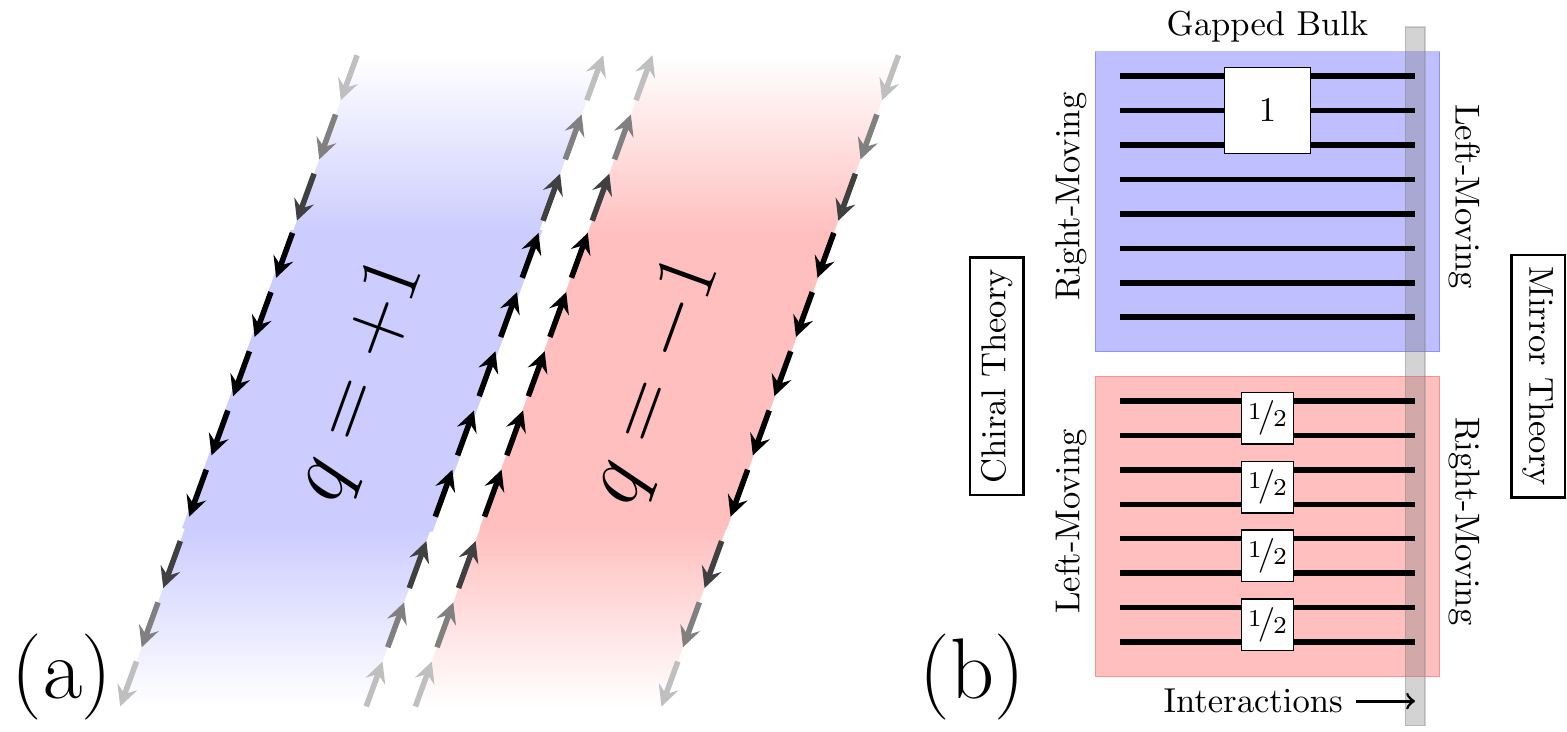}
\caption[Schematic of the $SU(2)$ Model]{\co. \emph{(a)} Layers of fermion hoping model with Chern number $+1$ and $-1$, which give rise to right-hand and left-hand chiral fermions on the 1+1D boundary.  \emph{(b)} General schematic of our model. Each black line represents a layer of our hopping model on a $2+1$D dimensional lattice.  We stack $8$ Chern number $+1$ layers (blue) with $8$ Chern number $-1$ layers (red). Next, we organize the fermions on those layers into $SU(2)$ representations. Each representation is illustrated by a white square with the spin label except the trivial spin-0 representations which we omit. This results in a $1_{R}\oplus (0_{R})^{5}\oplus (\nicefrac{1}{2}_{L})^{4}$ theory on the left edge and its mirror $(L\leftrightarrow R)$ theory on the right edge. We couple the fermions on the right edge to a Higgs field that gaps out the right edge, leaving only the chiral theory on the left edge.}\label{ch4:fig:modelschem}
\end{figure}

Here we present the details of our lattice model. Our model does not directly come from a discretization of the Weyl or Dirac Lagrangian; instead, we create a spacetime lattice description of states with nonzero Chern number, like those discussed in Chapter \ref{chap:CMQFT}.

We first create a hopping model with two edges. One edge will be described by a gapless chiral theory, with the other edge described by the mirror conjugate gapless theory and the bulk otherwise gapped. For $1+1$D chiral theories, the left- and right-handed excitations simply become spinless left- and right-moving complex fermions. In our model, the 8 left- and 8 right-moving fermions carry the following $SU(2)$ representations $1_{R}\oplus (0_{R})^{5}\oplus (\nicefrac{1}{2}_{L})^{4}$, which, as discussed in the previous Section, are the simplest which satisfy the ABJ anomaly cancellation conditions. 

Our approach begins by creating a fermion hopping model on a 2+1D space-time
lattice which contains 16 layers. Each layer has a finite gap in the 2+1D bulk
and one chiral complex fermion mode on the 1+1D boundary.  8 layers have
right-handed chiral fermions on the edge, and the other 8 layers have left-handed
chiral fermions.  We choose the 2+1D space-time lattice to be a thin slab of
size $L_t\times L_x\times L_w$. We will fix $L_{w}$ while taking
$L_t=L_x\equiv L \to \infty$, so the system is effectively $1+1$D.  One surface
of the slab is the normal sector and the other surface is the mirror sector
(see Fig. \ref{ch4:fig:modelschem}). We will add interactions between the fermions
in the mirror sector ({\it i.e.} only on one surface). The interactions are
induced by an $SU(2)$ Higgs field in the fundamental representation, which breaks the
$U(8)\times U(8)$ symmetry of the non-interacting system down to $SU(2)$.  The 8
right-moving and 8 left-moving chiral fermions form the following $SU(2)$
representations: $1_{R}\oplus (0_{R})^{5}\oplus (\nicefrac{1}{2}_{L})^{4}$ in
the mirror sector (and $1_{L}\oplus (0_{L})^{5}\oplus
(\nicefrac{1}{2}_{R})^{4}$ in the normal sector, see Fig.
\ref{ch4:fig:modelschem}).

\begin{figure}
\centering
\includegraphics[width=.6\textwidth]{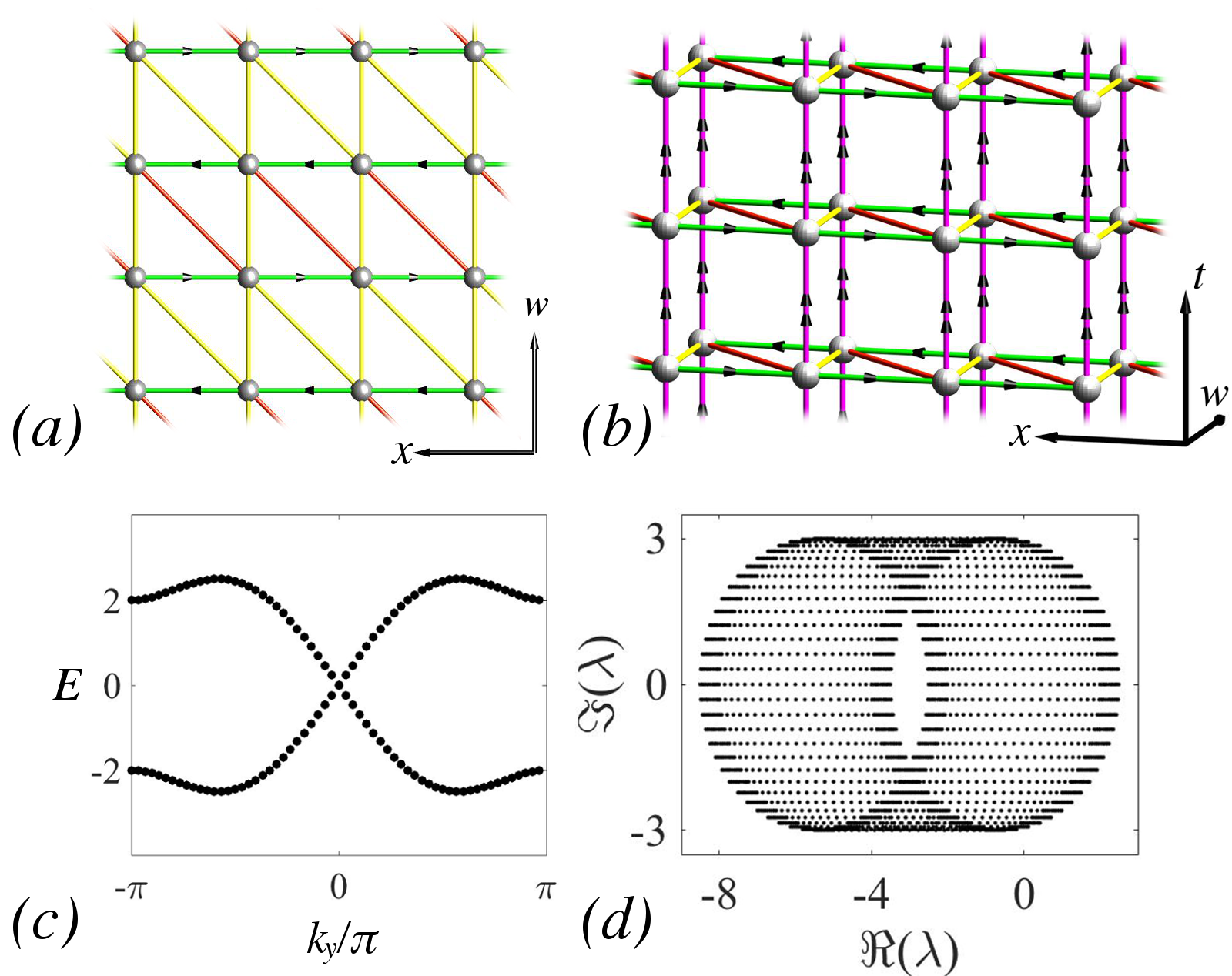}
\caption[Hopping Model for the $SU(2)$ Mirror Approach]{\co. \emph{(a)}: Chern number $+1$ hopping model. Fermion sites are shown as spheres, with hopping terms as links. A yellow link indicates a hopping of $+1$, a red link a hopping of $-1$, and a green link a hopping of $+i$ in the direction of the arrow and $-i$ in the opposite direction. Hopping around any plaquette generates a phase of $\pi$, hence with $1$ fermion per site this is a Chern number $1$ (Integer Quantum Hall) state. \emph{(b)}: Spacetime lattice with $L_{w}=2$. Each spatial component is just a slice of a IQH state shown in \emph{(a)}, while the purple links represent a Hopping of $t$ in \emph{only} the direction of the double arrows. In addition, each site is given the onsite chemical potential $-t\psi^{\dagger}_{i}\psi_{i}$; we later set $t=3$. The Hopping matrix corresponding to this model is our spacetime Lagrangian. \emph{(c)} Dispersion relation for the spatial lattice with $L_{w}=2$. \emph{(d)} Complex Eigenvalues of the spacetime hopping model. }\label{ch4:fig:LatticeDetail}
\end{figure}

We split our hopping model into spatial and temporal hopping, i.e. splitting the imaginary time Lagrangian into `$\partial_{\tau}$' and Hamiltonian terms as $\mathcal{L}=\partial_{\tau}-H$. The spatial (Hamiltonian) terms, detailed below, are motivated by Chern number states discussed previously, whereas the temporal hopping is provided by a doubling-free hopping term. Our approach breaks discrete rotational (Lorentz) symmetry, which only reappears at low energies. While all fermions will still have linear velocities, it is possible that, after the mirror edge is gapped, fermions on the chiral edge may have differing velocities. 

The spatial component of our lattice is provided by the Chern number $\pm 1$ states shown in Figure \ref{ch4:fig:LatticeDetail}a, and leads to an $L_{w}L_{x}\times L_{w}L_{x}$ matrix. We can write this matrix explicitly as:
\begin{multline}
\footnotesize
H_{L, (i_{x}, i_{w}, i_{\tau}) , (j_{x}, j_{w}, j_{\tau})}=\\\delta_{i_{\tau}, j_{\tau}}\left((-1)^{w}(i\delta_{i_{x}, j_{x}+1}\delta_{i_{w}, j_{w}}+\delta_{i_{x}, j_{x}+1}\delta_{i_{w}, j_{w+1}})\right.\\\left.+\delta_{i_{x}, j_{x}}\delta_{i_{w}, j_{w}+1}\right)+\text{H.C.}
\end{multline}
However, it is easier to understand either using Figure \ref{ch4:fig:LatticeDetail}a. The essential feature is that the hopping clockwise around any triangular half-plaquette yields a phase of $\frac{\pi}{2}$, so that at the lattice model describes particles hopping in an applied magnetic field. The physics of particles hopping on a lattice with varying fluxes is rich; here we only study particles with $\pi$ flux per plaquette. At half filling (ie filling only the lower band), there is one flux quantum per fermion. The model we have chosen leads to a lattice version of Landau levels: semiclassically, particles will travel in closed cyclotron orbits, leading to level quantization. In the bulk (i.e. with periodic boundary conditions in both the $x$ and $w$ directions) the model we have chosen has two bands, at energies $\pm 1$, with the chemical potential set to zero. 

In condensed matter physics, our lattice model describes a state known as a Chern number $+1$ state, due to the quantized curvature of the eigenstates of the lower (occupied) band. Crucially, states with nonzero Chern number cannot be smoothly deformed without closing the band gap to a state with zero Chern number, such as the vacuum. If we take open boundary conditions in the $w$ direction, there must be gapless chiral edge modes localized at the edges of the lattice. In this case, there is is a right mover at $w=1$, which we call the chiral edge, at a left mover at $w=L_{w}$, which we call the mirror edge. In this way, we have a condensed matter-inspired realization of the `Domain Wall' idea; instead of a standard lattice Dirac operator with a region with positive mass term situated next to a region with negative mass term, we have a Chern number $+1$ state next to the (zero Chern number) vacuum.

The essential features of this model can also be understood in momentum space. Taking the Fourier transform in the $x$-direction, with open boundary conditions in the $w$ direction, the Hamiltonian becomes an $L_{w}\times L_{w}$ matrix of the form:
\begin{equation*}
H_{R}=
{
\footnotesize
\left(\begin{array}{ccccc}2\sin k_{x} & 1-e^{ik_{x}} & 0 & 0 & ... \\1-e^{-ik_{x}} & -2\sin k_{x} & 1+e^{ik_{x}} & 0 & ... \\0 & 1+e^{-ik_{x}} & 2\sin k_{x} & 1-e^{-ik_{x}} & ... \\0 & 0 & 1-e^{-ik_{x}} & -2\sin k_{x} & \ddots \\\vdots & \vdots & \vdots & \ddots & \ddots\end{array}\right)
}
\end{equation*}
Near $k_{x}=0$, this becomes:
\begin{equation*}
H_{R}\approx
{
\footnotesize
\left(\begin{array}{cccccc}2k_{x} & 0 & 0 & 0 & 0 & ... \\0 & -2k_{x} & 2 & 0 & 0 & ... \\0 & 2 & 2k_{x} & 0 & 0 & ... \\0 & 0 & 0 & -2k_{x} & 2 & ... \\0 & 0 & 0 & 2 & 2k_{x} & \ddots \\\vdots & \vdots & \vdots & \vdots & \ddots & \ddots\end{array}\right)
}
\end{equation*}
This Hamiltonian has two low energy modes: a right mover at $w=1$ (the chiral edge), which appears above, and a left mover other at $w=L_{w}$ (the mirror edge), which would appear in the lower-right corner above (at $k_{x}=0$ for even $L_{w}$). Otherwise, one can check that the system has a gap of approximately $2$ for all $k_{x}$. If we were to implement periodic boundary conditions in the $w$ direction, we would add a term $1\pm e^{ik_{x}}$ in the upper right corner and $1\pm e^{-ik_{x}}$ in the lower left, and again one can check that this leads to a completely gapped Hamiltonian. We also define a left moving Hamiltonian $H_{L}=H_{R}^{*}$ in real space or $H_{L}(k_{x})=H_{R}(-k_{x})$ in momentum space, which has a left mover at $w=1$ and a right mover at $w=L_{w}$.

The above Hamiltonian, with open boundary conditions in the $w$ direction, provides the spatial hopping $H$ in our Lagrangian ``$\partial_{\tau}-H$''. Now we need to add the ``$\partial_{\tau}$'' term. In lattice gauge theory, there are many standard ways to construct fermion
hopping models, but many of them have fermion doubling in time direction. We use the trick in Ref. \cite{negele1988quantum} to prevent 
time-direction fermion doubling, which introduces a hopping in frequency space of $3(1-e^{-i\omega})$. The total (one-flavor) hopping model is then given by
\begin{equation}
\label{ch4:eq:MR}
M^{1+2d}_{L, (i_{x}, i_{w}, i_{\tau}) , (j_{x}, j_{w}, j_{\tau})}=3(\delta_{i_{\tau}, j_{\tau}+1}-\delta_{i_{\tau}, j_{\tau}})\delta_{i_{x}, j_{x}}\delta_{i_{w}, j_{w}}-H_{L, (i_{x}, i_{w}, i_{\tau}) , (j_{x}, j_{w}, j_{\tau})}
\end{equation}
and similarly for $M^{1+1d}_{R}$. For example, for $L_{w}=2$, the Lagrangian Hopping matrix in frequency momentum space becomes:
\begin{equation}
M_L=
{\footnotesize
\delta_{\alpha, \beta}\left(\begin{array}{cc}2\sin k_{x} +3(1-e^{-i\omega}) & 1-e^{ik_{x}} \\1-e^{-ik_{x}} & -2\sin k_{x}+3(1-e^{-i\omega})\end{array}\right)
}
\end{equation}
where $M_{L}$ acts diagonally on the flavor indices $\alpha, \beta$. The right-handed version acts as $ M_{R}(k_x,\omega)= M_{L}(-k_x,\omega)$. These define the matrices $M^{1+2D, R}$ and $M^{1+2D, L}$ shown in eq. (\ref{ch4:eq:L3D}). 
 
 \subsection{Higgs Field and Yukawa Coupling}

The mass term that gaps out all the mirror fermions can be generated by an
$SU(2)$-Higgs field $\phi$ in the fundamental representation that breaks
$G=SU(2)$ down to $G_\text{grnd}=1$.  Topologically, $G/G_\text{grnd} =
SU(2)\simeq S^{3}$, and so while $\pi_{1}(SU(2))=\pi_{2}(SU(2))=0$,
$\pi_{3}(SU(2))=\mathbb{Z}$.  Fortunately, this simply reflects the possibility
of a Wess-Zumino-Witten (WZW) \cite{WESS197195} term, and corresponds to the
perturbative ABJ anomalies which are absent in our model by design. 

Let us use a 8-component $\psi_{R}$ to describe the 8-layers of fermions with
Chern number 1, and $\psi_{L}$ for the 8-layers with Chern number $-1$.  The
Higgs coupling on the mirror surface is given by
\begin{equation}
\mathcal{L}_{\text{int}}=\psi_{R, i_{\tau} i_{x} i_{w}, \alpha}^{\dagger}M^{1+2D, RL}_{i_{\tau} i_{x}i_{w}; \alpha \beta} \psi_{L, i_{\tau} i_{x} i_{w}, \beta}
\end{equation}
where 
\begin{equation}
M^{1+2D, RL}_{i_{\tau} i_{x}i_{w}; \alpha \beta} =
g\delta_{i_w, L_{w}}\delta_{i,j}\Theta_{L}[\phi(i_{\tau}, i_{x})]\Theta_{R}[\phi(i_{\tau}, i_{x})]^{\dagger}
\end{equation} 
where $\delta_{i_w, L_{w}}$ constrains the action to the mirror surface. We have assumed the Higgs field $\phi$ to have a unit
ampltude $|\phi|=1$. In this case, the Higgs field can be viewed as an element
in the $SU(2)$ group.  $\Theta_{R}[\phi]$ and $\Theta_{L}[\phi]$ (as well as
$\Theta_{L}[\phi]\Theta_{R}^{\dagger}[\phi]$ and
$\Theta_{R}[\phi]\Theta_{L}^{\dagger}[\phi]$) are representation
matrices of $SU(2)$ that make the above Higgs coupling $SU(2)$ invariant.  The
explicit $8\times 8$ matrices $\Theta_{R}[\phi]$ and $\Theta_{L}[\phi]$ are
listed below:
\begin{align}
\Theta_{R}[\phi] &=
\Theta_{1}[\phi] \oplus I_{5\times 5},
\nonumber\\
\Theta_{L}[\phi] &=
\Theta_{\nicefrac{1}{2}}[\phi] \oplus
\Theta_{\nicefrac{1}{2}}[\phi] \oplus
\Theta_{\nicefrac{1}{2}}[\phi] \oplus
\Theta_{\nicefrac{1}{2}}[\phi].
\end{align}
where $\Theta_s(\phi)$ is the spin-$s$ representation
(with dimension $2s+1$). Because $\psi_{R}$ contains three fermions in the spin-$1$ representation and five singlets, $\Theta_{R}^{\dagger}\psi_{R}$ is invariant under $SU(2)$, and similarly for $\Theta_{L}^{\dagger}\psi_{L}$. Hence the coupling Lagrangian, composed of $\psi_{L}^{\dagger}\Theta_{L}[\phi]\Theta_{R}^{\dagger}[\phi]\psi_{R}$ and its conjugate, is as well.

With the above Higgs coupling on the mirror surface, the total Lagrangian is given by eq. (\ref{ch4:eq:L3D}), reproduced below:
\begin{align*}
&\ \ \ \ \psi_{ia}^\dag  M^{1+1D}_{ia,jb}(\phi) \psi_{jb}
\nonumber\\
&=
\psi_{R,i_\tau i_x i_w,\alpha}^\dag  
M^{1+2D,R}_{i_\tau i_x i_w;j_\tau j_x j_w}
\psi_{R,j_\tau j_x j_w,\alpha} 
\nonumber\\\tag{\ref{ch4:eq:L3D}}
&
+
\psi_{L,i_\tau i_x i_w,\alpha}^\dag  
M^{1+2D,L}_{i_\tau i_x i_w;j_\tau j_x j_w}
\psi_{L,j_\tau j_x j_w,\alpha}
\nonumber\\
&
+[
\psi_{R,i_\tau i_x i_w,\alpha}^\dag  
M^{1+2D,RL}_{i_\tau i_x i_w;\alpha\beta}(\phi)
\psi_{L,i_\tau i_x i_w,\beta} + h.c.].
\end{align*}

Schematically, this is:
\begin{equation}
\psi^{\dagger}\slashed{D}\Psi=\Psi^{\dagger}\left(\begin{array}{cc}M_{R} & g\delta_{w, L_{w}}\Theta_{R}^{\dagger}\Theta_{L} \\ g\delta_{w, L_{w}}\Theta_{L}^{\dagger}\Theta_{R} & M_{L} \end{array}\right)\Psi
\end{equation} 
where $\Psi^{\dagger}=(\psi_{R}^{\dagger}, \psi_{L}^{\dagger})$.
On the mirror surface and in the continuum limit, we have
\begin{equation}
\Psi^{\dagger}\slashed{D}\Psi=\Psi^{\dagger}
\left(\begin{array}{cc}
\partial_t - \text{i} \partial_x & g\Theta_{R}^{\dagger}\Theta_{L} \\ 
g\Theta_{L}^{\dagger}\Theta_{R} & \partial_t + \text{i} \partial_x
\end{array}\right)\Psi \label{chap4:eq:mirror2}
\end{equation} 
which has the familiar field theory form. 
On the normal surface, we have
\begin{equation}
\label{ch4:eq:NS}
\Psi^{\dagger}\slashed{D}\Psi=\Psi^{\dagger}
\left(\begin{array}{cc}
\partial_t + \text{i} \partial_x & 0 \\ 
0 & \partial_t - \text{i} \partial_x
\end{array}\right)\Psi
\end{equation} 
which is the gapless chiral theory we seek. Our task now is to ensure that the Higgs interaction gaps out the mirror theory \eqref{chap4:eq:mirror2} while preserving \eqref{ch4:eq:NS} Denoting the full content of eq. (\ref{ch4:eq:L3D}) as $\Psi^{\dagger}\slashed{D}\Psi$, the full partition function of our system is now:
\begin{equation}
\begin{aligned}
Z=\int D\phi e^{-S_{\text{H}}[\phi]}\int D\Psi^{\dagger}D\Psi\exp[-\Psi^{\dagger}\slashed{D}\Psi]\\
=\int D\phi e^{-S_{\text{H}}[\phi]}\det(\slashed{D})\label{ch4:eq:PartFun}
\end{aligned}
\end{equation}
where $S_{H}[\phi]=-U[\phi]$ is the action for the Higgs and we have neglected
a proportionality constant. 

We choose the Higgs action $S_{H}[\phi]$ such that the Higgs field is in a
disordered $SU(2)$ symmetric state with correlation length $\xi>1$ (with unit
lattice spacing). If we took $\xi\to\infty$, i.e. if the Higgs field $\phi$ were a
constant (which breaks $SU(2)$ symmetry), the mirror sector would be fully
gapped and all eigenvectors of $\slashed{D}$ with eigenvalues $\lambda$ with
$|\lambda|< \Delta$ would be localized near the normal surface. The central
result of this chapter is that the mirror edge is still gapped for a finite
correlation length, and the normal sector is hardly affected by the
gapping process.

Performing the full integral of $\phi$ disordered Higgs
phase is intractable. Instead, we study the above path integral in two steps.
In the next Section, we consider a typical random space-time Higgs fields with a finite
correlation length $\xi$ which preserves $SU(2)$ symmetry on average.
We will show that the mirror sector can be fully gapped by these random Higgs
fields when $\xi \gtrsim 7$ for $g=1$ and system sizes up to $L=80$. Then we can safely integrate out the mirror fermions
and obtain a non-linear $\sigma$-model for the Higgs field. Lastly, we argue
that the non-linear $\sigma$-model can be in the $SU(2)$ symmetric disordered
phase with a gap, the only low energy excitations coming from the massless
chiral fermions in the normal sector which is unaffected by the gapping process
in the mirror sector.

The choice of dynamical Higgs field is of central importance. If at any point $\phi$ fluctuates too rapidly, a low-energy fermion mode may be localized there. An ideal configuration would have $|\nabla\phi(x)|=\text{const.}>0$. In the lattice model, we first choose a random $\phi(x)$ and then smooth it, taking care to apply the most smoothing in regions of largest $|\nabla\phi|$. This nonlinear smoothing process leads to a $\phi$ with nearly constant but nonzero $|\nabla \phi|$. In a continuum sense, this can be achieved by a an action of the form $S(\phi)=-\int d^{2}x(-|\nabla\phi|^{2}+|\nabla\phi|^{4}+...)$. Since we cannot yet perform the integral over Higgs configurations, we do not specify a lattice action at this point.

\section{Gapping Process}\label{ch4:sec:Numerical}

We first show that introducing a Higgs field opens a spectral gap in the fermion determinant. Next, we confirm that this gap remains in a thermodynamic limit and check that the gapped mirror edge is indeed decoupled from the chiral edge. Finally, we address the resulting non-linear sigma model for the Higgs field obtained by integrating out the mirror fermions. 

\subsection{Numerical Results}

\begin{figure}
\centering
\includegraphics[width=\textwidth]{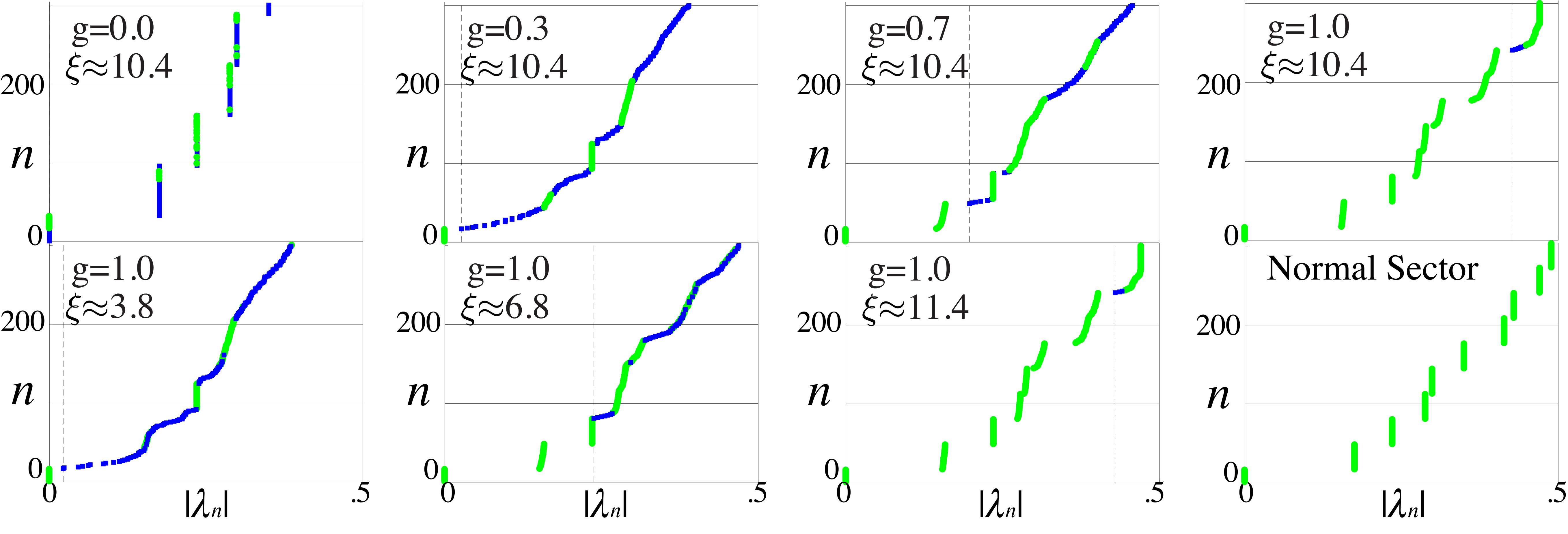}
\caption[Integrated Density of States in the $SU(2)$ Model]{\co. Integrated Density of States (IDOS) for various choices of the
coupling strength $g$ and the correlation length of the Higgs field $\xi$ with
$L=80$ and $L_w=2$. The horizontal axis is the eigenvalue magnitude, while the vertical axis is the eigenvalue number (in order of increasing magnitude). We first find the low-magnitude eigenvalues of the matrix $M^{1+1D}$ 
in eqn. (\ref{ch4:eq:model}), and order them
$|\lambda_{n}|\leq |\lambda_{n+1}|$. Each plot shows $n$ vs.  $|\lambda_{n}|$.
States localized on the normal surface (the weight of the eigenvector on the
normal surface is larger than $0.8$) are denoted by green circles, others are
denoted by blue squares. A black dashed line indicates the magnitude of the
lowest mirror-surface eigenvalue. In the upper panels, we use the same
configuration of the Higgs field with $\xi\approx 10.4$ and different Higgs
couplings $g$.  For $g=0$, the IDOS is two copies of our chiral edge theory. As
the interaction strength increases, states not localized to the normal surface
are gapped out until at $g=1$ only the normal sector remains (below
$|\lambda|\lesssim.45$). In the lower panels, we fix $g=1$ and smooth the Higgs
field, increasing the correlation length from $\xi\approx .6$ until $\xi\approx
11.4$. For $\xi\approx 3.8$, there are many low-lying mirror states. 
As we increase $\xi$, the mirror gap increases until at $\xi\approx 11.4$ no
mirror-edge states remain (below $|\lambda|\lesssim.45$). Normal sector: In the
lower-right panel, we calculate the IDOS for the normal sector
using \eqref{ch4:eq:NS}.  This IDOS matches that of our model with Higgs coupling
$g=1$ and $\xi>10$. Thus the low lying modes of our model are exactly described
by the normal sector only, and the mirror sector is fully gapped. The slight $\lambda_n$ renormalization can be mitigated by increasing $L_{w}$.}\label{ch4:fig:IDOS}
\end{figure}

To see what the process of gapping out the mirror
theory looks like, let us first fix the system size $L_{x}=L_{t}=L=80$ and Higgs
correlation length $\xi$. Figure \ref{ch4:fig:IDOS} (upper panels) show the
integrated density of states (the number of eigenvalues whose absolute values
are less then $|\lambda|$) as we turn on the interaction from $g=0$ to $g=1$.
At $g=0$, there are 32 gapless modes---16 from the chiral theory and 16 from
the mirror conjugate. As we turn on the interaction, the fluctuating Higgs
field smoothly gaps out the mirror theory modes, leaving only the chiral theory
at low energies. Using eq. (\ref{ch4:eq:MR}), we can calculate the integrated density of states (IDOS) for the normal sector alone, which we present in the lower-right panel of Figure \ref{ch4:fig:IDOS}. Comparing the normal sector IDOS to the lattice results, we see that the low-lying modes of the lattice models are indeed described by the unaffected normal sector alone.

We can determine if a given eigenstate $\ket{v}$ is localized on the normal edge by examining the expectation value of the $w$ position operator with respect to $\ket{v}$. If a state were indeed perfectly localized at $w=1$, then we would have $\braket{w}_{v}=1$. For our purposes, we say that a state is localized on the normal edge if $\braket{w}_{v}<1.2$. While the precise value of this cutoff is arbitrary, for larger $L_{w}$, this is a very strong condition, as it implies that the wavefunction of $\ket{v}$ vanishes except for very near the $w=1$ edge and implies that $\ket{v}$ has essentially no probability near the $w=L_{w}$ edge. 

Next, we fix $g=1$ and instead vary the Higgs field correlation length $\xi$ in
Figure \ref{ch4:fig:IDOS} (lower panels). For $\xi \lesssim 4$, the mirror edge
modes appear at small magnitude, though their momentum structure is wiped out
by the rapidly fluctuating Higgs field. As $\xi$ is increased, the mirror
theory modes again are driven to large magnitude, leaving only the chiral
theory at small eigenvalues. 

\begin{figure}
\centering
\includegraphics[width=\textwidth]{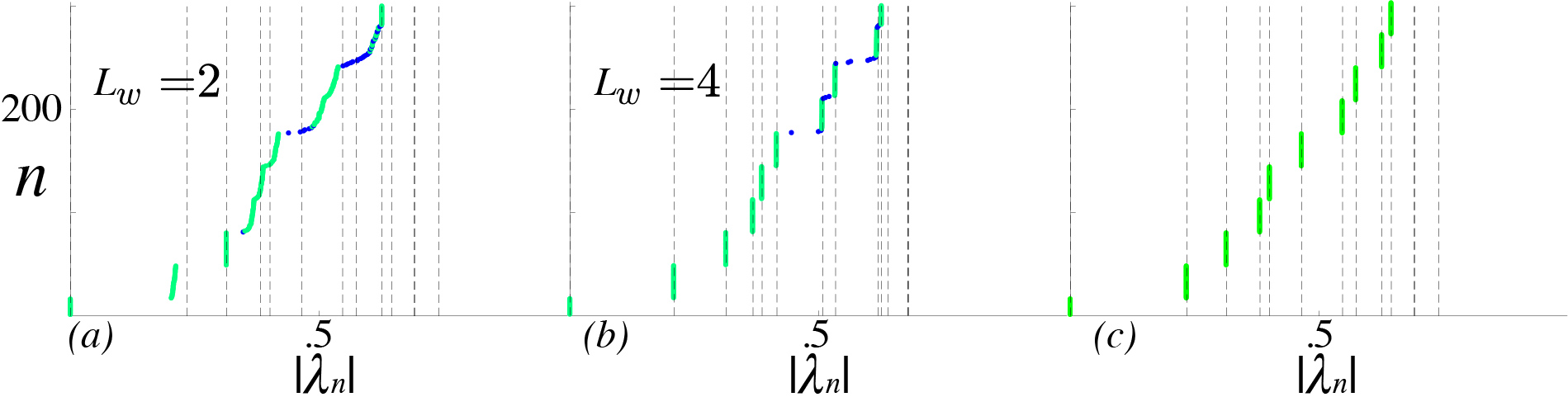}
\caption[Strip Width and the Integrated Density of States]{\co.  \emph{(a-b)}: Integrated Density of States (IDOS) for two values of the width $L_{w}$. The horizontal axis is the eigenvalue magnitude, while the vertical axis is the eigenvalue number (in order of increasing magnitude). We fix $L_{x}=L_{t}=60$, choose a single Higgs configuration $\phi$ with correlation length $\xi \approx 8.1$, and examine the low-energy IDOS as we vary $L_{w}$. States localized on the normal edge (i.e. with $\braket{w}<1.2$) are denoted by green circles, others are denoted by blue squares. Vertical dotted lines indicate the magnitude of the free (no Higgs) eigenvalues. For $L_{w}=2$ \emph{(a)}, the gapless modes remain but the momentum structure is substantially affected by the gapping of the mirror edge. For $L_{w}=4$, the low-lying states match with their free (Higgs-less) eigenvalues very well. No mass, and minimal momentum renormalization, is transmitted to the normal sector by the gapping of the mirror sector. The computation can be extended to larger values of $L_{w}$ at high computational cost. \emph{(c)} IDOS for the normal sector, calculated artificially using the momentum-space Lagrangian. The $L_{w}=4$ lattice IDOS matches this artificial normal sector IDOS, though higher-magnitude modes are shifted upwards by the lowest-lying gapped mirror modes.}\label{ch4:fig:Decouple}
\end{figure}

The mirror edge is effectively decoupled from the chiral edge. To see this, we fix $L_{x}=L_{t}=60$, choose a single Higgs configuration $\phi$ with correlation length $\xi \approx 8.1$, and examine the low-energy IDOS as we vary $L_{w}$. In Figure \ref{ch4:fig:Decouple}(a-b), we plot the IDOS for $L_{w}=2,4$. States localized on the normal edge are marked in green while all others are marked in blue. Further, we have plotted the $g=0$ (no Higgs coupling) eigenvalues as vertical dotted lines. For $L_{w}=2$, we see that while the mirror sector is indeed gapped out, the low-lying excitation energies are smeared away from their $g=0$ values. For $L_{w}=4$, the normal sector excitation energies are almost exactly their $g=0$ eigenvalues, decoupling from the gapped mirror sector with no residual mass. One can compute this for higher values of $L_{w}$ (though at significant computational cost) and find that the low-lying eigenvalues almost exactly match their $g=0$ values. Additionally, we can calculate `by hand' what the integrated density of states for the normal sector alone and finite lattice size would look like. We plot this in Figure \ref{ch4:fig:Decouple}c. Comparing the lattice results (a-b) to the `by-hand' calculation (c), we see that at low magnitude, the lattice results reproduce the expected `by-hand' result very well for larger $L_{w}$, thus demonstrating that the normal sector decouples from the mirror sector with no residual mass. Surprisingly, we find that even just $L_{w}=2$ is enough to capture the gapping process and allows us to reach large system sizes with modest computational resources. In Figures \ref{ch4:fig:IDOS} and \ref{ch4:fig:Decouple} we took $L_{w}=2$, as we will in the rest of this paper.

\begin{figure}
\centering
\includegraphics[width=.4\textwidth]{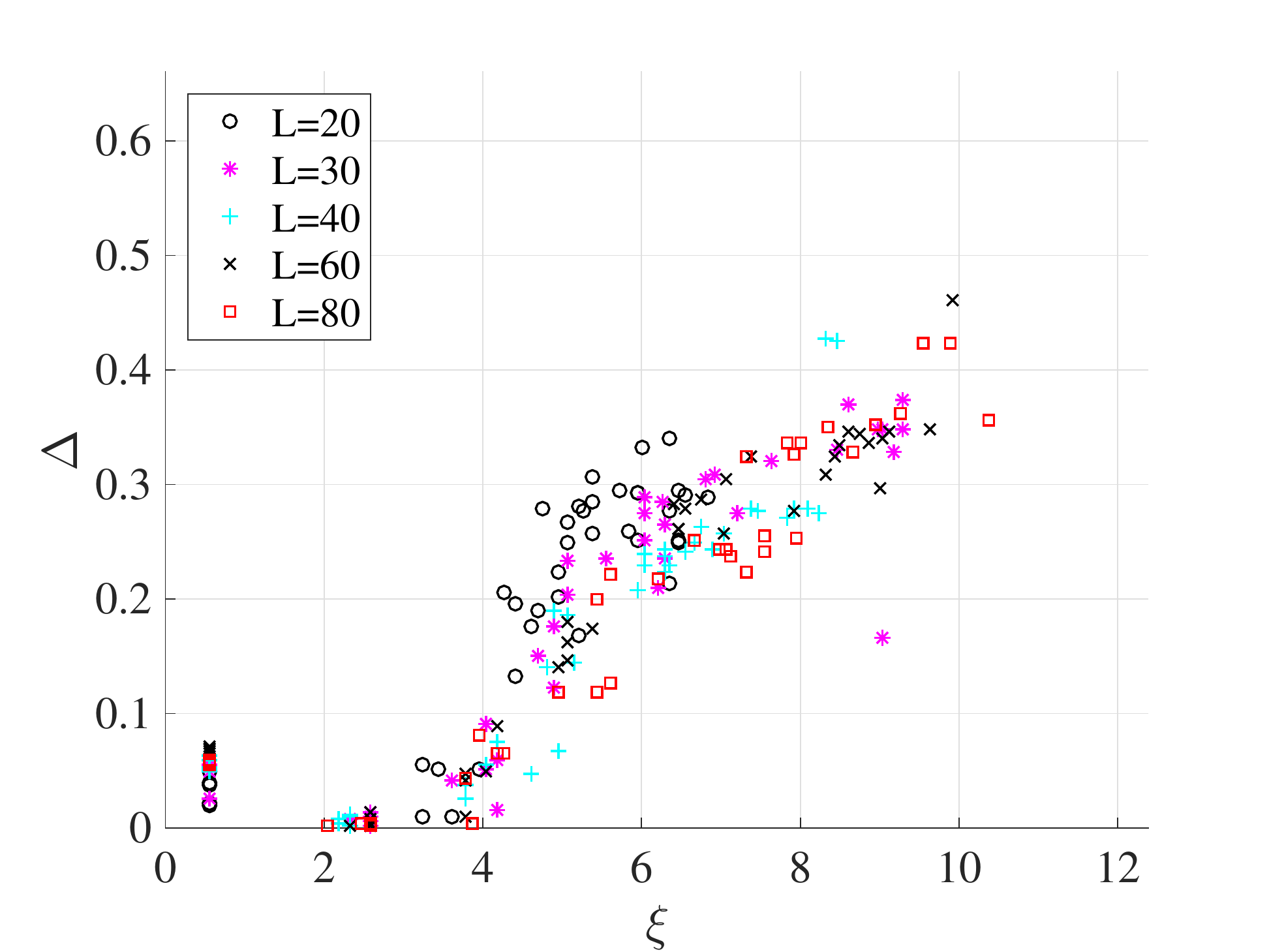}
\caption[Mirror Theory Gap]{Mirror theory gap as a function of correlation length $\xi$ for
various choices of $L$. We couple the fermions on the mirror surface to a
Higgs field fluctuating with correlation length $\xi$ and look for the smallest
magnitude eigenvalue for states not localized on the chiral theory edge.
$\xi=L$ would correspond to the usual symmetry-breaking Higgs mechanism. We
choose a finite $\xi$, independent of $L$. For $\xi \gtrsim 8$, the mirror edge
is gapped with $\Delta\approx .35$, and this gap is independent of $L$ for
$L\gtrsim 40$, indicating thermodynamic behavior. Note that, in the
$\xi=L\to\infty$ `symmetry-breaking' limit, $\Delta=g=1$.}
\label{ch4:fig:thermo}
\end{figure}

The Higgs interaction in the mirror sector could leak
into the normal sector and cause residual couplings in the normal sector. 
The only relevant couplings in the normal sector would be mass terms, but by
design the $SU(2)$ chiral fermions in the normal sector have no $SU(2)$
symmetric mass terms. Hence residual couplings from the mirror sector can generate only exponentially small $SU(2)$ symmetric interactions on the chiral theory edge. Since there is no $SU(2)$ symmetric mass term, the effect of the exponentially small $SU(2)$ symmetric interactions is weak and the gapless chiral theory is effectively decoupled from the gapped mirror theory.

Now we turn to the thermodynamic behavior, fixing $g=1$ and examining the gap
given to the mirror modes. Figure \ref{ch4:fig:thermo} shows how this mirror-mode
gap scales with the system size $L$ and the correlation length $\xi$. At
$\xi\approx 8$, for system sizes $L\gtrsim40$, we see that the gap is roughly
constant at $\Delta\approx.35$, and largely independent of system size. This
indicates that we are indeed seeing thermodynamic-limit behavior and that we
have successfully gapped out only the mirror theory.

\subsection{Effective Higgs Action}

The spectral gap of $\slashed{D}$ in the mirror sector allows us to safely
integrate out the mirror fermions and obtain a 1+1D non-linear $\sigma$-model
of the Higgs field $\phi$ on the mirror surface.  Since $\slashed{D}$ is not
Hermitian, this determinant may be complex and its phase can fluctuate.  The
complex phase could give rise to the topological terms in the non-linear
$\sigma$-model which would make the Higgs effective theory gapless.  However, the
target space of the $SU(2)$ Higgs non-linear $\sigma$-model is $SU(2)=S^3$
which has no topological defects in 1+1D space-time.  The only potential
topological term in the 1+1D $S^3$ non-linear $\sigma$-model is the WZW term
from $\pi_3(S^3)=\mathbb{Z}$.  This topological term would correspond to
the ABJ $SU(2)$ anomaly.  Such a topological term does not appear since our
model is free of ABJ anomalies by design.  It is believed that a
non-linear $\sigma$-model with no topological defects or topological term can
be a gapped disordered phase that does not break any symmetry.  This is the phase we believe our model to be in, though further numerical study of nonlinear $\sigma$-models is needed to conclusively confirm this.

We have demonstrated that coupling the mirror edge to a disordered Higgs field introduces a gap in the fermion spectrum on the mirror edge, and have argued that the resulting non-linear $\sigma$-model for the Higgs field is gapped. Because the chiral theory is effectively decoupled from the mirror edge, all that remains at low energy is our low energy, chiral, $SU(2)$ theory. 

\section{Anomaly Matching and Application to a Condensed Matter System}\label{ch4:sec:CM_gen}

The model we considered in this Chapter needs a significant modification to be considered as a true condensed matter system, as it violates the spin statistics theorem (SST). In particular, the SST states that fermions have half-integer spin, while bosons have integer spin. However, we have considered a fermion model with one spin-$1$ mode. Here we propose several other models that escapes this issue. However, these new models result in larger on-site matrices which introduce further technical challenges. We also include $U(1)$ symmetry, and ensure that, as holds in condensed matter systems, all fermions have odd change.

\begin{table}[]
    \centering
    \begin{tabular}{|c|c|c|}\hline
    Representation & Left Anomaly & Left Chiral Central Charge \\ \hline
    $\nicefrac{1}{2}_L\oplus \nicefrac{9}{2}_L^2\oplus\nicefrac{3}{2}_R\oplus \nicefrac{5}{2}_R \oplus \nicefrac{11}{2}_R$ & $662$ & $22$ \\
    $\nicefrac{1}{2}_L \oplus \nicefrac{17}{2}_L^2 \oplus \nicefrac{7}{2}_R^2 \oplus \nicefrac{21}{2}_R$ & $3878$ & $38$ \\ 
    $\nicefrac{1}{2}_L \oplus \nicefrac{3}{2}_L \oplus\nicefrac{13}{2}_L\oplus\nicefrac{17}{2}_L\oplus\nicefrac{5}{2}_R^3 \oplus \nicefrac{19}{2}_R$ & $2870$ & $38$\\ \hline
       \end{tabular}
    \caption[Anomaly-Free Chiral $SU(2)$ Representations for Fermionic Excitations.]{A Few Anomaly-Free Chiral $SU(2)$ Representations for Fermionic Excitations. We ensure that the Adler-Bell-Jackiw anomalies cancel for every $U(1)$ subgroup of $SU(2)$. We assume that all excitations are fermionic, and spin-statistics requires that the fermionic excitations have odd spin.}
    \label{ch4:tab:SU2_Reps}
\end{table}

\begin{table}[]
    \centering
    \begin{tabular}{|c c c|}\hline
    \multicolumn{3}{|l|}{$(\nicefrac{1}{2} , 3)_L \oplus (\nicefrac{5}{2}, 3)_L\oplus (\nicefrac{9}{2},3)_L \oplus (\nicefrac{9}{2}, 11)_L (\oplus\nicefrac{3}{2}, 7)_R\oplus (\nicefrac{5}{2}, 7)_R^2\oplus (\nicefrac{11}{2}, 7)_R\hspace{2cm}$} \\ \hline
    Left $SU(2)$ Anomaly & Left $U(1)$ Anomaly & Left Chiral Central Charge \\ 
     $732$ & $1372$ & $28$ \\ \hline
     \multicolumn{3}{|l|}{$(\nicefrac{1}{2}, 3)_L \oplus (\nicefrac{9}{2}, 3)_L^2 \oplus (\nicefrac{9}{2}, 11)_L \oplus (\nicefrac{3}{2}, 3)_R\oplus (\nicefrac{5}{2}, 7)_R\oplus (\nicefrac{9}{2}, 7)_R \oplus (\nicefrac{11}{2}, 7)_R$} \\ \hline
     Left $SU(2)$ Anomaly & Left $U(1)$ Anomaly & Left Chiral Central Charge \\ 
     992 & 1408 & 32\\\hline
    \end{tabular}
    \caption[Anomaly-Free Chiral $SU(2)\times U(1)$ Representations for Fermionic Excitations.]{A Few Anomaly-Free Chiral $SU(2)\times U(1)$ Representations for Fermionic Excitations. We ensure cancellation of the Adler-Bell-Jackiw anomalies independently for the $U(1)$ and $SU(2)$ symmetries. In addition to spin-statistics requiring that fermionic excitations have odd spin, in condensed matter systems all fermions will have odd charge.}
    \label{ch4:tab:SU2_U1_Reps}
\end{table}

Recall that in order for the $U(1)\leqslant SU(2)$ ABJ anomalies to cancel, we require that:
\begin{equation}
\sum_{L}\frac{4}{3} s_i(s_i+1)(2s_i + 1) = \sum_{R}\frac{4}{3} s_i(s_i+1)(2s_i + 1)
\end{equation} 
We also require the cancellation of all gravitational anomalies, which is equivalent to:
\begin{equation}
\sum_L (2s_i+1) = \sum_R (2s_i+1)
\end{equation}
Combining these, and ensuring that we only use half-integer $SU(2)$ representations, we can search for chiral representations, several of which are shown in in Table \ref{ch4:tab:SU2_Reps}.
If we also include $U(1)$ symmetry, with charges $q_i$, then we have a further anomaly cancellation condition:
\begin{equation}
\sum_{L}(2s_i + 1) q_i^2 = \sum_{R}(2s_i + 1) q_i^2
\end{equation} 
Including this condition, we can search for chiral $SU(2)\times U(1)$ theories, two of which are given in Table \ref{ch4:tab:SU2_U1_Reps}. 

Finally, let us note the mathematical perspective on these spin-statistics restrictions. In any fermionic system, there is an additional fermionic parity symmetry $\Z_2^f$. A system with symmetry group $G$ would naively have the actual symmetry group $g\times \Z_2^f$.  However, in reality $\Z_2^f$ is realized as a subgroup of $G$. In the case of $SU(2)$, $\Z_2^f$ is given by the subgroup $\{I, -I\}$, while for $SU(2)\times U(1)$, the subgroup appears as $\{(I, 1), (-I, 1) = (I, -1)\}$.

\section{Summary of the $SU(2)$ Model}\label{ch4:sec:Summary}
The method that we have demonstrated in this chapter leads to a lattice regularization for chiral QFTs. Both fermions and
the gauge symmetry $G$ are defined on on a space-time lattice with no extra
dimensions, and the lattice model is entirely local.  Our approach should work
for any $d+1$D chiral fermion theory that satisfies: (1) there exist mass terms
that make all the chiral fermions massive and break the $G$ symmetry down to
$G_\text{grnd}$; and (2) $\pi_n(G/G_\text{grnd})=0$ for $n\leq d+2$.  

While this work presents progress towards a solution to the long standing chiral fermion problem, three important considerations remain. First, although all low-energy modes in our model are linear, the velocities of fermions in different $SU(2)$ representations may be different; in the future we will try to include a `shift' symmetry that fixes the velocities to be equal. Second, the integral over Higgs considerations can be performed numerically in an approximate way. This is not a trivial task, as the system suffers from a sign problem. Finally, coupling to a weak gauge field (with $U=\exp[iA_{ij}]$ near unity) should be studied as well.

These remaining challenges are considerable. Because the problem requires interactions, it will be almost always be difficult to solve. This rather severe technical challenge both limits the future utility of the approach, and obscures the proof of its concept. In the next Chapter, we will move up a dimension and consider a lattice regularization of Chern-Simons theories, which describe the $2+1$d bulk that a chiral theory might live on. The Chern-Simons theory will lead to an ungauged SPT in that we study in Chapter \ref{chap:U1SPT}, which, as an almost trivial consequence, will yield a much more tractable chiral edge theory in Chapter \ref{chap:Chiral}.

\chapter{A Lattice Rotor Model for $2+1$d Bosonic $U^\kappa(1)$ Chern-Simons Theory}\label{chap:KMatrix}

In the last chapter, we saw a lattice model for a fermionic chiral field theory. That model faced significant challenges in its implementation due to strong interactions. In this chapter, we attack a related problem in one higher dimension: the description of Abelian topological order through $U^\kappa(1)$ Chern-Simons theory. We will develop a lattice model which describes these theories at low energy. This will be a theory of $U(1)$ lattice rotor variables, and is to our knowledge the first of its kind. The semiclassical approximations we use to extract this low-energy behavior are more reliable and under better control than the Higgs integration of the previous chapter; in turn, this theory itself will lead us back to the chiral edge in the following chapter. 

In the rest of this thesis, we will use the simplicial cochain formalism detailed in Appendix \ref{sec:CaC}. We will also work in units where the flux of field variables, and the total angle of a circle, is quantized to unity rather than $2\pi$. 

We wish to develop a lattice model for the Chern-Simons theories described by an action:
\begin{equation}
  S = \pi i \sum_{I,J} K_{IJ}\int a_{I}da_J\label{chap5:eq:naiveCS}
\end{equation}
where, as in Chapter \ref{chap:CFTQM}, $K_{IJ}$ is a symmetric integer matrix (note that we have adjusted the flux quantization condition relative to the standard version we used previously). For simplicity, we assume that the theory is bosonic, so that $K_{IJ}$ has even diagonal entries. The simplest guess for a lattice model is to transpose \eqref{chap5:eq:naiveCS} to the lattice, exchanging forms for cocycles, the wedge product for the cup product, and $d$ for the lattice differential:
\begin{equation}
  \pi i \sum_{I,J} K_{IJ}\int a_{I}da_J + \frac{1}{g}\sum_{\Box} \sum_{I} (da_I)^{2}
\end{equation}
where the integral $``\int''$ is interpreted as evaluation against a generator of the top cohomology of the spacetime manifold and $\Box$ labels plaquettes. This action has a similar `doubling' problem as lattice fermions at momenta $\pm \pi$, and we have introduced a Maxwell term to suppress those points. However, this action has a more fundamental issue as well: the field $a$ is naturally $\R$-valued and cannot be interpreted as a $U(1)$ variable. In getting around this, we will discover a new Chern-Simons action. 

These technical issues with Chern-Simons lattice actions are well known, but they have not stopped considerable work on putting the theory on the lattice. One approach has been to attempt to construct local lattice models for Chern-Simons theory whose many-body Hilbert
space admits a tensor product decomposition $ \cV = \bigotimes_i \cV_i$, where
$\cV_i$ is the local Hilbert space on site-$i$. The key is to find a proper
local Hamiltonian $H$ acting on $\cV$ such that the low energy properties of
$H$ are fully described by a Chern-Simons field
theory \cite{ZHK8982,WWZ8913,PhysRevLett.63.322,W9102,FK9169,WZ9290,Lopez:1994vw,PhysRevB.97.125131,PhysRevB.90.174409}.
However, those lattice models are usually not solvable. Given a lattice model,
we usually do not know if it is in a quantum-Hall topologically ordered phase, if the lattice model produce a Chern-Simons theory at low energy, or even which Chern-Simons theory it produces. Here we are looking
for a better controlled result, where we can derive, under a controlled approximation, the
low energy effective Chern-Simons field theory from the lattice model.

A second approach \cite{ELIEZER1992118,PhysRevB.92.115148,BERRUTO2000366,BIETENHOLZ2003935} also tried to construct lattice gauge models that produce
 Chern-Simons field theory at low
energies. The many-body
Hilbert space $\cV_\text{gauge}$ for the lattice gauge theory in these cases is formed by
gauge invariant states, which are not local, as the $\cV_\text{gauge}$ does not
admit the tensor product decomposition $ \cV_\text{gauge} \neq \bigotimes_i
\cV_i$. References \cite{ELIEZER1992118,PhysRevB.92.115148} proposed lattice gauge models with
compact $U(1)$ gauge group. However, in those models the gauge field in each link is not
compact; rather the compactness is enforced at global level. In contrast, the link
variables in this paper are already in compact $U(1)$ groups.
References \cite{BERRUTO2000366,BIETENHOLZ2003935} proposed lattice gauge models with an entirely non-compact
$U(1)$ gauge group (i.e. $\R$), which is quite different from the
compact $U(1)$ gauge theory we describe here.

Our goal is to realize the most general bosonic $U(1)$
Chern-Simons theory via a local bosonic lattice model with \emph{compact} degrees of
freedom on each link. In contrast to previous lattice models of Chern-Simons theory, we want our local lattice model to be semiclassically solvable, in the
sense that we can reliably determine its low energy effective theory. 

We will find local bosonic model on spacetime lattice satisfying exactly these conditions.
Under a controlled semi-classical approximation for small $g$ in \eqref{CSlatt},
we show that our spacetime lattice model can produce any even-$K$-matrix CS
field theory \cite{WZ9290} of \emph{compact} $U(1)$'s in continuum limit (see
\eqref{CSKIJ}). 

\section{Lattice Model}
Now we can turn to the main lattice model. Let $\cM^3$ be a three-dimensional simplicial complex, with (possibly empty) boundary $\cB^2 = \partial \cM^3$. Our field variables $a$ will be defined on the links of 
Since $a_I$ is $\R/\Z$-valued, all physical quantities to be invariant under the following ``gauge'' transformation
\begin{align}
\label{gaugeZ}
 a_I \to a_I + n_I,
\end{align}
where $n_I$ are arbitrary $\Z$-valued 1-cochains. In this language, the na\"{i}ve action amplitude
$e^{i 2\pi \sum_{I\leq J} k_{IJ} \int_{\cM^3} a_{I} d a_{J}}$
is not locally a $U(1)$ theory because it is not gauge invariant.

We can motivate our final action by going to to one-higher dimension, as in Chapter \ref{chap:CFTQM}. Instead of considering Chern-Simons theory on a three manifold $\cM^3$, we go to a four manifold $\cN^4$ bounded by $\cM^3$ (i.e. $\cM^3 = \partial \cN^4$) and evaluate:
\begin{equation}
 i\pi \sum_{I, J} K_{IJ}\int_{\cN^4} f_I\cup f_J
\end{equation}
The cup product, as well as general conventions for lattice cocycles, is defined in Appendix \ref{sec:CaC}. The question is now how to define $f_I$. We might expect to have $f_I = da_I$, but this cannot be correct: the field strength is a physical quantity and so must be a periodic function of $a$. Instead, we choose the field strength:
\begin{equation}
 f_I = da_I - \toZ{da_I}
\end{equation}
where $\toZ{x}$ denotes the nearest integer to $x$. On any plaquette $\braket{ijk}$, we have:
\begin{equation}
f_I = \frac{1}{2\pi i} \log\left[e^{2\pi i a_I(ij)} e^{2\pi i a_I(jk)} e^{2\pi i a_I(ki)}\right]
\end{equation}
where the logarithm is taken with a branch cut along the negative real axis. Hence $f$ is manifestly invariant under \ref{gaugeZ}. The resulting action is:
\begin{equation}
 i \pi \sum_{I, J} \int_{\cN^4}(da_I - \toZ{da_I}) (da_{J} - \toZ{da_J})
\end{equation}
Let us first assume that the field strength is weak, which implies that $da_I-\toZ{da_I} \approx 0$ and hence $d\toZ{da_I} = 0$. Then we may rewrite the action as:
\begin{equation}
 2i \pi \sum_{I\leq J} \int_{\cN^4}d\left[ a_{I}(da_{J} - \toZ{da_J}) - \toZ{da_I} a_J\right]
\end{equation}
which would lead us to the surface term action that we desired. However, we need to add back in a term to account for cases when $d\toZ{da}\neq 0$. The key here is to preserve a certain $1$-form symmetry that will be examined in Section \ref{ch5:sec:Symmetries}. This leads to another term in the action $-2\pi i \sum_{I\leq J} \int \cM^3 a^J \cup_1 d\toZ{da^J}$. The 1-cup product $\hcup{1}$ \cite{S4790} is defined in Appendix \ref{sec:CaC}. All told, we consider the following partition function:
\begin{align}
\label{CSlatt}
& Z =\int [\prod d a_I]\ 
e^{i 2\pi \sum_{I\leq J} k_{IJ} \int_{\cM^3} d \big(a_I(a_J-\toZ{a_J} )\big) }
\nonumber\\
&
 e^{i 2\pi \sum_{I\leq J} k_{IJ} \int_{\cM^3} a_{I} (d a_{J} -\toZ{d a_J})-\toZ{d a_I}a_J }
\\
&
e^{-i 2\pi \sum_{I\leq J} k_{IJ} \int_{\cM^3} a_J\hcup{1}d \toZ{d a_I}} 
e^{- \int_{\cM^3} \frac{|d a_I - \toZ{d a_I}|^2}{g}},
\nonumber 
\end{align}
To see that the path integral \eqref{CSlatt} is invariant under gauge
transformation \eqref{gaugeZ}, even when the spacetime manifold $\cM^3$ has a boundary, we first note that
$e^{-i 2\pi \sum_{I\leq J} k_{IJ} \int_{\cM^3} a_J\hcup{1}d
\toZ{d a_I}}$ and $d a_I - \toZ{d a_I}$ are invariant
under \eqref{gaugeZ}. Under \eqref{gaugeZ}, the term 
\begin{equation}
e^{i 2\pi \sum_{I\leq J}
k_{IJ} \int_{\cM^3} a_{I} (d a_{J} -\toZ{d a_J})-\toZ{d
a_I}a_J }
\end{equation}
changes by a factor
\begin{equation}
e^{i 2\pi \sum_{I\leq J} k_{IJ} \int_{\cM^3} n^{I} d a_{J} -d n^Ia_J }
=
e^{-i 2\pi \sum_{I\leq J} k_{IJ} \int_{\partial\cM^3} n^Ia_J }
\end{equation}
This is exactly canceled by the change of the term 
\begin{equation}
 e^{i 2\pi \sum_{I\leq J} k_{IJ} \int_{\cM^3} d \big(a_I(a_J-\toZ{a_J})\big) } 
=
 e^{i 2\pi \sum_{I\leq J} k_{IJ} \int_{\partial \cM^3} \big(a_I(a_J-\toZ{a_J})\big) }
\end{equation}
Hence the action amplitude of the above path integral is indeed invariant under $\eqref{gaugeZ}$, even when $\cM^3$ has a boundary.

Now we show that the bosonic lattice model \eqref{CSlatt} realizes a
topological order described by $U^\ka(1)$ Chern-Simons topological quantum
field theory in the limit of small $g$. Taking this limit forces $d a_I$ to be nearly integer-valued, i.e. $da_I - \toZ{da_I} \approx 0$. In turn, this imples that $d\toZ{da_i} = 0$ since if $d a_{I} =
\epsilon +\toZ{d a_{I}}$ where $\epsilon$ is small, then
\begin{align}
 d \toZ{d a_{I}} = -d \epsilon +d d a_I =-d \epsilon.
\end{align}
As $d \toZ{d a_{I}}$ is quantized to be an integer, we must have $d \toZ{d a_{I}}=0$. This $\Z$-valued 2-cocycle $\toZ{d
a_I}$ characterize the $U^\ka(1)$
principle bundle on the spacetime, since
\begin{align}
 \int_{\cM^2} (d a_I-\toZ{d a_I})=
 -\int_{\cM^2} \toZ{d a_I}
\end{align}
for any closed $\cM^2$. Note that $\int_{\cM^2} (d a_I-\toZ{d
a_I})$ is the magnetic flux through $\cM^2$ which is always quantized to be an integer. In other words, $-\int_{\cM^2} \toZ{d a_I}$ is the Chern number. Furthermore, on any local patch of spacetime, we can use the
gauge transformation \eqref{gaugeZ} to set $\toZ{d a_I} = 0$ on
the patch. In this case, the action amplitude in the path integral
\eqref{CSlatt} becomes quadratic (i.e. non-interacting) 
\begin{align}
e^{i 2\pi \sum_{I\leq J} k_{IJ} \int_{\cM^3} a_{I} d a_{J}} \label{quadratic}
\end{align}
Since $d a_I$ is close to zero,
we can use a 1-form $A^I$ to describe the 1-cochain $a_I$:
\begin{align}
 \int_i^j A^I = 2\pi(a_I)_{ij}
\end{align}
Then the above action amplitude can be rewritten as
\begin{align}
\label{CSKIJ}
&
e^{i 2\pi \sum_{I\leq J} k_{IJ} \int_{\cM^3} a_{I} d a_{J}}
\approx e^{i \sum_{I J} \frac{K_{IJ}}{4\pi} \int_{M^3} A^{I} d A^{J}}
\nonumber\\
&\ \ \ \
K_{IJ} = K_{JI} \equiv \begin{cases}
2 k_{IJ}, & \text{ if } I=J,\\
k_{IJ}, & \text{ if } I<J,\\
\end{cases}
\end{align}
when $A^I$ is nearly constant on the lattice
scale. Hence the low energy dynamics of our lattice bosonic model are described
by a $U^\ka(1)$ Chern-Simons field theory \eqref{CSKIJ} at low energies.

When $f_I = da_I - \toZ{da_I} \approx 0$, i.e. when $da_I$ is nearly an integer, we also have gauge invariance under the usual gauge symmetry:
\begin{equation}
a_I \to a_I + d\theta_I
\end{equation}
for $\theta_I$ an $\RZ$-valued one-cochain. This is apparent both in the formulation \eqref{quadratic} and its continuum counterpart \eqref{CSKIJ}. In order for this gauge invariance to emerge, we had to assume that spacetime $\cM^3$ had no boundary and that $d\toZ{da_I} = 0$, i.e. that the field configurations had no monopoles. That gauge invariance should break on the boundary is, as discussed Chapter \ref{chap:CFTQM}, no surprise in Chern-Simons theories. Furthermore, we should understand the monopole-free requirement in the same way: the monopole should be thought of as trapping a defect. In topological quantum field theories, defects can be realized as punctures in the manifold, which therefore are boundaries. In effect, each monopole traps a region of non-topologically ordered matter, which therefore carries a gauge-symmetry breaking boundary.

Due to the boundaries induced by monopoles, it is not clear what the ground state is for large $g$, and it  may have a different topological
order from the one described by the $K$-matrix Chern-Simons theory. However, the higher-form symmetries we discuss in the next section always hold.

\section{Higher Symmetries and Anomalies}\label{ch5:sec:Symmetries}
The most powerful aspect of the action we have constructed is its 1-symmetries. First, consider the model on a closed manifold, so that we may
ignore the surface term. Then under the shift:
\begin{equation}
a_{I} \to a_{I} + \tilde\beta_{I} \hspace{1cm} \sum_I \tilde\beta_{I}K_{IJ} \in \Z\label{1symm}
\end{equation}
where $\tilde\beta_{I}$ are $\RZ$-valued 1-cocycles, the exponentiated action
is invariant so long as $\sum_I \tilde\beta_{I}K_{IJ}$ are $\Z$-valued
1-cochains. These transformations \eqref{1symm} are the 1-symmetries of
lattice model \eqref{CSlatt}. 

To see this result, we first note that, under the transformation
\eqref{1symm}, the action amplitude in \eqref{CSlatt} on a closed manifold changes
by a factor
\begin{equation}
e^{i 2\pi \sum_{I\leq J} k_{IJ} \int_{\cM^3} \tilde\beta^{\RZ}_{I} (d a^{\RZ}_{J} -\toZ{d a^{\RZ}_J})-\toZ{d a^{\RZ}_I}\tilde\beta^{\RZ}_J -i 2\pi \sum_{I\leq J} k_{IJ} \int_{\cM^3} \tilde\beta^{\RZ}_J\hcup{1}d \toZ{d a^{\RZ}_I}} 
\end{equation}
Because we may integrate by parts on a closed manifold and $d\tilde\beta^\RZ_{I} = 0$,
the change is of the form (see Appendix \ref{sec:CaC}):
\begin{align}
& e^{-i 2\pi \sum_{I\leq J} k_{IJ} \int_{\cM^3} \tilde\beta^\RZ_{I} \toZ{d a^{\RZ}_J} + \toZ{d a^{\RZ}_I}\tilde\beta^\RZ_J +\tilde\beta^{\RZ}_J\hcup{1}d \toZ{d a^{\RZ}_I}}
\nonumber\\
&= e^{-i 2\pi \sum_{IJ} K_{IJ} \int_{\cM^3} \tilde\beta^\RZ_{I} \toZ{d a^{\RZ}_J} }
\end{align}
which remains unity for all $\toZ{d a^{\RZ}_{J}}$ iff $\sum_I
\tilde\beta^{\RZ}_{I}K_{IJ}$ are $\Z$-valued cochains. We see that, on a fixed link
$ij$, the allowed values of $(\tilde\beta^{\RZ}_{I})_{ij}$ form the rational lattice
$K^{-1}$. The 1-symmetries are given by the rational lattice $K^{-1}$ modulo
the integer lattice, which is same as the integer lattice modulo the lattice $K$. In
other words, the 1-symmetries are $Z_{k_1}\times Z_{k_2} \times \cdots$
1-symmetries, with $k_i$ being the diagonal entries of the Smith normal form of
$K$.

For example, for $U(1)$ Chern Simons theory with $\kappa = 1$ and $K_{11}
=2k_{11} =k$, we have a $\mathbb{Z}_{k}$ 1-symmetry. For mutual Chern-Simons theory (that
describes a $Z_n$ gauge theory), with 
$(K_{IJ}) = \left(\begin{array}{cc}0 & n\\ n & 0\end{array}\right)$
we have a ${Z}_{n}\times {Z}_{n}$ 1-symmetry. 

Some of the above 1-symmetries are anomalous. To determine which, we need examine which of the transformations in \eqref{1symm} changes
the action amplitude when the spacetime has a boundary.
Under the transformation \eqref{1symm}, the action amplitude in \eqref{CSlatt} only
changes by a factor defined on the boundary $\partial \cM^3$:
\begin{multline}
e^{i 2\pi \sum_{I\leq J} k_{IJ} \int_{\partial \cM^3} 
a^{\RZ}_I(\tilde\beta^{\RZ}_J-\toZ{\tilde\beta^{\RZ}_J})
+\tilde\beta^{\RZ}_I(a^{\RZ}_J-\toZ{a^{\RZ}_J})
+\tilde\beta^{\RZ}_I(\tilde\beta^{\RZ}_J-\toZ{\tilde\beta^{\RZ}_J})
}...
\\
\times e^{i 2\pi \sum_{I\leq J} k_{IJ} \int_{\partial \cM^3} 
\tilde\beta^\RZ_J \hcup{1} \toZ{d a^{\RZ}_I}-\tilde\beta^\RZ_I a^\RZ_J
}
\\
=
e^{i 2\pi \sum_{I\leq J} k_{IJ} \int_{\partial \cM^3} 
a^{\RZ}_I(\tilde\beta^{\RZ}_J-\toZ{\tilde\beta^{\RZ}_J})
+\tilde\beta^{\RZ}_I(\tilde\beta^{\RZ}_J-\toZ{\tilde\beta^{\RZ}_J})
+\tilde\beta^\RZ_J \hcup{1} \toZ{d a^{\RZ}_I}
}...
\\
\times
e^{-i 2\pi \sum_{I\leq J} k_{IJ} \int_{\partial \cM^3} 
\tilde\beta^{\RZ}_I\toZ{a^{\RZ}_J}
}
\end{multline}
We see that the transformations leave the action amplitude invariant if
$\sum_{I\leq J} k_{IJ} \tilde\beta^\RZ_J = 0$ and $\sum_{I\leq J}
k_{IJ} \tilde\beta^\RZ_I = \text{integer}$. We note that $\tilde\beta^\RZ_I$ satisfy the
condition $\sum_{I J} K_{IJ} \tilde\beta^\RZ_I \in \Z $, and so the first
equation implies the second. We find that the 1-symmetry transformations in
\eqref{1symm} are anomaly-free if
\begin{align}
 \label{AFc}
 \sum_{I\leq J} k_{IJ} \tilde\beta^\RZ_J = 0, \ \forall I
\end{align}

For the level $k=K_{11}$ Chern-Simons theory with a single $U(1)$ gauge field, this is
simply the fact that the only $Z_k$ 1-symmetry is anomalous and must break at the boundary. For the case of mutual Chern-Simons theory (ie the $Z_n$ gauge theory)
with $Z_n\times Z_n$ 1-symmetry, this implies that one of the $Z_n$ 1-symmetries is anomalous and
must break at the boundary, while the other $Z_n$ 1-symmetry
is anomaly-free. Note that the choice of lattice model automatically selects
which of the $Z_n$ 1-symmetry is anomalous; one can select the opposite by
replacing all $\sum_{I\leq J}$ with $\sum_{I\geq J}$. 


\subsection{Framing Anomaly}
It is well known that the Chern-Simons theory has a framing
anomaly \cite{W8951,GF14106812}. After integrating out the
physical degrees of freedom $a^\RZ_I$ in \eqref{CSlatt} in the small $g$ limit, we
should get a partition function given by the 2+1D gravitational Chern-Simons term:
\begin{align}
 Z(M^3,g_{\mu\nu}) \propto
e^{i \frac{2\pi c}{24} \int_{M^3} \Omega_3} 
\end{align}
where the 3-form $\Omega_3$ satisfies $d \Omega_3 = p_1$ and $p_1$ is the first
Pontryagin class for the tangent bundle. Here $c$ is the chiral central charge
-- the difference between the numbers of positive and negative eigenvalues of
the $K$-matrix. There is a framing anomaly when $c \neq 0$ mod 24.


This framing anomaly might prevent a local lattice
realization of chiral Chern-Simons theory with a non-zero central charge $c\neq 0$. However, our
construction shows that chiral $U(1)$ Chern-Simons theory can always be realized on any
2+1D spacetime lattice. We believe that this is possible because our spacetime
lattice has extra structure, namely the branching
structure \cite{C0527,CGL1314,CGL1204}, which encodes the ordering of lattice vertices and appears in the definition of the cup product. If we take different branching structures on the same spacetime lattice, the resulting
partition function $Z$ may be different. In turn, this branching structure dependence
of partition function may represent the framing anomaly.

\section{Summary of the Bosonic Lattice Chern-Simons Model}

We have constructed a local bosonic model which realizes any Abelian $2+1$d topological order, represented as a lattice model of $K$-matrix Chern-Simons theory. Moreover, this model has a local Hilbert space, and is manifestly a function of compact rotor variables. This is a significant improvement on previous models. Even stronger evidence for the phase produced by this model is given by the exactly realized $1$-symmetries and their anomalies. We now have a lattice model for any $2+1$d Abelian topological order.

Nonetheless, it is not clear how to use this achievement to get back towards a chiral $1+1$d field theory. While these topological orders carry edge modes, it is not clear how this could be extracted from the actions we have constructed. In the next section, we will `ungauge this model,' turning a $2+1$d topological order into a $2+1$d SPT state. Moreover, these SPT models will be exactly solvable, and that will point the way back towards the edge theory.

\chapter{A $U(1)$ SPT and Discontinuous Lattice Topological Terms}\label{chap:U1SPT}

In this chapter, we develop an exactly solvable model for SPT states by `ungauging' the Chern-Simons model of the previous chapter. In this new, exactly solvable SPT model, we will find examples of almost every phenomenon discussed in Chapters \ref{chap:CMQFT} and \ref{chap:CFTQM}, from chiral anomalies to the short-range entanglement that characterizes SPTs. This will lead to a $1+1$d chiral lattice field theory as an almost trivial corollary in the next chapter.

We begin in Section \ref{ch6:sec:Ungauging} where we make use of an ungauging process to turn the lattice gauge theory model of topological order from Chapter \ref{chap:KMatrix} into a lattice non-linear sigma model describing SPT states. This allows us to get our first glimpse of the larger model, and by ungauging the model in the presence of a background gauge field we will establish the first proof of the Hall conductance of this model. In Section \ref{chap6:sec:general}, we integrate this model into a two-dimensional phase diagram describing the superfluid-Topological Mott Insulator transition (SF-tMI). The original `ungauged' action then corresponds to the tMI fixed points, and we build an understanding of these phases as condensates of charged vortices. We then examine the phase transitions for these models in Section \ref{chap6:sec:Fermions}, and present a brief lattice renormalization group argument in Section \ref{sec:LatticeRG}.

\section{Ungauging the Chern-Simons Theory}\label{ch6:sec:Ungauging}
Here we ungauge the Chern-Simons lattice action to build our SPT model, both as-is and in the presence of a background gauge field. Let us begin with the Chern-Simons action reduced to just a single mode:
\begin{equation}
S = 2\pi i k \int \left[ a(da - \toZ{da}) - \toZ{da}a - a\cup_1 d\toZ{da} + d \big(a(a-\toZ{a}  )\big) \right]\label{ch6:eq:CSsingle}
\end{equation}
where we have dropped the Maxwell term. 
Recall that, because we are working with angular variables $a_{ij}$, and not the rotor variables $U_{ij} = e^{2\pi i a_{ij}}$, we must impose a gauge redundancy on all physical quantities. In order to to ensure that the degrees actually remain rotors, we require a ``rotor redundancy'', with all physical quantities being invariant under the replacement
\begin{equation}
a_{ij} \to a_{ij} + m_{ij} 
\end{equation}
Put more simply, we require that all physical quantities must be periodic functions of $a_{ij}$. This redundancy ensures that the action \eqref{ch6:eq:CSsingle} is a function of $\RZ\simeq U(1)$ variables, and in turn is the source of level quantization, ensuring that $k\in \Z$. 

The Chern-Simons action is formulated in terms of a one-cochain $a$, i.e. variables defined on the links of the lattice. To ungauge this model we set
\cite{PhysRevB.99.205139}:
\begin{align}
 a_{ij} = \phi_{i} - \phi_{j}
,
\ \ \text{or }\ \
a = d \phi
\end{align}
where $\phi$ is zero-cochain defined on the vertices. 
The partition function is now given by:
\begin{equation}
\label{U1spt}
 Z =\int [\prod d \phi] 
 e^{i 2\pi k \int_{\cM^3} 
d \phi d \toZ{d \phi} 
}
\end{equation}
where $\cM^3$ may have boundaries, and the measure is taken to be integration
over all sites, $ \int \prod d \phi = \prod_{i}
\int_{-\frac{1}{2}}^\frac{1}{2}d \phi_i$. 

This action has a similar ``rotor redundancy'' as the Chern-Simons action. Note that we may shift:
\begin{equation}
\phi_i \to \phi_i + n_i \label{eq:gaugeredun}
\end{equation}
where $n_i$ is a $\Z$-valued zero-cochain. The term $d\toZ{d\phi}$ is invariant because it transforms as:
\begin{equation}
d\toZ{d\phi} \to d\toZ{d\phi + dn} = d\toZ{d\phi} + d^2n = d\toZ{d\phi}
\end{equation}
and so the whole action is invariant modulo $2\pi$. As in the case of the Chern-Simons theories, this redundancy ensures that this action is a function of the $\phi_i$ as $\RZ\simeq U(1)$ variables. Beyond this rotor redundancy, our models will also possess two global symmetries. The first is the $U(1)$ symmetry of boson number conservation, which acts as 
\begin{equation}
\phi_i \to \phi_i + \theta\label{eq:U1sym}
\end{equation} 
where $\theta$ is a constant.
We also have a charge conjugation symmetry:
\begin{equation}
\phi_i \to -\phi_i, \label{eq:CCsym}
\end{equation}
which sends $g_i = e^{2\pi i \phi_i} \to g_i^*$. Note that eq. \eqref{U1spt} enjoys all three  of these symmetries. In particular, the level $k$ is quantized as a direct result of the gauge redundancy \eqref{eq:gaugeredun}. 

\begin{figure}
\begin{center}
\includegraphics[width = .75\textwidth]{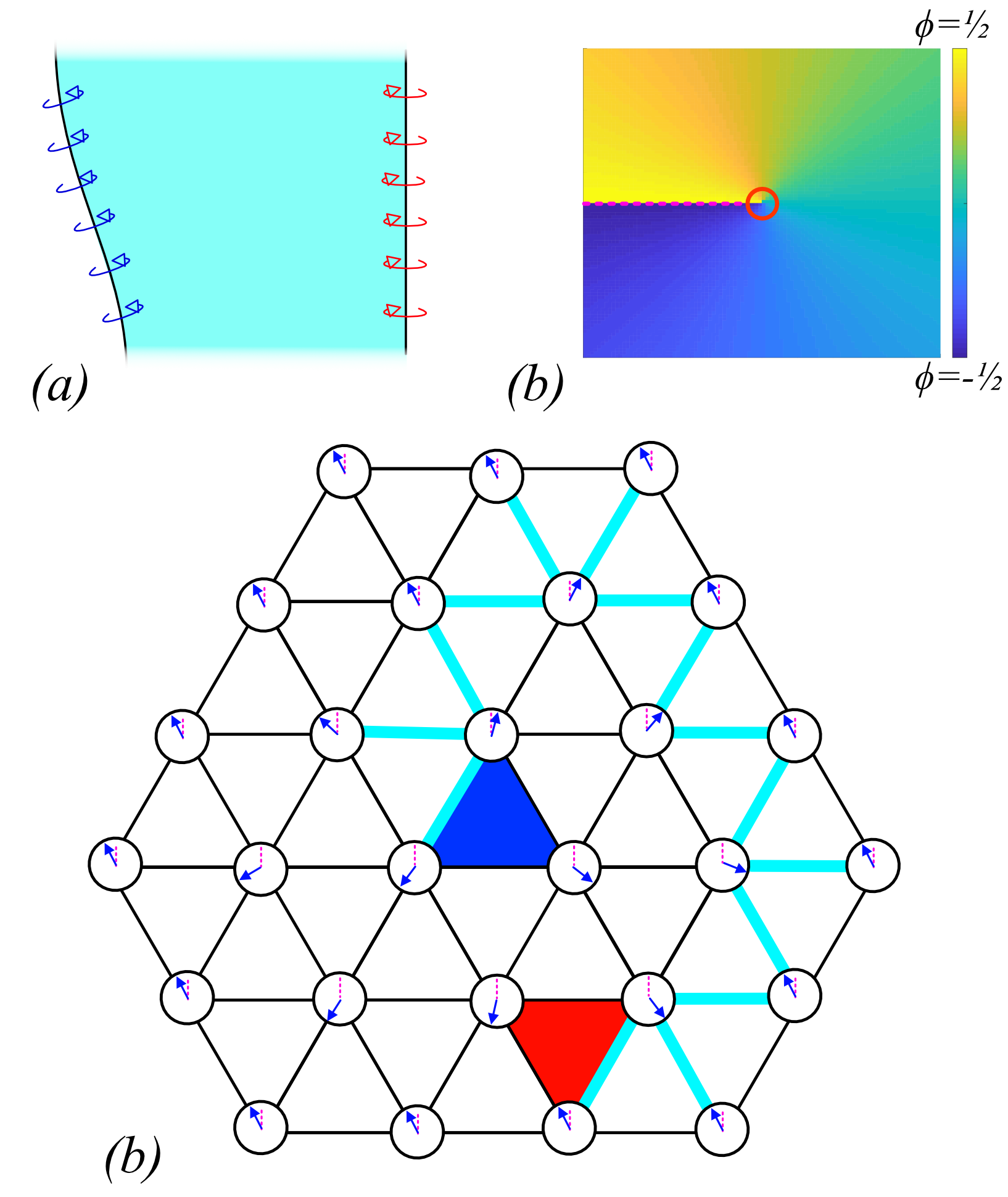}
\caption[Lattice Vortex Term]{\emph{(color online). (a)} In $2+1$ dimensions, $\toZ{d\phi}$ is dual to a surface where $\phi$ has a branch cut, and $\jmath_v = -d\toZ{d\phi}$ is dual to the line where a branch cut ends, i.e. a vortex \emph{(b)}. \emph{(c)} Working in two dimensions for convenience, we can see on a microscopic level that  $-d\toZ{d\phi}$ is the vortex number on a plaquette. The field variables are shown as clocks, while links with nonzero $\toZ{d\phi}$ are marked in turquoise. Plaquettes with $-d\toZ{d\phi} \neq 0$ are marked in blue (vortex) and red (anti-vortex).}\label{fig:vortex}
\end{center}
\end{figure}

To understand this action, note that $\jmath_v\equiv \star(- d\toZ{d\phi})$ is the vortex current. To see this, first note that $\toZ{d\phi}$  is a one-cochain which, in three dimensions, is dual to a surface where $\phi$ has a branch cut. Hence $d\toZ{d\phi}$ is dual to the line where the branch cut surface ends (Figure \ref{fig:vortex}a). Branch cut surfaces end at vortex lines; a $2$d slice of a vortex line and branch cut are shown in Figure \ref{fig:vortex}b, where we see that the branch cut terminates in a vortex. On a more microscopic level, consider the $2$d lattice in figure \ref{fig:vortex}c. The $U(1)$ variables on sites are shown as clocks with zero marked as a vertical line. The branch cut links with $\toZ{d\phi} = 0$ are marked in turquoise, and the branch cut terminates in a vortex (blue) and anti-vortex (red).  Hence we see that $- d\toZ{d\phi}$ is indeed the vortex number on a plaquette, with the sign fixed so that $- d\toZ{d\phi} = +1$ at a plaquette hosting a vortex. Note that conservation of vortex number means that $d^\dagger \jmath_v = 0$, where $d^\dagger \equiv \star d \star$. 

To gain intuition, we can also rewrite the action in terms of the group variables $g_i = e^{2\pi i \phi_i}$. The vortex current an a plaquette $p$ becomes:
\begin{equation}
\star \jmath_v =-d\toZ{d\phi} = -\frac{1}{2\pi i} d \log(g_i g_j^*) = -\frac{1}{2\pi i}(\log(g_i g_j^*) + \log(g_j g_k^*) + ...)
\end{equation}
where the logarithm is taken with a branch cut along the negative real axis. In terms of these variables, the exponentiated action becomes:
\begin{equation}
e^{iS} = \prod_{\braket{i, j} \in \text{ links}} (g_i g_j^*)^{\star \jmath_v(\Box(i, j))}
\end{equation}
where $\Box(ij)$ is the plaquette dual to the link $\braket{i,j}$ under the cup product. This formulation is not as useful as working in terms of the $\phi_i$, but we do wish to establish that this is truly a $U(1)$ theory. 

The topological action suggests that this model realizes an SPT state. To see the SPT order, we
repeat the ungauging in the presence of a weak background gauge field $\bar
a$ and evaluate the effective action for $A$. In the
presence of background $U^\ka(1)$ background gauge field $A$, the
ungauing is done via 
\begin{equation} 
a = A + d \phi.
\end{equation} 
Now the model is given by
\begin{multline}
\label{U1sptG}
 Z =
 e^{i 2\pi  k \int_{\cM^3} A (d A -\toZ{d A})-\toZ{d A}A -i 2\pi  k \int_{\cM^3} A\hcup{1}d \toZ{d A}}  
\\
\int [\prod d \phi]\ 
e^{-i 2\pi  k \int_{\cM^3} d\phi\hcup{1}d \toZ{d A} + i 2\pi  k \int_{\cM^3} d \phi (d A -\toZ{d A})-\toZ{d A}d \phi  + i 2\pi  k \int_{\partial \cM^3} 
(A +d \phi)(A+d \phi-\toZ{A +d \phi}  ) }
\nonumber 
\end{multline}
Note that this model has a rotor redundancy for $A$:
\begin{equation}
A \to A + m \label{Zredundancy}
\end{equation}
in addition to the usual one for $\phi$.
If $\cM^3$ is closed, this can be simplified to
\begin{multline}
 Z =
 e^{i 2\pi  k \int_{\cM^3} A (d A -\toZ{d A})-\toZ{d A}A -i 2\pi  k \int_{\cM^3} A\hcup{1}d \toZ{d A}}  
\\
\int [\prod d \phi]
e^{-i 2\pi  k \int_{\cM^3} d\phi\hcup{1}d \toZ{d A} -i 2\pi  k \int_{\cM^3} d \phi \toZ{d A} + \toZ{d A}d \phi }
\end{multline}
Now we assume that the background field is ``weak.'' Because we are working with $\RZ$ valued fields, a weak field means that $dA$ is nearly an integer, i.e. $|dA - \toZ{dA}| < \epsilon$. Noting that this implies that $d\toZ{d A} = 0$, the path integral becomes: 
\begin{align}
\label{U1sptInv}
 Z =
 e^{i 2\pi  k \int_{\cM^3} A (d A -\toZ{d A})-\toZ{d A}A } 
\end{align}
This is the Chern-Simons response on lattice. When $\cM^3$ is closed, the
action is invariant under gauge transformations of the background gauge field
$A \to A + d\varphi$. If $\cM^3$ is a disk, the
redundancy (\ref{Zredundancy}) can be used to set $\toZ{dA} = 0$,
and the response becomes:
\begin{align}
\label{U1sptInv2}
 Z =
 e^{i \pi \sum k \int_{\cM^3} A d A}
\end{align}
This Chern-Simons response, in terms of the level $k$,
describes the Hall conductance and is the SPT invariant for our model. We will use this action, coupled to a background gauge field $A$ in what follows below, and will see the Hall conductance from several different perspectives.

\section{Generalization to Arbitrary $k$ and the Topological Mott Insulators}\label{chap6:sec:general}

With the model \eqref{U1spt} in hand, we can actually pursue a generalization to non-integer $k$. The key to this is to note that we can replace the action by:
\begin{equation}
S = 2\pi i k \int d\phi d\toZ{d\phi} = 2\pi i k \int (d\phi - \toZ{d\phi}) d (d\phi - \toZ{d\phi})
\end{equation}
Now, this action is rotor redundant under \eqref{eq:gaugeredun} for any $k$. In this section, we elaborate this model, adding to it a second term which reveals the character of this action as a topological Mott Insulator. 

\subsection{The Topological Mott Insulators}

The phase transitions that occur in lattice boson systems between Mott insulators \cite{Mott_1949} and superfluids (SF) have been extensively studied \cite{SFMI_Greiner, 2010Sci...329..547B, 2010Sci...329..523D, 2013CRPhy..14..712P,  PhysRevLett.97.060403, PhysRevLett.98.080404, 2010Sci...329..547B}, in part because they embody the competition between kinetic energy and interactions which underlies much of condensed matter physics. If the kinetic energy 
dominates, the bosons condense into a superfluid. On the other hand, repulsive interactions tend to favor the bosons being localized in real space; if they dominate the system forms a Mott Insulator. 

In a seminal work \cite{PhysRevB.40.546}, Fisher et. al. showed that the superfluid-Mott insulator (SF-MI) quantum phase transition generically falls into one of two universality classes. 
In the Mott insulating phase, the average density is pinned to a fixed integer; if the the chemical potential is varied, then the extra bosons (or holes) may condense into a SF, leading to an `ideal' mean-field transition with dynamical critical exponent $z=2$. On the other hand, if the chemical potential is chosen so that particle-hole symmetry is maintained, then the SF-MI transition is actually a multicritical point and lies in the XY universality class, with $z=1$. The XY transition in particular in $2+1$d is in the same universality class as the condensation of superfluid helium at a finite temperature, and so has seen impressive theoretical \cite{Campostrini:2000iw, Gottlob:1993zd, Hasenbusch:1999cc, Guida:1998bx} and experimental \cite{PhysRevLett.76.944, PhysRevLett.84.4894} study.

This story becomes more complex in systems where time-reversal symmetry is broken. In $2+1$d, the insulating state can develop a quantized Hall conductance, 
with such states being known variously as $U(1)$ symmetry protected topological phases \cite{CGL1314},
topological Mott Insulators (tMIs), or the Bosonic Integer Quantum Hall Effect \cite{LV1219,SL1301, GERAEDTS2013288}. 
These models generated significant excitement; recently a fermionic analogue \cite{2008PhRvL.100o6401R} has been proposed as the mechanism behind the Quantum Anomalous Hall state in twisted bilayer graphene \cite{2020arXiv201107602C} (though we will focus on bosonic tMIs in this chapter).
The topological Mott insulating states are labeled by their Hall conductance, which for the bosonic systems we consider is always an even integer multiple of $e^2/h$, viz. $\sigma_{xy} = (2k) \frac{e^2}{h}$ for $k\in\mathbb{Z}$. 

As for the SF-MI transition, we can ask about the nature of the phase transition that occurs between a SF and a tMI. 
What should the the critical exponents of this transition be? 
Due to the strong charge density and current fluctuations, the answer is not immediately clear. 

We will focus on the particle-hole symmetric $z=1$ multicritical point, where we will be able to write down a lattice field theory well-suited for describing the phase diagram at fixed integer boson density. We show that the critical exponents of the 
SF-tMI $XY$ transition are exactly the same as the regular
SF-MI $XY$ transition. Moreover, we will see that the bulk dynamics of all local excitations are identical at or even away from the critical point, unless there is an applied background gauge field. We do this by constructing a well defined lattice model and showing that the bulk dynamics are invariant under a ``level-shift symmetry'' induced by adding a 
topological $\theta$-term, that changes the Hall conductance by $2k \frac{e^2}{h}$, hence connecting the dynamics of the tMIs to the $k=0$ trivial MI. As our model is well-regulated on the lattice, we can show that this level-shift symmetry is exact, i.e. valid at all relevant energy scales. Crucially, this means that the extraordinary numerical, theoretical, and experimental study of the $2+1$d XY transition is applicable to the SF-tMI transition as well. 

In the presence of a background gauge field, level-shift symmetry is broken and the topological term leads to a Chern-Simons response and quantized Hall conductance. In particular, the topological term causes the vortices which proliferate in the insulating phase to carry charge. We discuss how charged vortices lead directly to the quantized Hall response. This quantized Hall response characterizes the tMI phase and we argue that it persists in the vicinity of the SF-tMI transition.

For two transitions related by the level-shift symmetry ({\it i.e.} for two models differing only by the topological term), the correlations of any local operators are identical, at and away from the transition point, even though the two transition points have different Hall conductance. This is possible since the level-shift symmetry, {\it i.e.} the topological term, changes the definition of current operators.

Beyond the description of the topological Mott insulators, these models also represent an advance on a purely theoretical front. It has been known for some time \cite{CGL1314,Chen1604, PhysRevB.95.205142} that topological terms for $d+1$-dimensional spacetime lattice models are labeled by elements of the group cohomology $H^{d+1}(G, U(1))$. The cocycles of these models provide actions that are lattice analogs of continuum $2\pi$-quantized topological $\theta$-terms ({\it i.e.} with $\theta=0$ mod $2\pi$), and some of our analysis parallels continuum work \cite{Xu:2011sj}. It has also been known that in order to obtain a nontrivial group cohomology class for continuous groups, one must consider discontinuous cocycles. However, the first explicit expression of these discontinuous cocycles was only found very recently; the $U(1)$ models discovered in \cite{DeMarco:2021erp} and suitably generalized here are the first examples. They follow considerable work placing quantum Hall physics on lattices \cite{Chen:2019mjw, Sun:2015hla}, and join similar works aiming to describe $2+1$d systems with \cite{Wang:2021smv} or without \cite{Bauer:2021fvc} gappable boundaries. They are one of several approaches demonstrating ways around \cite{2021arXiv210702817H, DeMarco:2021erp} a related no-go theorem \cite{2019CMaPh.373..763K, 2021arXiv210710316Z} preventing Hall conductance on the lattice, which does not hold in our case due to the infinite-dimensional on-site $U(1)$ rotor Hilbert space. These discontinuous cocycles are the key to understanding the SF-tMI transition, and also pave the way for studying more complicated transitions involving nonabelian Lie groups.

\section{Model Overview}\label{sec:ModelOverview}
We will be concerned with lattice systems of bosons in $2+1d$ at fixed average integer density.
As in the traditional $XY$ model, the field variables in our theory are $U(1)$ rotors living on the sites $i$ of a three-dimensional Euclidean spacetime lattice. We denote the field variables by $\phi_i$, which correspond to the phase of the microscopic boson operator on site $i$ (as we are working at fixed average density, we are allowed to work solely with the phase modes $\phi_i$). As before, we let the $\phi_i$ be periodic under shifts by unity, rather than by $2\pi$. Hence a $U(1)$ angle $\phi_i$ will take values in $[0, 1)$, while a group element is given by $g_i = e^{2\pi i \phi_i}$. This will avoid numerous factors of $2\pi$ below, and one may always convert back to the usual notation by replacing $\phi \to \tilde\phi/2\pi$.

The trivial (i.e. non-topological) $XY$ model Lagrangian satisfies all of these symmetries. It is
\begin{equation}
S = -\frac{1}{g} \sum_{\braket{i,j}} \cos( 2\pi  (d\phi)_{ij})\label{eq:XYLag}
\end{equation}
Here the sum runs over all links $\braket{i, j}$ in the $2+1$d lattice, and $(d\phi)_{ij} = \phi_i - \phi_j$. The phase of the model is controlled by $g$. As $g\to 0$, the strong `kinetic' term sets $d\phi = 0$ and so $\phi = \text{const.}$, confining all vortices and resulting in the $U(1)$ symmetry breaking superfluid phase. As $g\to\infty$, the fluctuations of $\phi$ overpower the kinetic suppression and vortices proliferate, destroying long-range order and leading to the Mott insulating phase. 

To get the tMIs from a lattice model, we simply add our topological term:
\begin{equation}
S_{k} = -2\pi i k\int (d\phi - \toZ{d\phi}) d (d\phi - \toZ{d\phi})\label{eq:Sk}
= 2\pi i k\int(d\phi - \toZ{d\phi})  d  \toZ{d\phi}
\end{equation}
as $d^2 = 0$. The full model combines the kinetic energy with the theta term:
\begin{equation}
S_{g, k}[\phi] = -\frac{1}{g}\sum_{\text{links}}\cos (2\pi d\phi) 
+2\pi i k \int (d\phi - \toZ{d\phi})d\toZ{d\phi}\label{eq:action}
\end{equation}
and the complete partition function is:
\begin{equation}
Z_{k, g} = \int D\phi e^{ \frac{1}{g}\sum_{\text{links}}\cos 2\pi d\phi- 2\pi i k \int (d\phi - \toZ{d\phi})d\toZ{d\phi}}
\end{equation}
where the measure $\int D\phi = \left[\prod_{i}\int_{-\frac{1}{2}}^{\frac{1}{2}}d \phi_i \right]$ has already gauge-fixed the rotor redundancy (\ref{eq:gaugeredun}).

We thus have a two parameter phase diagram, defined in terms of a rotor-redundant model with global $U(1)$ and particle-hole symmetries. Our task in the remainder of this section will be to understand this phase diagram.

Let us first fix $k=0$, in which case we have reduced to the traditional $XY$ model for the SF-MI transition. This model has been extremely well studied, and it is known that there are two phases: for small $g$, the system spontaneously breaks $U(1)$ symmetry, setting $d\phi = 0$ and resulting in a superfluid phase. For large $g$, the system is in a disordered symmetry preserving phase, corresponding to a Mott insulator. The transition between the two occurs at some non-universal $g_c$, with $g < g_c$ corresponding to the superfluid phase and $g>g_c$ corresponding to the Mott Insulator phase. Physically, we can understand this in terms of spin-zero bosons on a lattice at unit filling. When the system is a superfluid, phase rigidity $\phi \approx \text{const.}$ develops. On the other hand, when the system is a Mott insulator, the phase of each Boson fluctuates independently, thereby preserving $U(1)$ symmetry and destroying any long-range order. 

Adding the topological term (and thereby breaking time-reversal symmetry) turns the one-dimensional phase diagram into a two-dimensional one. In the superfluid phase, vortices are suppressed, and we expect that the topological term should be unimportant, leading to just a single SF phase. On the other hand, vortices proliferate in a disordered phase, and so we expect the topological term to be important in the disordered phase. As we will soon see, the tMI fixed points are given by $g\to \infty$, $k\in \mathbb Z$. For generic $k$ and large $g$, we will see that the system flows to attractive fixed points at $k\to \toZ{k}$ and $g\to \infty$ which are studied in Section \ref{sec:IntegerPhases}. For large $g$ and half-odd-integer $k$ we have the tMI-tMI transition discussed in Section \ref{chap6:sec:Fermions}.

In terms of this vortex current $\jmath_v$, we may rewrite the action as:
\begin{equation}
    S_{g, k}[\phi] = -\frac{1}{g}\sum_{\text{links}}\cos( 2\pi d\phi )- 2\pi i k \int (d\phi - \toZ{d\phi})\star \jmath_v,
\end{equation}
where we see that the effect of the topological term is to couple $\phi$ to the vortex current. Moreover, the topological term actually gives the vortices mutual statistics. Written in terms of $\jmath_v$, the topological term on a closed manifold is
\begin{equation}
    S_{k}[\phi] =   -2\pi i k \int  \toZ{d\phi}\star \jmath_v = 2\pi i k  \int \jmath_v \cup  \frac{d}{\square }\jmath_v,
\end{equation}
where $\square$ is the Laplacian. That this term gives statistical phases to braided vortex lines can be seen by noting that it is of the same form as the response of a Chern-Simons gauge field coupled to a background current. More directly, current conservation $d^\dagger \jmath_v = 0$ allows us to write $\jmath_v = \star dL$, where the Poincare dual of $dL$ marks the vortex lines in spacetime. In terms of $L$, 
\beq 2\pi i k \int \jmath_v\cup \frac{d}{\square}\jmath_v = 2\pi i k \int L \cup dL,\eeq 
which indeed gives $2\pi i k$ times the linking number of the vortex worldlines.
Hence we see that the topological term changes the statistics of vortices, with vortex lines having mutual statistics of $e^{2\pi i k}$.  

Given that the topological term charges the statistics of vortices to be $e^{2\pi i k}$, we expect that the bulk dynamics are only sensitive to the value of $k$ mod 1. This is indeed true, and this fact is responsible for constraining the critical exponents of the SF-tMI transition to be equal to those of the usual SF-MI transition. To see this, note that
\begin{equation}
    S_{g, k+1}[\phi] - S_{g, k}[\phi] = 2\pi i \int\Big(d\phi d\toZ{d\phi} - \toZ{d\phi} d\toZ{d\phi}\Big)
\end{equation}
The first term is a surface term, while the second is an integer multiple of $2\pi i$ and may be dropped. Hence, away from a boundary, the bulk dynamics at $k$ and $k+1$ are identical. All correlation functions of local operators, and hence all critical exponents, are identical (see Appendix \ref{app:LS} for a physical interpretation). We will call the bulk invariance under $k\to k+1$ the ``level shift symmetry''.

Continuum topological $\theta$ models have similar level shift symmetries. From a theoretical perspective, shifting the level in this model corresponds to adding an SPT order. In this particular case, it is remarkable that doing so changes the action by only a surface term and therefore does not change the bulk dynamics of the system for any local operator and hence leads to identical critical behavior of local operators at the SF-MI and SF-tMI critical points. More generally, this may be true not just for the SF-tMI transition but for generic symmetry breaking transitions. If two topological phases have the same symmetry, differ only by the addition of an SPT or invertible topological order, and undergo symmetry-breaking phase transitions to the same symmetry-breaking phase, similar arguments may imply that the critical behavior of the two models is identical --- but more study will be required in this area.  

Returning to the model at hand, the level-shift symmetry 
under $k\to k+1$ has important implications for the action of time-reversal symmetry. 
Note that time-reversal reverses the orientation of the lattice and so exchanges the representative of the top cohomology, effectively changing the sign on the integral in the topological term $S_k[\phi]$, which consequently is odd under time reversal. This implies that if $2k \in \mathbb Z$, then the bulk dynamics are time-reversal symmetric. 
\begin{figure}
    \centering
    \includegraphics[width=.45\columnwidth]{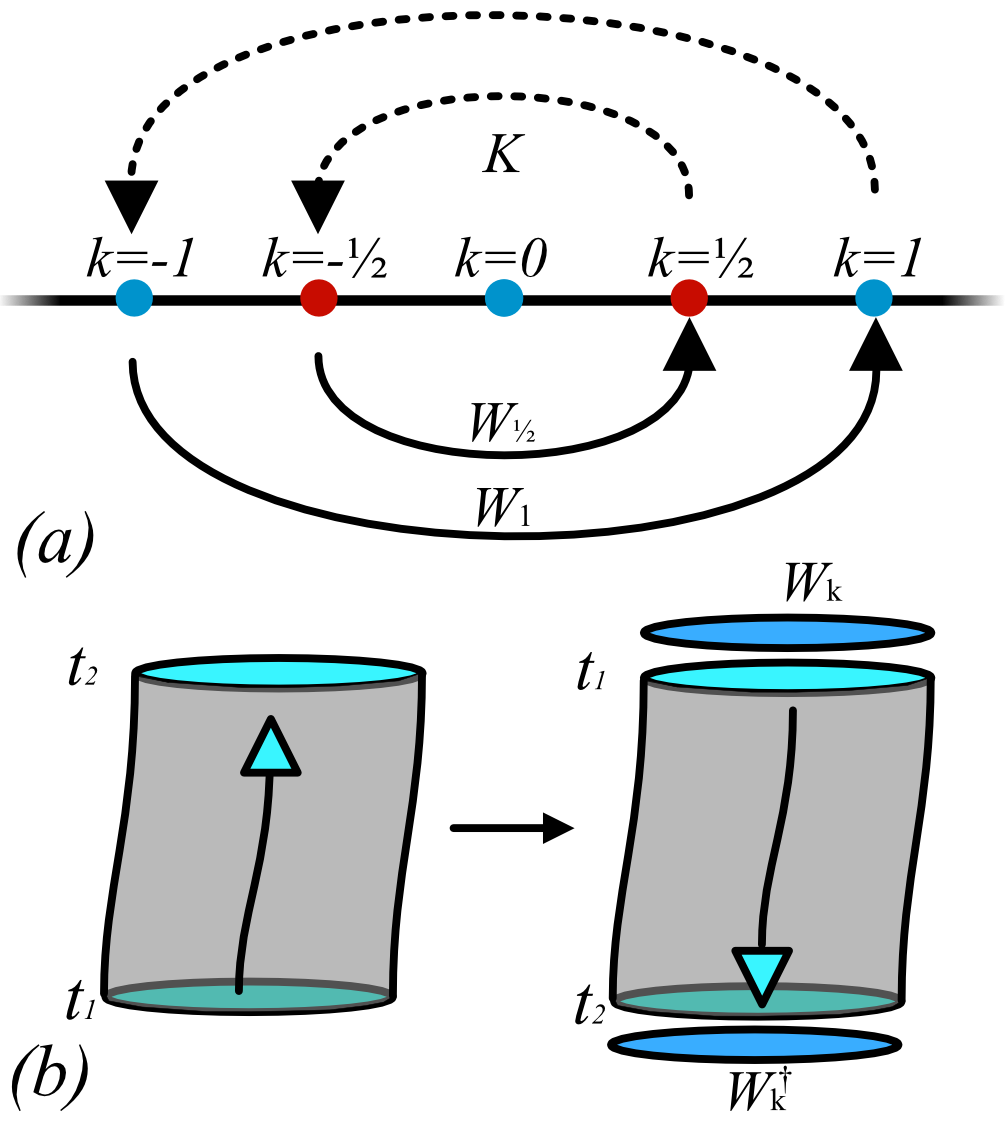}
    \caption[Time-reversal and Boundary Effects]{\emph{(color online)} For $2k\in \Z$, the action of the $W_k$ operators cancels out the effects of time-reversal on the boundary of an open manifold.}
    \label{fig:time_reversal}
\end{figure}

While the standard time reversal symmetry does not hold on the boundary, a modified time-reversal symmetry does. Time-reversal symmetry holds for $2k \in \Z$ because $S_{g, k}$ and $S_{g, -k}$ differ only by a surface term:
\begin{equation}
S_{g, k} - S_{g, -k} = 2\pi i (2k) \int d(\phi d\toZ{d\phi})
\end{equation}
On a closed manifold, this surface term vanishes, and the theory at $k$ and its time-reversed conjugate at $-k$ are identical. To extend this symmetry to open manifolds, we need to cancel the leftover effects on the boundary (See Figure \ref{fig:time_reversal}). Let us consider a spacetime manifold $\cM^3$ with a spatial boundary $\cB^2$. For $2k\in \mathbb Z$, we define the operators $\hat\E_{2k} = e^{- 2\pi i (2k)\int_{\cB^2} \phi d\toZ{d\phi}}$, and let $\hat K$ be the complex conjugation operator. We define the time-reversing operators:
\begin{equation}
    \T_k = \hat K \hat \E_{2k}
\end{equation}
For $k=0$, this reduces to the usual time-reversal operator. For other integer or half-odd-integer $k$, the operator $\hat \E_k$ corresponds to shifting the level by $2k$. This is precisely the the boundary change of the bulk level under shifting $k$ by an integer. For each $k$ with $2k\in \Z$, the model is invariant under the $\T_k$ operator.

\begin{figure}
    \centering
    \includegraphics[width = .55\columnwidth]{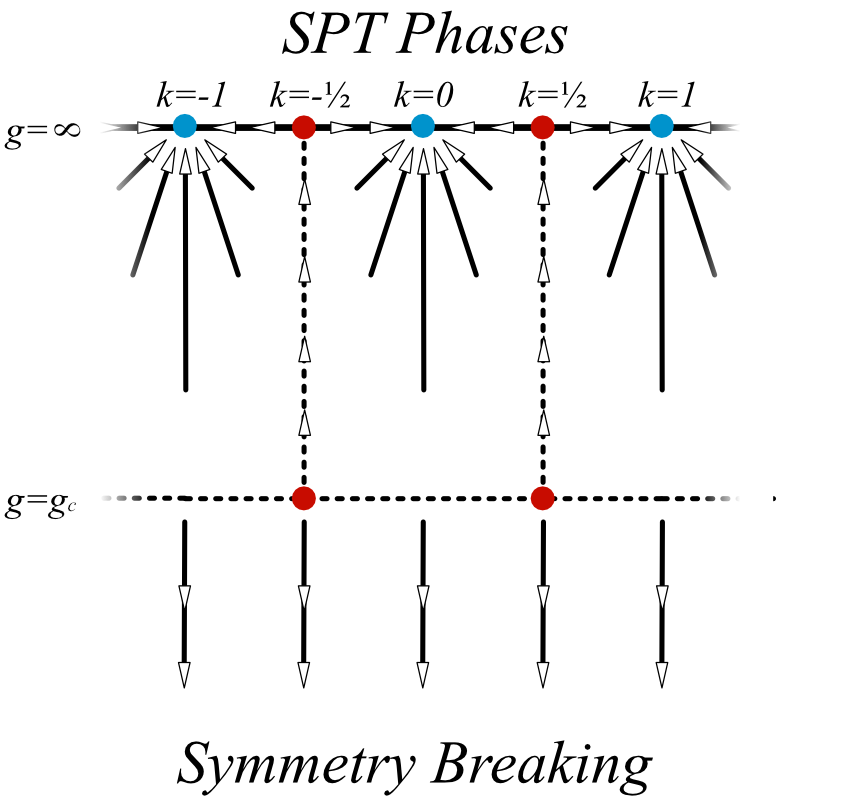}
    \caption[Renormalization Group Flow for the $U(1)$ Topological $\theta$-term Model]{\emph{(color online.)} Because the bulk dynamics are invariant under $k\to k+1$, we may construct a sketch of the RG flow diagram. Along the $g=\infty$ axis are the $\T_k$ fixed points, while at $g=g_c$ we have the SF-(t)MI transitions.}
    \label{fig:Phase_Diagram_Sketch}
\end{figure}

Together, these observations lead to the RG flow diagram shown in Figure \ref{fig:Phase_Diagram_Sketch}. The line at $g_c$ separates the disordered phase from the symmetry breaking phase. As there are few vortices in the symmetry breaking phase, we expect that the topological term will be only be relevant in the MI phase. The novel parts of this model are the disordered fixed points with modified time-reversal symmetry with $2k \in \Z$ at $g\to \infty$. Later, we will see that the integer fixed points with $k\in \Z$ describe gapped SPT phases with Hall conductance, while the half-odd-integer fixed points describe the topological transition between tMIs at $k$ and $k+1$.

In the next section, we will study the integer $k$ phases in detail. We will discuss in detail how to see that they describe the topological Mott insulators, and will describe the physical mechanisms behind the SF-tMI phase transition and the tMI Hall conductance. In Section \ref{chap6:sec:Fermions}, we will study the half-odd-integer phases, and argue that they are the transitions between the tMIs.

\section{The Integral Phases}\label{sec:IntegerPhases}
We now consider the fixed point phases with $g\to\infty$ and $k\in \Z$. In this case, the action reduces to:
\begin{equation}
2\pi i k \int d\phi d\toZ{d\phi} = 2\pi i k \int d(\phi d\toZ{d\phi})
\label{eq:IntAction}
\end{equation}
This is the action of the exactly solvable model in \cite{DeMarco:2021erp}, where it is shown that this action creates a ground state which has a nontrivial Chern number $2k$ and Hall response of $2k \frac{e^2}{h}$. In what follows, we will see that the ground state consists of a condensate of charged vortices, and that this condensation leads to an SPT phase with Hall conductance.

Before proceeding to that ground state physics, it will be useful to examine the model coupled to a background gauge field. In the SF phase\footnote{Up to one-cup products, the minimally coupled gauged action and the one derived by ungauging the lattice Chern-Simons action differ by a term:
\begin{equation}
    4\pi i k \int (d\phi - A)(\toZ{dA} + d\toZ{d\phi - A}) = 2\pi i k \int (d\phi - A)\toZ{\star j}
\end{equation}
which vanishes in the SF phase, since there $\toZ{\star j} = 0$. In the tMI phase, we take the action derived by ungauging. Note that the ungauged action is only rotor redundant for $k\in \Z$, which means that we do not know how to couple the model to a background gauge field in the tMI phase for non-integer $k$.}, we may minimally couple the action to get:
\begin{multline}
    S_{g, k}[\phi; A] = \frac{1}{g}\sum_{\braket{i,j}}\cos 2\pi ((d\phi)_{ij} - A_{ij}) 
    \\
+2\pi i k \int (d\phi - A - \toZ{d\phi-A})d(A + \toZ{d\phi-A})\label{eq:gauged}
\end{multline}
where we have used the form of the action in eq. (\ref{eq:Sk}) to create the minimal coupling. As with $\phi$, we have also taken $A$ to have a period of unity, rather than $2\pi$. Correspondingly we have two rotor redundancies
\begin{align}
    \phi_i \to \phi_i+n_i \\
    A_{ij}\to A_{ij} +m_{ij}
\end{align}
with $m, n \in \Z$, in addition to the usual gauge symmetry:
\begin{align}
    \phi_i \to \phi_i + \varphi_i \nonumber \\ 
    A_{ij} \to A_{ij} + \varphi_i - \varphi_j
\end{align}
and global $U(1)$ symmetry. With $k\in \mathbb{Z}$, the topological part of the action in the superfluid phase becomes:
\begin{equation}
    S_{k}[\phi; A] = 
+2\pi i k \int 
\Big((d\phi-A) dA  -\toZ{d\phi - A}dA \\+ (d\phi - A)d\toZ{d\phi - A}\Big),\label{eq:int_gauged}
\end{equation}
where the term $\int \toZ{d\phi-A} d \toZ{d\phi-A}$ has been dropped on account of it being an integer.
With the background field, the vortex current becomes:
\begin{equation}
    \star \jmath_v = d(d\phi - A - \toZ{d\phi - A}) = -(dA + d\toZ{d\phi - A})
\end{equation}
Evaluating (\ref{eq:int_gauged}) on a closed manifold and substituting in the vortex density, the topological part of the action becomes:
\begin{equation}
    S_{k}[\phi; A] = 
    2\pi i k \int \big(AdA + A\cup(\star \jmath_v) + (\star\jmath_v)\cup  A\big)\label{eq:chargedVortices}
\end{equation}
Thus we see a second effect of the topological term: in addition to giving the vortices mutual statistics at non-integer $k$, at integer $k$ it gives vortices a charge of $2k$. In turn, we will see next that it is the condensation of these charged vortices that gives rise to the tMI phase.

\subsection{Ground State and SPT Order}\label{sec:IntegerPhases:GroundState}

A closer look at the action (\ref{eq:IntAction}) above reveals an apparent paradox. The action would seem to be trivial, as the Lagrangian $d\phi d\toZ{d\phi} = d(\phi d\toZ{d\phi})$ is a total derivative, and yet we will see \cite{DeMarco:2021erp} that the state is a stable, gapped phase with Hall conductance. The resolution to this reflects the SPT nature of the tMI phase. In particular, the term which the Lagrangian is a total derivative of, viz. $\phi d\toZ{d\phi}$, is itself not locally $U(1)$ symmetric. On a closed manifold, one may rewrite the Lagrangian as a total derivative of $-d\phi [d\phi]$, but in this case the rotor redundancy does not hold locally, and the model thus fails to even be well-defined. There is no way to write the Lagrangian total derivative of a term which is locally both $U(1)$ symmetric and rotor redundant. 

Thus the apparent paradox is not a paradox at all. The Lagrangian is indeed a total derivative, but it is not a derivative of anything $U(1)$ symmetric and rotor redundant. If we break $U(1)$ symmetry, then the model is trivial, and the action can be canceled by an allowed total derivative term. On the other hand, so long as $U(1)$ symmetry is preserved (along with rotor redundancy, which is always required) then the action is nontrivial. 

At the same time, the model is easily solved because it is a total derivative. In particular, the bulk dynamics are trivial: if $\partial \cM^3 = 0$, then the action vanishes and the partition function simply becomes unity. However, the ground state of the model, exposed on a spatial boundary of spacetime, contains nontrivial physics.

We can see this behavior directly from the ground state wavefunction, by which we mean the wavefunction created on the boundary. Specifically, we obtain the ground state on a $2$d manifold $\cB^2$ by evaluating the action \eqref{eq:IntAction} on a spacetime $\cM^3$ with (spacelike) boundary $\cB^2 = \partial \cM^3$. Because the action is a surface term, we can immediately write down the $g\rightarrow\infty$ ground state:
\begin{equation}
\ket{\psi_k}=\int D\phi e^{2\pi i k \int_{\cB^2} \phi d\toZ{d\phi}}\ket{\{\phi\}}
\label{eq:GroundState}
\end{equation}
where $\ket{\{\phi\}} = \otimes_i \ket{\phi_i}_i$ and again where we gauge-fix the measure as $\int D\phi = \prod_{i\in \cB^2} \int_{-\frac{1}{2}}^{\frac{1}{2}} d\phi_i$.

To understand the ground state further, let us first examine the $k=0$ case. In that case, the $g\rightarrow \infty$ ground state is 
\begin{equation}
\ket{\psi_{k=0}}=\int D\phi \ket{\{\phi\}} = \otimes_i \ket{0}_i\label{eq:keq0gs}
\end{equation}
where $\ket{0} = \int_{-\frac{1}{2}}^{\frac{1}{2}} d\phi \ket{\phi}$ is the $U(1)$ symmetric state with eigenvalue $1$. This is the ground state of a trivial Mott insulator. Recall that $\phi_i$ should be thought of as the phase of a particle at site $i$. In the Mott insulating phase, the particle wavepackets do not overlap and their phases fluctuate independently, as in eq. (\ref{eq:keq0gs}) This should be compared to deep in the symmetry breaking phase, where phase rigidity develops and where all the $\phi_i$ are equal in the ground state. For nonzero $k$, the MI wavefunction is `twisted' by the operator
\begin{equation}
    \mathcal{\hat E}_k = e^{2\pi i k \int \hat \phi d\toZ{d\hat \phi}} = e^{-2\pi i k \int \hat \phi \hat \rho_v}
\end{equation}
so that 
\begin{equation}
    \ket{\psi_k} = \mathcal{\hat E}_k \ket{\psi_{k=0}}
\end{equation}
Thus $\mathcal{\hat E}_k$ is the operator that provides the SPT entanglement to twist the trivial MI into a tMI. Note that on a closed manifold, $\hat{\mathcal{E}}_k$ is $U(1)$ invariant, as under $\phi \to \phi+\theta$,
\begin{equation}
    \hat \E_k \to e^{2\pi i k \theta \int \hat \rho_v}\hat \E_k
\end{equation}
On a closed manifold, the total vortex number must vanish, and so $\int \hat \rho_v = 0$. This is however not necessarily true on a manifold with boundary; hence the $U(1)$ symmetry is anomalous and breaks in the presence of a boundary.

We can also see this behavior locally. Let us split the $\hat \E_k$ into operators on each plaquette $\Delta$ that transform the $k=0$ wavefunction into a nontrivial $k$ ground state, namely: 
\begin{equation}
M_k[\Delta] = e^{-2\pi i k \int_{\Delta} \phi \rho_v}
\end{equation}
so that 
\begin{equation}
    \ket{\psi_k} = \prod_\Delta M_k[\Delta] \ket{\psi_{k=0}} = \hat \E_k \ket{\psi_{k=0}}
\end{equation}
On any plaquette $\Delta$, $M_k[\Delta]$ is not $U(1)$ invariant. However, when multiplied together over a closed manifold, the $M_k$ are indeed $U(1)$ invariant, as
\begin{equation}
    \prod_{\Delta\in \cB^2} M_k[\Delta]
    = \mathcal{\hat E}_k
\end{equation}
which, as discussed above, is $U(1)$ invariant. This behavior is how SPTs are often characterized: if we relinquish $U(1)$ symmetry, then a finite set of local unitary transformations (the $M_k[\Delta]$) suffices to transform the trivial ground state $\ket{\psi_{k=0}}$ into a general $\ket{\psi_k}$. On the other hand, if we require $U(1)$ symmetry, then only the nonlocal operator $\mathcal E_k = \prod_{\Delta} M_k[\Delta]$ suffices.

\subsection{Hall Conductance}

\begin{figure}
    \centering
    \includegraphics[width=.25\columnwidth]{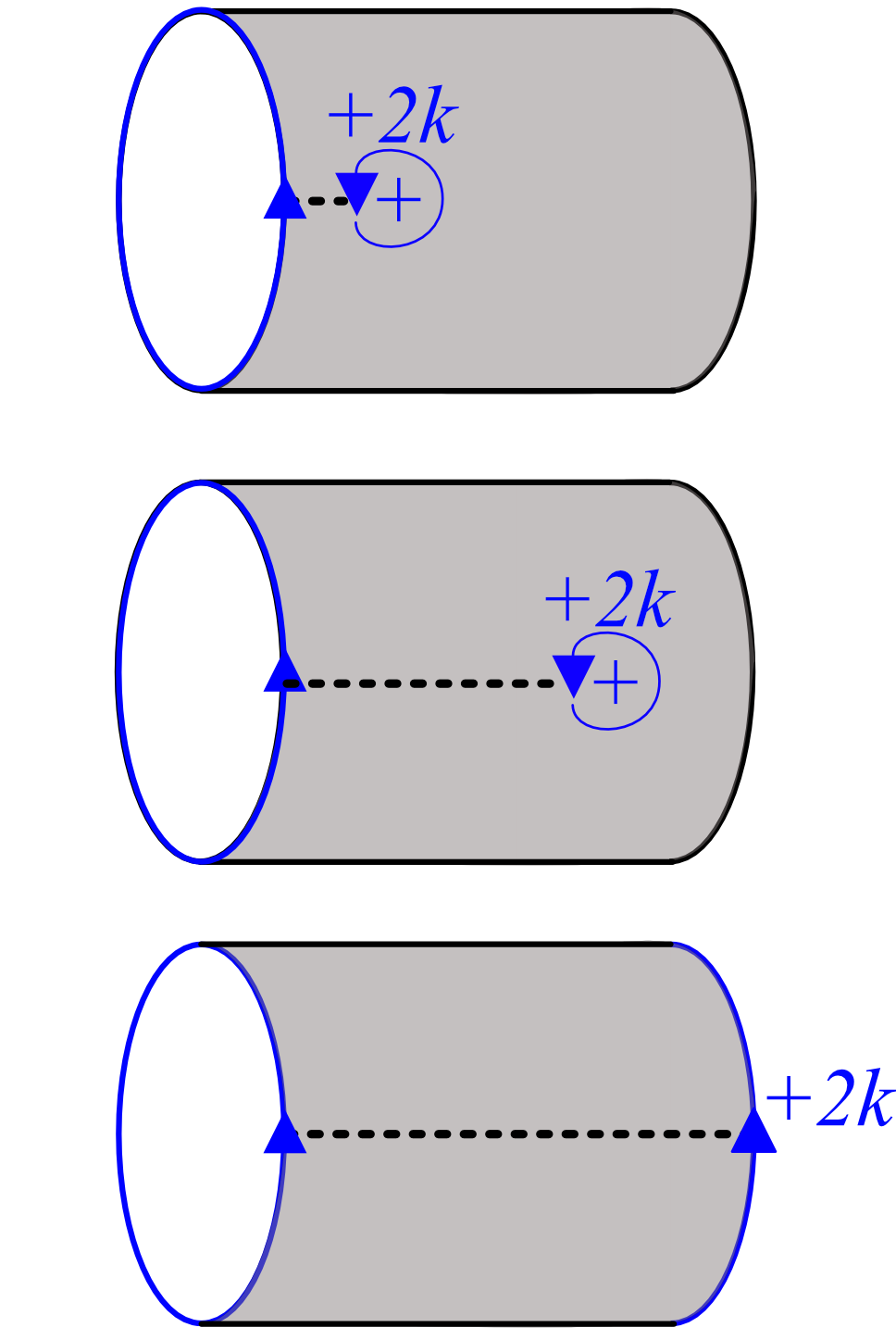}
    \caption[Vortices and the Laughlin Thought Experiment]{\emph{(color online).} In the Laughlin thought experiment, twisting the boundary conditions of an open cylinder by a flux quantum transfers charge from one edge to another. We may understand the twisting boundary conditions as the tunneling of a vortex from one edge to another. Because the vortices carry charge $2k$, twisting boundary conditions transfers charge $2k$ from one edge to another and the Hall conductance is $2k$.}
    \label{fig:HallConductance}
\end{figure}

The topological invariant for these phases is the Hall conductance. In Section \ref{sec:ComProj}, the Chern number of the integer ground states will be calculated explicitly by considering the model with twisted boundary conditions, and we will find it to be $2k$. Physically, we can understand this result in the context of the Laughlin thought experiment. Consider the system on an open cylinder, as shown in Fig. \ref{fig:HallConductance}. We wish to twist the boundary conditions around the open direction by unity. Let $\ket{\psi_k, \theta = 0}$ be the ground state at level $k$ before twisting of boundary conditions and $\ket{\psi_k,\theta = 2\pi}$ be the ground state after an adiabatic twisting process. The two are related by the the creation of a vortex/anti-vortex pair at opposite edges of the sample, with the resulting branch cut providing the $2\pi$ phase jump:
\begin{equation}
    \ket{\psi_k,\theta=2\pi } = \hat V_{\Delta, \Delta'}\ket{\psi_k, \theta = 0},
\end{equation}
where $\hat V_{ij}$ creates a vortex at plaquette $\Delta$ and an anti-vortex at plaquette $\Delta'$. 
We can physically imagine this as dragging a vortex across the cylinder in order to create the branch cut. However, we have seen that the vortices have charge $2k$, and so dragging a vortex across the sample transfers charge from one side of the sample to the other, resulting in a hall conductance of $2k\frac{e^2}{h}$.

This Hall conductance is measurable through the tMI phase, and should persist in the vicinity of the SF-tMI phase transition. Taking into account (\ref{eq:chargedVortices}) near a phase transition, an effective field theory becomes:
\begin{equation}
    S_{\text{eff}}=\alpha(\rho_{SF}) \int (\Pi_T A)^2 + 2\pi i k \int AdA 
    + \beta(\rho_{SF}) \int (\Pi_T A) d(\Pi_T A) + ...
\end{equation}
where $\Pi_T$ is the transverse projector. The functions
$\alpha(\rho_{SF}), \beta(\rho_{SF})$
must vanish with $\rho_{SF} \to 0$, since they may only arise from integrating out gapless modes. In particular, as the transition is continuous, no term may suddenly cancel the transverse conductance due to the Chern-Simons term.

We have now examined the tMI integer phases of the model, examined the the entanglement structure of the tMIs, understood the SF-tMI transition as the proliferation charged vortices, and explained how this leads to a Hall conductance in the disordered phase and near the transition. In the next section, we create a commuting projector model for the tMI phase.

\subsection{Commuting Projector Model}\label{sec:ComProj}
Given that the bulk behavior of the path integral is trivial, we expect that we
should be able to create an exactly solvable Hamiltonian model to describe the
time-evolution. Furthermore, because the path integral defines a wavefunction
on any spatial boundary independently of the bulk dynamics, we expect that this
Hamiltonian model should be a commuting projector onto a ground state. As we
shall now see, both are true. 

Commuting projector models have been a central tool for understanding the new
zoo of theories. Employed most famously by Kitaev \cite{Kitaev:1997wr} to
provide an exactly solvable model for the previously proposed 2+1d $Z_2$
topological order \cite{RS9173,W9164} with emergent fermions and anyons, they
now describe models for a wide class of string-net topological order
\cite{Levin:2004mi}, recently unleashed a flurry of research on fractons
\cite{Cc0404182,Vijay:2015mka}, and continue to underlie our microscopic
understanding of exotic phases.

It is quite surprising then that no commuting projector model has been
discovered for gapped phases with non-zero Hall conductance. It was commonly
believed that none could exist, and recently a no-go theorem has been proposed
\cite{Kapustin:2020aa}, ruling out a large class of potential theories with a
finite Hilbert space on each site. 

\begin{figure}
\centering
\includegraphics[width = .6\columnwidth]{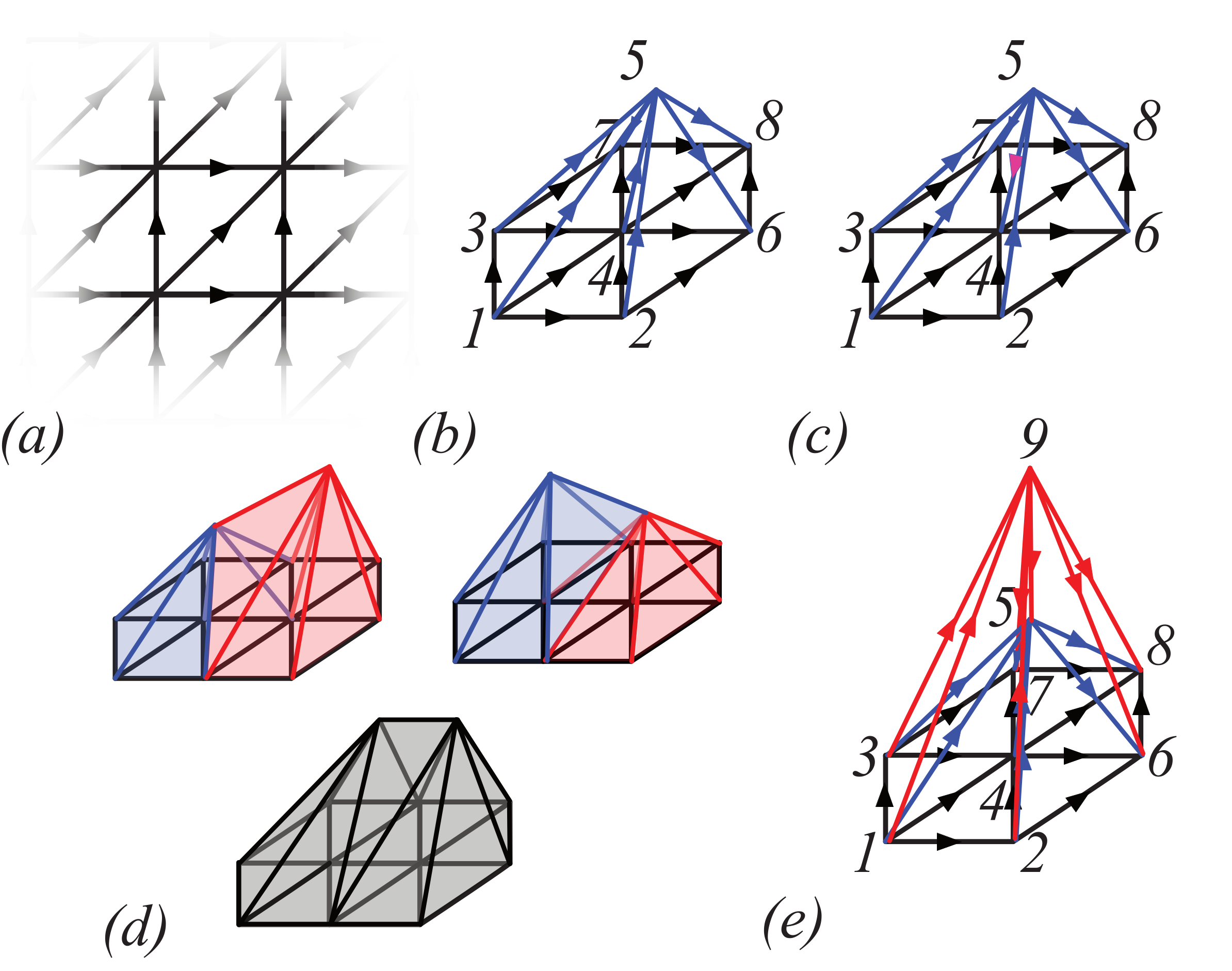}
\caption[Commuting Hermitian Projectors]{\emph{(Color Online).} We construct a commuting projector model for
the lattice in $(a)$ by evaluating the spacetime path integral for the complex
in $(b)$ and turning the amplitude into operators. Because the path integral
contains only a surface term, we can show that $(c)$ the matrix is hermitian,
(d) the operators commute, and $(e)$ they are projectors.}
\label{fig:Hamiltonian_Lattice}
\end{figure}

We can use the spacetime formalism to
construct a commuting projector Hamiltonian on a triangular lattice of the sort
shown in Fig. \ref{fig:Hamiltonian_Lattice}a.  To do so, we consider the time
evolution of a single site, $\phi_4 \to \phi_5$, while preserving the
orientation of lattice links as shown in fig \ref{fig:Hamiltonian_Lattice}b.
Evaluating the path integral on the complex shown yields the matrix elements
for the transition $\phi_4 \to \phi_5$ as a function of the surrounding
$\phi_i$:
\begin{multline}
M_{\phi_{4} \to \phi_{5}}(\phi_{1}, ..., \phi_{8})
=
\exp\Big\{
2 \pi i k\Big[ \\
  \phi_{0}\Big(\toZ{\phi_{5} - \phi_{2}} + \toZ{\phi_{3} - \phi_{5}}  
  + \toZ{\phi_{4}-\phi_{3}} + \toZ{\phi_{2} - \phi_{4}}\Big)
  \\
+ \phi_{2}\Big(\toZ{\phi_{5} - \phi_{6}} + \toZ{\phi_{2} - \phi_{5}}  + \toZ{\phi_{4} - \phi_{2}} + \toZ{\phi_{6} - \phi_{4}}\Big)
\\
+ \phi_{3}\Big(\toZ{\phi_{7} - \phi_{5}} + \toZ{\phi_{4} - \phi_{7}}  + \toZ{\phi_{3} - \phi_{4}} + \toZ{\phi_{5} - \phi_{3}}\Big)
\\
+ \phi_{5}\Big(\toZ{\phi_{6} - \phi_{5}} + \toZ{\phi_{8} - \phi_{6}}   + \toZ{\phi_{7} - \phi_{8}} + \toZ{\phi_{5} - \phi_{7}}\Big)\\ 
+ \phi_{4}\Big(\toZ{\phi_{7} - \phi_{4}} + \toZ{\phi_{8} - \phi_{7}} + \toZ{\phi_{6} - \phi_{8}} + \toZ{\phi_{4} - \phi_{6}}\Big)
\Big]
\Big\}\label{eq:M_itoj}
\end{multline}
We will interpret this transition amplitude as the matrix element for an
operator $\hat M_4$ acting on site-4. However, eq. (\ref{eq:M_itoj}) is
somewhat daunting. Let us set consider $\hat M_4$ as an operator acting only on
the Hilbert space on site-4. If $k=0$, then $\bra{\phi_{4}'}\hat
M_4\ket{\phi_{4}} = 1$, and $\hat M_4$ is simply the projector onto the
state with zero angular momentum in each $U(1)$. For nonzero $k$, we may
rewrite the transition amplitude as $M_{\phi_{4} \to \phi_{5}} =
\exp(2\pi i (f(\phi_{5}) -f(\phi_{4}))$ (note that this implies hermiticity) where $f(\phi)$ is a
function defined in the appendix that depends on $\phi$ and takes as parameters
$ \phi_{1}...\phi_{8}$, but not $ \phi_{4}$ or $\phi_{
5}$. This implies that, up to an overall phase, the $\hat M_i$ act as
\begin{equation}
\hat M \ket{\phi} \propto \int d\phi e^{2\pi i f(\phi)} \ket{\phi}
\end{equation}
We see then that these projectors may be thought of as `twisting' the zero
angular momentum state by a phase function $f(\phi)$ which depends on the
surrounding values of $\phi_{i}$. The phase itself is determined by the
cocycle of the action in (\ref{U1spt}). We may construct $\hat M_i$ for an entire lattice.

The $\hat M_i$ inherit a number of remarkable properties from the fact that the
$2+1$d path integral action contains only a surface term. First, they mutually
commute: consider the three spacetime complexes in figure
\ref{fig:Hamiltonian_Lattice}c, which addresses the only nontrivial case of two
adjacent $\hat M_i$. The two colored complexes correspond to time evolving
either the blue site followed by the red or the red followed by the blue,
respectively. Because the action contains only a surface term, and the surfaces
are identical, it assigns the same amplitude to both cases. Back in the
Hamiltonian picture, this implies that $[\hat M_i, \hat M_j] = 0$.

For the same reason, the $\hat M_i$ are projectors. The fundamental mechanism
is illustrated in Fig. \ref{fig:Hamiltonian_Lattice}d, where we see the
effect of time-evolving twice. In the language of eq. \ref{eq:M_itoj}, this is
the expression $M_{\phi_{5}\to\phi_{
9}}M_{\phi_{4}\to\phi_{5}}$; in the Hamiltonian picture this
is $\hat M_i^2$. However, because the path integral action depends only on the
values of $\phi$ on the surface, we could equally well drop site $5$ and
its associated links; the amplitude will not change. This implies that 
$\hat M_i^2 = \hat M_i$. 

The aforementioned hermiticity of the  $\hat M_i$ is also due to this reason, as reversing the orientation
of the link $\braket{4,5}$ in Fig. \ref{fig:Hamiltonian_Lattice}c changes the amplitude by complex conjugation.

There is a simple way to view $\hat M_i$.
We note that the groundstate wave function \eqref{eq:GroundState}
on a spatial lattice $\cM^2$ is a phase factor, which defines the unitary
operator 
\begin{align}
\hat{\mathcal{E}}_{k} =
e^{- 2\pi i \sum_{I,J} k\int_{\cM^2} \hat\phi_I^\RZ \hat{\rho}^v_J}, 
\end{align}
where $\hat{\rho}^v_J = -d \toZ{d\hat{\phi}^\RZ_J}$ is the vortex density operator defined previously. Let $\hat P_i$ be the projector onto the state of zero momenta at site $i$,
with matrix elements $\braket{\phi'_I(i) | \hat P_i | \phi_I(i)} = 1$.  Then we find:
\begin{equation}
\label{MUP}
\hat M_i =  \hat{\mathcal{E}}_{k} \hat P_i  \hat{\mathcal{E}}_{k}^\dagger
\end{equation}
Thus the $\hat M_i$ operators are essentially projectors onto zero angular
momentum states, twisted by the phases of the unitary operators
$\hat{\mathcal{E}}_{k}$.  On a spatial lattice without boundary,
the groundstate wave function \eqref{eq:GroundState} has $U(1)$ symmetry, i.e. is
invariant under $\phi_{i} \to \phi_{i} + \theta + n_{i}$.
Thus the $\hat M_i$ also have $U(1)$ symmetry. Note also that the commuting, projector, and Hermitian properties of $\hat M_i$ immediately follow from this description. The  ground state
wavefunction is given by \eqref{eq:GroundState} and can be written as $\ket{\psi_{k}} =
\hat{\mathcal{U}}_{k} \ket{\psi_0}$, where $\ket{\psi_0} = \otimes_{i}\ket{\ell = 0}_{i}$ is the product state of the zero-angular-momentum
states $\ket{\ell=0}$ on each site $i$. 

Using the mutually commuting projectors $\hat M_i$, we can define a Hamiltonian 
\begin{equation}
H = -g\sum_i \hat M_i
\end{equation}
to obtain a system with ground state $\ket \psi$ from eq. (\ref{eq:GroundState}) and gap $g$.

With a model producing the ground state eq. (\ref{eq:GroundState}) now in hand, we return to the Hall
conductance. We have already argued from the path integral that coupling the
system to a background gauge field leads to the expected Chern-Simons response
function, but here we appeal directly to the wavefunction of the system on a
torus to confirm the hall conductance. 

\begin{figure}
\centering
\includegraphics[width = .6\columnwidth]{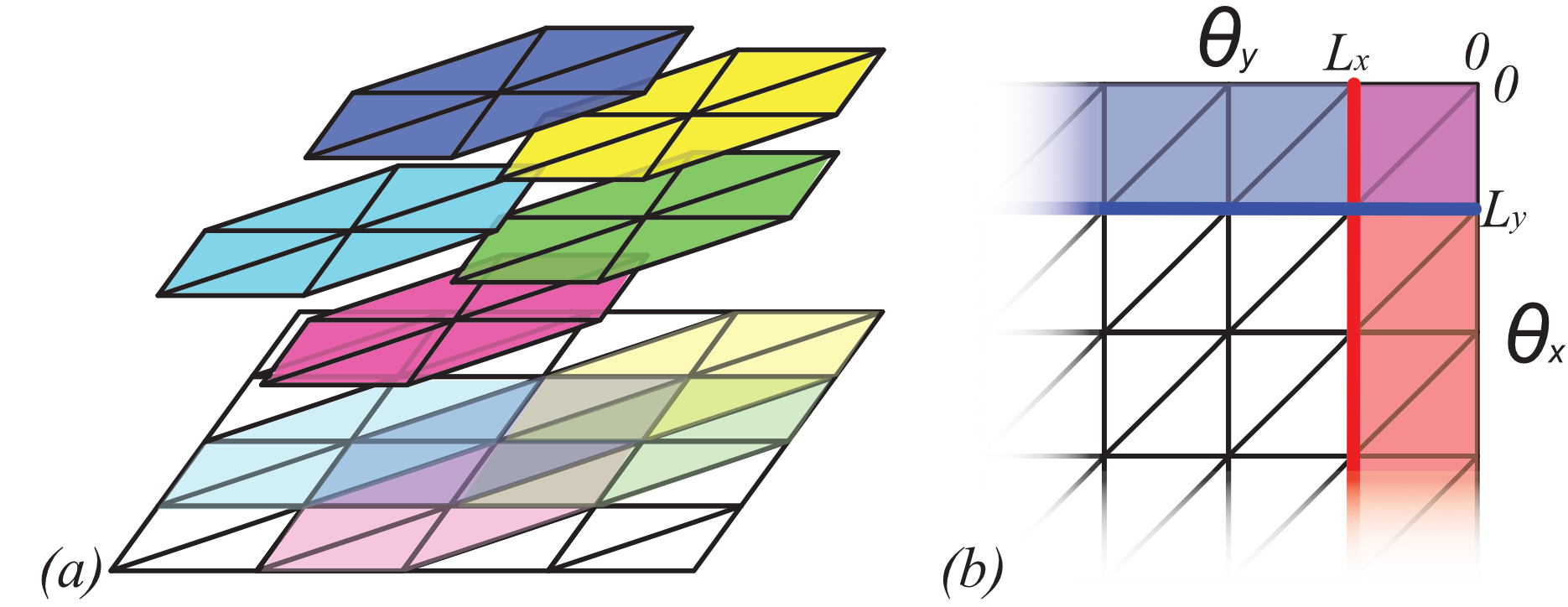}
\caption[Commuting Projectors and the Hall Conductance]{\emph{(Color Online)}. $(a)$ The commuting projectors act on a hexagon to define a ground state. $(b)$ To calculate the Chern number of the ground state wavefunction over the holonomy torus, we twist the boundary conditions by $\theta_x n$ and $\theta_y m$ in the $x$ and $y$ directions, respectively.}
\label{fig:HCSU}
\end{figure}

Consider the wavefunction on the lattice shown in Fig. \ref{fig:HCSU}. We twist the boundary conditions by $\theta_x n$, $\theta_y
m$, 
so that $\phi_{x=0, y}= \phi_{x=L_x, y} - \theta_x n$ and $\phi_{x, y=0} = \phi_{x, y=L_y} - \theta_y m^I$,
where the integral vectors $n_m$ allow us
to describe both direct ($n = m$) and mixed hall responses. Denote the
ground state with boundary conditions $\theta_x, \theta_y$ by $\ket{\theta_x,
\theta_y}$. 
We are interested in the phase accumulated when we twist the boundary
conditions by an integer, sending $\theta_x \to \theta_x + \ell_x, \theta_y \to
\theta_y + \ell_y$ with $\ell_x, \ell_y \in \mathbb Z$. 

Consider first the plaquettes marked in red. The contribution of these plaquettes to the path integral is of the form:
\begin{multline}
\exp\Big\{
-2\pi i k 
\Big[
(\phi_{x=L, y} + n \theta_x - \phi_{x=L_x - 1, y})
\toZ{\phi_{x=L_x, y+1} - \phi_{x=L_x, y}}
\\\nonumber
 -
 \Big(
(\phi_{x=L_x - 1, y+1} - \phi_{x = L_x-1, y})
\toZ{\phi_{x=L_x, y+1} + n\theta_x - \phi_{x=L_x -1, y+1}} 
\Big)
\Big]
\Big\}
\end{multline}
Incrementing $\theta_x$ by $\ell_x$ changes the first term by an integer, while the second term changes by $(\phi_{x=L_x - 1, y+1} - \phi_{x = L_x-1, y})\ell_x$, and so the multiplicative change on the path integral is:
\begin{equation}
e^{2\pi i k (\phi_{x=L_x - 1, y+1} - \phi_{x = L_x-1, y})n\ell_x }
\end{equation}
Similarly, the plaquettes marked in blue change by:
\begin{equation}
e^{-2\pi i k (\phi_{x+1, y=L_y-1} - \phi_{x, y=L_y-1})m\ell_y }
\end{equation}
On the purple plaquettes at the corner, the contribution to the path integral is:
\begin{multline}
\exp\Big\{
-2\pi i k 
\Big[
\Big((\phi_{x=L_x, y=L_y-1} + n \theta_x - \phi_{x=L_x-, y=L_y-1})
\toZ{\phi_{x=L_x, y=L_y} + m\theta_y - \phi_{x=L_x, y=L_y-1}} 
\Big)
\\
-\Big(
(\phi_{x=L_x-1, y=L_y} + m\theta_y - \phi_{x=L_x-1, y=L_y-1})
\toZ{\phi_{x=L_x, y=L_y} + n\theta_x - \phi_{x=L_x-1, y=L_y}} 
\Big)
\Big]
\Big\}
\end{multline}
which will change by:
\begin{multline}
\exp\Big\{
-2\pi i k 
\Big[
(\phi_{x=L_x, y=L_y-1} + n \theta_x - \phi_{x=L_x-, y=L_y-1}) m\ell_y
\\
-(\phi_{x=L_x-1, y=L_y} + m\theta_y - \phi_{x=L_x-1, y=L_y-1})n\ell_x 
\Big]
\Big\}
\end{multline}
Combining the change on the red, blue, and purple plaquettes, the overall change to the wavefunction is:
\begin{equation}
e^{-2\pi i k \left[
n \theta_x m\ell_y
-m\theta_y n\ell_x
- n\ell_x \int_{\gamma_1}d\phi 
+ m\ell_Y\int_{\gamma_2} d\phi
\right] }
\end{equation}
where $\gamma_1, \gamma_2$ are the red and blue loops in Fig. \ref{fig:HCSU}, respectively. As that $\gamma_1, \gamma_2$ are closed, the sums along those curves vanish. Hence the wavefunction transforms as:
\begin{equation}
\ket{\theta_x+\ell_x, \theta_y+\ell_y} =
e^{-2\pi i k \left[
n \theta_x m\ell_y
-m\theta_y n\ell_x
\right] }
\ket{\theta_x, \theta_y}
\label{eq:ChernPhase}
\end{equation}
To make sense of this, apply the gauge transformation:
\begin{equation}
\ket{\theta_x, \theta_y} \to e^{-2\pi i k n m \theta_x \theta_y} \ket{\theta_x, \theta_y} \
\end{equation}
In this gauge, one may replace the phase in eq. (\ref{eq:ChernPhase}) by:
\begin{equation}
e^{-2\pi i k
(n  m + n m)\theta_x\ell_y }
\end{equation}
We should recognize this as the boundary conditions for a particle on a torus with flux:
\begin{equation}
2k mn
\end{equation}
We see then that our Hamiltonian system has the (mixed) Hall conductance
$n\cdot k \cdot m$, in agreement with the Chern-Simons response function
derived from the spacetime path integral. In the case of $\kappa = 1$ with $n =
m = 1$, this becomes a system with integer hall conductance $K = 2k$ wich is an even
integer. 

We must also understand how this model evades the no-go theorem. The infinite dimensional on-site Hilbert space of the rotors, combined with an
action that is not a continuous function of the field variables, is what allows
this model to exist. The discontinuous action is critical for commuting
projectors; if it were continuous, then one could truncate the Hilbert space to
low-angular momentum modes and render the on-site Hilbert space finite while
retaining the full $U^\ka(1)$ symmetry and commuting-projector property, hence
running afoul of the no-go theorem \cite{Kapustin:2020aa}. Conversely, the
no-go theorem assumes that the ground state wavefunction is a finite Laurent
polynomial in $e^{i\theta_x}, e^{i\theta_y}$, an assumption which is violated
in our model (See the details of the Chern number calculation in the Appendix
for an example). This commuting projector model represents a fixed-point theory for $U^\kappa(1)$ SPT phases with nonzero Hall conductance; it may be that any such fixed-point theory requires an infinite on-site Hilbert space.

\section{Topological Transitions at Half-Integer $k$}\label{chap6:sec:Fermions}
We have now established that the phases at integral $k$ are gapped SPT states. We also understand the $XY$ transition caused by increasing $g$ along $k=0$ and, by level-shift symmetry, for all integer $k$. In this section, we examine a different phase transition, that obtained by changing $k$ along the line $g=\infty$. We argue that that this must be a topological transition, and map it to a model of superconducting fermions. Whether the resulting order parameter is complex or real, and thus the superconductor breaks time-reversal symmetry or is gapless, corresponds to whether the transition is first-order or second order. We are, however, unable to discern which of the two occurs\footnote{The nature of the transition may depend on lattice details.}. 

Taking $g\to\infty$ and setting $k = m -\frac{1}{2}$, with $m\in\Z$, the action (\ref{eq:action}) becomes:
\begin{equation}
S_{g\to\infty, k=m - \frac{1}{2}}[\phi] = \pi i (2m - 1) \int_{\cM^3} (d\phi - \toZ{d\phi})d\toZ{d\phi}
\end{equation}
This action is invariant under the time-reversing operator $\T_{m+\frac{1}{2}}$ defined previously. If $\partial \cM = \emptyset$, then the action reduces to:
\begin{equation}
S_{g\to\infty, k=m - \frac{1}{2}}[\phi] = -\pi i (2m - 1) \int_{\cM^3} \toZ{d\phi}d\toZ{d\phi}. \label{eq:khalfbulk}
\end{equation}

As a first measure, note that we may rewrite the bulk action as the boundary of a time-reversal symmetric term in one higher dimension. For a four-manifold $\mathcal{N}^4$ such that $\partial \mathcal{N}^4 = \cM^3$,
\begin{equation}
    -\pi i (2m - 1) \int_{\cM^3} \toZ{d\phi}d\toZ{d\phi}
    = 
    -\pi i (2m - 1) \int_{\mathcal N^4} d\toZ{d\phi}d\toZ{d\phi}
\end{equation}
Note that this action, after exponentiation, is invariant under the usual time-reversal. 
For the our purposes, the critical fact is that time-reversal forces $m$ to be an integer in the four dimensional theory, and hence the four dimensional theory is a $U(1)\times \T$ SPT. Our three-dimensional theory, as the boundary of an SPT, must break the $U(1)$ or $\T$ symmetry, be gapless, or develop topological order.

We do not believe that these theories develop topological order. On the one hand, the integer $k$ phases are stable $U(1)$ SPTs; if the half-odd-integer phases were stable topological orders, then there would be to be another class of critical $k$ points, somewhere between the integer and half-odd-integer $k$ (recall Figure \ref{fig:Phase_Diagram_Sketch}). On the other hand, in Section \ref{sec:LatticeRG}, we show a numerical RG calculation that suggests that $k$ flows to the nearest integer, with the points where $k$ is a half-odd-integer separating different phases. We find no evidence of a stable phase at half-odd-integer $k$, nor do we find any fixed points besides the integer and half-odd-integer $k$.

Hence we are left with the options of the half-odd-integer phases breaking $U(1)\times \T$ symmetry (in which case the transition between the integer $k$ $U(1)$ SPTs would be first order) or gaplessness (in which case the transition is second order). We now argue that when $k$ is a half-odd-integer, the model admits a description in terms of  emergent fermions, which order in a way dependent on whether or not the transition is continuous.


\subsection{Emergent Fermions}

\begin{figure}
\begin{center}
\includegraphics[width = .45\columnwidth]{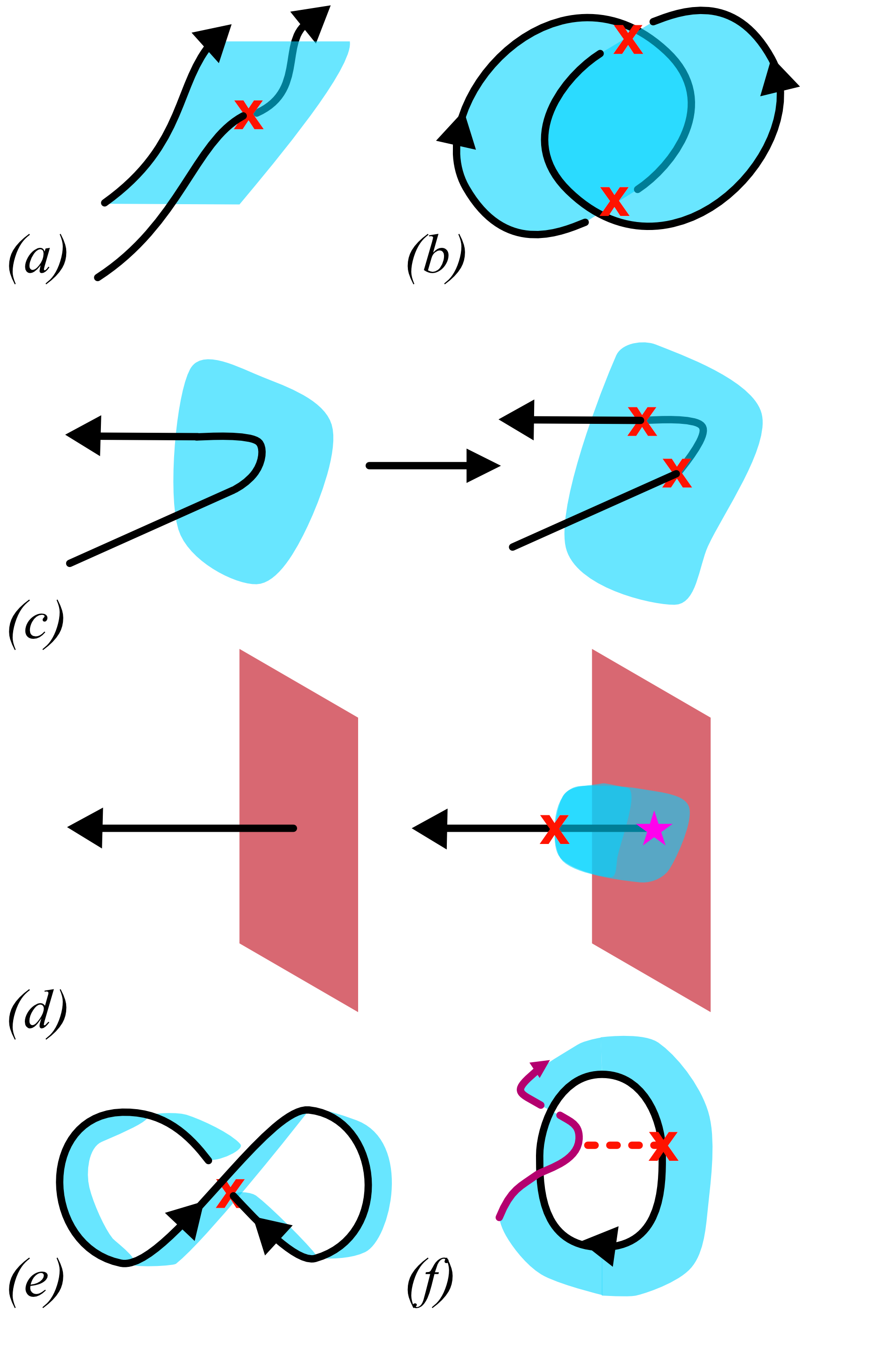}
\caption[Emergent Fermions at the Phase Transition]{\emph{(color online).} $(a)$ The topological term at half integer $k$ assigns a factor of $(-1)$ for each intersection (red $x$) of a vortex line (black) with a branch cut wall (blue). $(b)$ Disconnected lines must have two such intersections and so may not have anyonic statistics. $(c)$ Rotor redundancy $\phi\to \phi+n$ locally moves the branch cut walls, and always create intersections in pairs in the bulk. $(d)$ in the presence of a boundary (red), moving a domain wall can create a single intersection, and rotor redundancy is preserved by a charge at the end of the vortex line (magenta star) implemented by the $d\phi$ term.  A single vortex line describing the exchange of two particles has a single intersection, as does the (topologically equivalent) case of $(d)$ a $2\pi$ rotation of a single particle. Hence the vortices are fermions.}\label{fig:fermiStatistics}
\end{center}
\end{figure}

Let us work with the bulk action (\ref{eq:khalfbulk}). Recall that that the vortex current is given by $\jmath_v = -\star d\toZ{d\phi}$. Using this, we can rewrite this action as
\begin{equation}
    -\pi i (2m - 1) \int_{\cM^3} \jmath_v \frac{d}{\Box}\jmath_v
\end{equation}
Recall that in $2+1$d, the $1$-cochain $\toZ{d\phi}$ is dual to a two dimensional surface which describes the branch cut emanating from a vortex. The $2$-cochain $d\toZ{d\phi}$ is dual to a one-dimensional vortex line. The topological $\toZ{d\phi} d\toZ{d\phi}$ term counts the intersections of the vortex lines with the branch cut walls, i.e. twice the linking number (See figures \ref{fig:fermiStatistics}a, \ref{fig:fermiStatistics}b).

In the bulk, this action (\ref{eq:khalfbulk}) is invariant under the rotor redundancy because sending $\phi \to \phi+n$ corresponds to a moving a branch cut wall or creating/annihilating them in pairs. As shown in figure \ref{fig:fermiStatistics}, in the bulk, a local move of a branch cut wall must generate intersections in pairs, hence the action is invariant.

We can further understand the character of these vortices by examining their braiding. First, consider two disconnected, contractible lines. As shown in figure \ref{fig:fermiStatistics}b, these two lines must intersect in two places. In fact $\toZ{d\phi}d\toZ{d\phi}$ is twice their linking number; because the action assigns a factor of $(-1)$ to each intersection, these two lines have no braiding phase, and the vortices do not have anyonic statistics. 

However, a more complex situation arises when we consider just a single vortex line. In figure \ref{fig:fermiStatistics}e, we see a line representing the creation of two vortex-antivortex pairs, the exchange of two vortices, and the annihilation of the swapped pairs. Figure \ref{fig:fermiStatistics}f is topologically equivalent, and shows the case of a $2\pi$ rotation of a vortex. In either case, a single intersection is created, resulting in a factor of $(-1)$. Hence the vortices are fermions.

Now that we have identified the vortices as fermions, it may seem that the task is complete. The content of the action (\ref{eq:khalfbulk}) is exhausted, and a simple guess for the emergent behavior of the system may be completely free fermions. However, this misses a crucial factor. There is a Jacobian we neglected passing from the rotor field $\phi$ to the branch cut and vortex fields $\toZ{d\phi}$ and $\jmath = -\star d\toZ{d\phi}$. This Jacobian leaves a statistical imprint on the theory which will generate interactions for the fermions.

First, recall that we have defined the path integral measure as
\begin{equation}
\int D\phi = \left[\prod_{i}\int_{-\frac{1}{2}}^{\frac{1}{2}}d \phi_i \right]
\end{equation}
As each $\phi_i \in [-\frac{1}{2}, \frac{1}{2})$, the branch cut field $\toZ{d\phi}_{ij} = \toZ{\phi_i - \phi_j}$ may take values in $\{-1, 0, 1\}$. However, $\toZ{d\phi}$ is not distributed uniformly over these three values. If we let $\phi_i$ and $\phi_j$ be chosen randomly, the probability $p$ that $\toZ{d\phi}$ takes on a given value is 
\begin{equation}
\toZ{d\phi} = \begin{cases}
-1 & p=\nicefrac{1}{8} \\
~0 & p=\nicefrac{3}{4} \\
~1 & p=\nicefrac{1}{8}
\end{cases}
\end{equation}
This immediately leads to a probability for the observation of a vortex. On a given plaquette, the vortex number is given by $\rho_v = - d\toZ{d\phi}$, and on a triangular lattice must take values in $\{-3, -2, ..., 3\}$. One can extrapolate the probabilities from those for $\toZ{d\phi}$ to get:
\begin{equation}
\rho_v=-d\toZ{d\phi} = \begin{cases}
-3 & p=\nicefrac{1}{512}    \\
-2 & p=\nicefrac{18}{512}    \\
-1 & p=\nicefrac{111}{512}    \\
~0 & p=\nicefrac{252}{512}    \\
~1 & p=\nicefrac{111}{512}    \\
~2 & p=\nicefrac{18}{512}    \\
~3 & p=\nicefrac{1}{512}    \\
\end{cases}
\end{equation}
We see that when the $\phi$ are completely random, although having no vortex is the most likely single option, it is still less than $50\%$ probability. More than half of the time there is a vortex on a plaquette.

The details of the induced interactions are lattice dependent. However, the generic behavior can be described simply as follows. For random $\phi$, approximately one quarter of the lattice plaquettes will host a vortex, while another quarter will host an anti-vortex. Vortices and anti-vortices on adjacent sites have a weak attractive interaction and may annihilate, while both vortices and anti-vortices may hop to adjacent plaquettes with real amplitude. 

Recalling that the vortices and anti-vortices are statistically fermions for half-odd-integer $k$, we may convert this rough theory into an effective fermion theory. We consider a honeycomb lattice (the dual of the triangular lattice) with real hopping coefficients. We denote vortices as spin-up fermions, anti-vortices as spin down, and impose a strong on-site repulsive interaction to prevent overlap. In this context, conservation of vortex current density corresponds to $S^z$ symmetry, while vortex-antivortex annihilation results in superconducting pairing terms. A Hamiltonian for this model is accordingly
\begin{multline}
    H = \sum_{\braket{i,j}} \Big(t\sum_{\sigma}(c^\dagger_{i, \sigma}c_{j\sigma} + c^\dagger_{j, \sigma} c_{i, \sigma}) 
    + \Delta (c^\dagger_{i\uparrow}c^\dagger{j\downarrow} + c_{j\downarrow}c_{i\uparrow})
    -g c^\dagger_{i, \uparrow}c_{i, \uparrow}c^\dagger_{j, \downarrow}c_{j, \downarrow}
    \Big) 
    \\
    -\mu \sum_{i}\sum_{\sigma}c^\dagger_{i, \sigma}c_{i, \sigma}
    + U\sum_{i}c^\dagger_{i, \uparrow}c_{i, \uparrow}c^\dagger_{i, \downarrow}c_{i, \downarrow}
\end{multline}
Here the $t$ term represents the fermion hopping; $\Delta$ the vortex-antivortex annihilation; $g>0$ the nearest neighbor attractive interactions; $U\to \infty$ the strong on-site repulsion; and $\mu$ the chemical potential. The parameters should be tuned to roughly reproduce the fermion density, hopping, pairing, and attractive interactions from the statistical analysis. 

Within this framework, we can again see the question of symmetry breaking or gaplessness (or, equivalently, of a first-order or second-order transition) in the topological transition between $k$ and $k+1$ phases. The question is of the form of the pairing wavefunction $\psi_{ij} = \braket{c^\dagger_{i, \uparrow} c^\dagger_{j, \downarrow}}$. The strong on-site repulsion will suppress s-wave pairing, and we are left with p-wave or d-wave pairing. These must either break time-reversal symmetry or be gapless; whichever the dynamics favor will determine the nature of the topological transition. However, determining which case occurs (which may depend on the lattice), would have to be determined in a future work.

\section{Lattice Renormalization Group}\label{sec:LatticeRG}

\begin{figure}
\begin{center}
\includegraphics[width = .65\columnwidth]{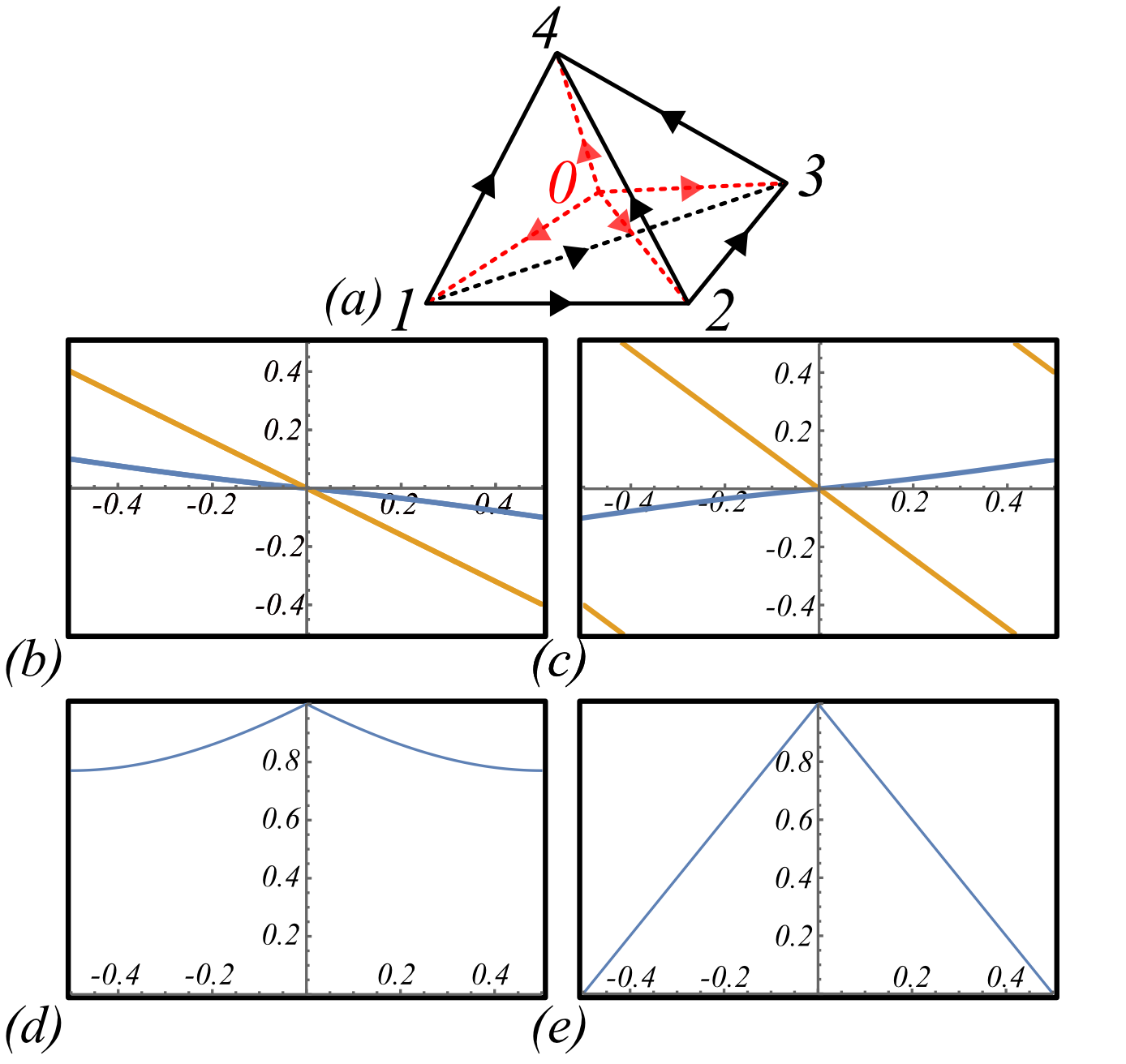}
\caption[Lattice Renormalization Group for the $U(1)$ $\theta$-term Model]{\co. \emph{(a)} We integrate out $\phi_0$ in the simplex shown to perform a simple Lattice RG calculation. \emph{(b), (c)}:  Phase of the quantity $\Phi_c$ calculated in eq. \eqref{eq:PhiC} (blue) and $\Phi_k$ from eq. \eqref{eq:PhiK} (orange) for \emph{(b)} $k = .8$ and \emph{(c)} $k=1.2$. \emph{(d), (e)}:  Magnitude of $\Phi_c$ for \emph{(d)} $k = .8$ and \emph{(e)} $k=1.2$. }\label{fig:LatticeRG}
\end{center}
\end{figure}

In this section we calculate the effects of integrating out a rotor on a given site, which reveals the flow of $k$ to integer fixed points. Taking $g\to\infty$ recall that the action in this limit is:
\begin{equation}
    S = 2\pi i k \int d\phi d\toZ{d\phi}.
\end{equation}
To evaluate this, we place it on a tetrahedral lattice shown in Figure \ref{fig:LatticeRG}a and perform a course-graning. Consider integrating out point $0$ in Figure \ref{fig:LatticeRG}a. We wish to calculate:
\begin{multline}
\int d\phi_0 \exp\Bigg\{2\pi i k \Bigg[\left(\phi_1 - \phi_0 - \toZ{\phi_1 - \phi_0}\right)
\Big[e
- \left(\toZ{\phi_2 - \phi_1} + \toZ{\phi_3 - \phi_2} - \toZ{\phi_3 - \phi_1}\right)
\\
+ \left(\toZ{\phi_2 - \phi_1} + \toZ{\phi_4 - \phi_2} - \toZ{\phi_4 - \phi_1}\right)
- \left(\toZ{\phi_3 - \phi_1} + \toZ{\phi_4 - \phi_3} - \toZ{\phi_4 - \phi_1}\right)
\Big] 
\\
+ \left(\phi_2 - \phi_0 - \toZ{\phi_2 - \phi_0}\right)\left[\toZ{\phi_3 - \phi_2} + \toZ{\phi_4 - \phi_5} - \toZ{\phi_4 - \phi_2}\right]
\Bigg]
\Bigg\}
\end{multline}
which can be rewritten as:
\begin{equation}
\int d\phi_0 \exp\Bigg\{-2\pi i k \Bigg[
(\phi_2 - \phi_1 - \toZ{\phi_2 - \phi_0} - \toZ{\phi_0 - \phi_1})
\left(
\toZ{\phi_3 - \phi_2} + \toZ{\phi_4 - \phi_3} - \phi_{4 - \phi_2}
\right)
\Bigg]
\Bigg\}
\end{equation}
Or, setting $\rho_{234} = - \toZ{\phi_3 - \phi_2} + \toZ{\phi_4 - \phi_3} - \toZ{\phi_4 - \phi_2}$, 
\begin{align}
\underbrace{e^{-2\pi i k (\phi_2 - \phi_1 - \toZ{\phi_2 - \phi_1})\rho_v}}_{\Phi_k}
\underbrace{\int d\phi_0 e^{2\pi i k 
( \toZ{\phi_2 - \phi_0} + \toZ{\phi_0 - \phi_1} - \toZ{\phi_2 - \phi_1})
\rho_{234}
}}_{\Phi_c}\label{eq:PhiK}
\end{align}
Now the amplitude has decomposed into the expected phase $\Phi_k$, times a correction term $\Phi_c$. Turning our attention to the integral in $\Phi_c$, we may use $U(1)$ symmetry to set $\phi_2 = 0$. Then The integral becomes:
\begin{equation}
\Phi_c = \int d\phi_0  e^{2\pi i k (\toZ{\phi_0 - \phi_1} + \toZ{\phi_1} - \toZ{\phi_0})}
=
1-|\phi_1 - \toZ{\phi_1}|+ |\phi_1 - \toZ{\phi_1}| e^{2\pi i k \rho_{234} \text{sgn}(\phi_1 - \toZ{\phi_1})}
\end{equation}
Restoring $U(1)$ symmetry, this is:
\begin{equation}
\Phi_c = 1- |\phi_2 - \phi_1 - \toZ{\phi_2 - \phi_1}|+ |\phi_2 - \phi_1 - \toZ{\phi_2 - \phi_1}| e^{-2\pi i k \rho_{234}\text{sgn}(\phi_2 - \phi_1 -\toZ{\phi_2 - \phi_1})}\label{eq:PhiC}
\end{equation}
To understand this expression, fix $\rho_{234} = 1$. The phase of both $\Phi_k$ and $\Phi_c$ functions of $\phi_2 - \phi_1$ are plotted in figures \ref{fig:LatticeRG}b and \ref{fig:LatticeRG}c for $k = .8$ and $k = 1.2$, respectively. For $k= .8$, the phase of $\Phi_k$ and $\Phi_c$ vary jointly, and $\Phi_c$ is pushing the effective $k$, back towards $k=1$. For $k= 1.2$, they vary oppositely, and $\Phi_c$ is pushing the effective $k$ down, again back towards $k=1$.

This trend continues across the spectrum. For $k$ neither integer nor half integer, $\Phi_c$ acts to push the effective $k$ back towards the nearest integer. For $k\in \Z$, $\Phi_c = 1$, and there is no correction, reflecting the fact that the integral theories are time-reversal-invariant, fixed point theories. For $k\in \Z + \frac{1}{2}$ $\Phi_c$ is real, reflecting the time-reversal invariance of the half-odd-integer theories. 

A further question arises because $|\Phi_c| < 1$ when $k\notin \Z$ and $|\phi_2 - \phi_1 -\toZ{\phi_2 -\phi_1}|>0$. Figures \ref{fig:LatticeRG}d and \ref{fig:LatticeRG}e show $|\Phi_c|$ for $k=.8$ and $k= .5$, respectively. For $ k=.8$, the damping effect is somewhat small, while for $k= .5$ it is extreme, and $|\Phi_c| = 0$ for $|\phi_2 - \phi_1 -\toZ{\phi_2 -\phi_1}| = \pm .5$. For general $k$ and small $|\phi_2 - \phi_1 - \toZ{\phi_2 - \phi_1}|$, this implies that there should be a logarithmic term in the action. Note that this only applies when $\rho_{234} \neq 0$; when $\rho_{234} = 0$, $\Phi_c = 1$ in all cases. 

\section{Discussion}

In this Chapter, we have used a new discontinuous cocycle model to shed light on the transition between a superfluid and a topological Mott Insulator. Without a background gauge field, the correlation functions of all local operators are identical at and away from the SF-tMI and SF-MI transitions, while introducing a background gauge field leads to different Hall responses for MI and tMI. This immediately implies that all critical exponents of correlation functions of $e^{i2\pi\phi_i}$ and its conjugate momentum must be the same in the SF-MI and SF-tMI phases, which allows the considerable numerical results of the XY model to be applied to the SF-tMI transition. 

We may also ask what other topological phases may be described by a similar framework. An obvious $U(1)\times \T$ phase in $1+1$d is one coming from the action:
\begin{equation}
    S = i\pi k\int_{\cM^2} d\toZ{d\phi}
\end{equation}
On the other hand, $d+1$ dimensional, with even $d>0$, generalizations of our model are given by:
\begin{equation}
    S = 2\pi i k \int_{\cM^{d+1}}(d\phi - \toZ{d\phi})(d\toZ{d\phi})^{\frac{d}{2}}
\end{equation}
where the exponent is taken using the cup product. If, as in $2+1$ dimensions, these are SPT states for integral $k$, they would exhaust all the $U(1)$ SPTs predicted by group cohomology \cite{PhysRevB.87.155114}. Many more possibilities arise with multiple $U(1)$ fields $\phi^I$. One could also consider extending this formalism to SPTs with nonabelian symmetry. Could the Haldane phase be described by a similar exactly solvable discontinuous path integral?
In every case, the existence of a model of the sort we have described here would carry the same implication of level-shift symmetry: in the absence of a gauge field, all correlations of local operators would be the same in the symmetry-breaking to topological transition as in a similar symmetry-breaking to trivial state transition. The SPT order would then by diagnosed by the response to an applied gauge field. 

Even just the model we have described has further applications. Because the edge theory may be treated alone, this proposes a solution to the $U(1)$ ``chiral fermion problem:'' The edge theory must describe pairs of counter-propagating modes with differing charges, so that the total chiral central charge of the edge theory vanishes, while the modes carry a chiral representation of $U(1)$. This is the subject of the next chapter.

\chapter{A Chiral Lattice Field Theory in $1+1$d}\label{chap:Chiral}

We can now create the promised simple chiral lattice field theory. All the ingredients for this theory are actually in the last chapter, where we developed an exactly solvable lattice model for a $U(1)$ SPT. Here we extract the bosonic chiral edge theory and examine its properties. This $1+1$d chiral boson edge theory also paves the way for simulation of more complicated chiral QFTs, including in higher dimensions and for nonabelian symmetries. The simplest extension of this theory is a chiral fermion theory that may be obtained by introducing a spin structure, which solves the Chiral Fermion Problem.

All of the properties we need in the SPT theory from the previous chapter arise because that theory is a fixed-point theory. In turn, the key to writing a fixed-point theory on the lattice was first discovered in the group cohomology formalism \cite{Chen:2011pg} but not implemented until the works described in this thesis: in order to capture topological actions, we must allow for actions which are not continuous functions of the field variables. Once we have a fixed-point topological action in $2+1$d, we have a gapped bulk and a gapless $1+1$d edge which decouple exactly, since the penetration of the gapless mode into the bulk must be zero at the fixed point. This is what enables us to write down the $1+1d$ model and study its properties.

In retrospect, that we should require physical quantities that are not continuous functions of the field variables in fixed-point theories follows from a simple argument. Consider, as we will shortly, a lattice QFT consisting of $U(1)$ variables on the sites of a lattice. The space of field configurations is $U(1)^{\text{n}_{\text{sites}}}$. Consider a function $\rho$ of the field variables which indicates the vortex number on each plaquette. In a fixed-point theory, the output of $\rho$ should be an integer for each plaquette, i.e. an element of $\mathbb Z^{\text{n}_{\text{plaquettes}}}$. As a function from a connected manifold, $U(1)^{\text{n}_{\text{sites}}}$, to a discrete space, $\mathbb Z^{\text{n}_{\text{plaquettes}}}$, $\rho$ must be either discontinuous or constant, and constant would be useless. Hence when describing vortices in a fixed point theory, we should allow for discontinuous physical quantities; this holds for topological defects in many theories.

\section{Lattice Model}

\begin{figure}
    \centering
    \includegraphics[width = .3\columnwidth]{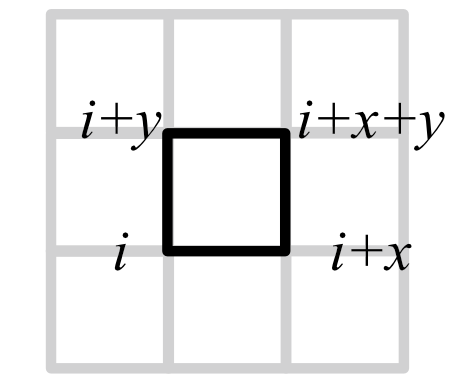}
    \caption[Square Lattice]{The model we study works on any lattice with a simplicial structure, but we will write down explicit expressions on this square lattice.}
    \label{fig:plaquette}
\end{figure}

The starting point for our $1+1$d chiral lattice model is actually the $2+1$d action from the last section, with $k\in \Z$ and $g\to \infty$. In this Chapter, we will be focused on its $1+1$d boundary, so we change notation and let $\cN^3$ be the $2+1$d lattice, with $\cM^2 = \partial \cN^3$ its $1+1$d boundary. Hence the action is:
\begin{multline}
    S_k[\phi] = -2\pi i k \int_{\cN^3} (d\phi - \toZ{d\phi}) \cup d(d\phi - \toZ{d\phi}) 
    \\
    = 2\pi i k \int_{\cN^3} d\phi \cup d\toZ{d\phi} = - 2\pi i k \int_{\cN^3} d\phi \cup \rho_v\label{eq:3dAction}
\end{multline}
where used the fact that $d^2 = 0$ and that the action appears as $e^{iS}$ to simplify the action and introduced $\rho_v \equiv -d\toZ{d\phi}$. Recall that we have a gauge redundancy which ensures that the theory is genuinely a function of $U(1)$ rotors:
\begin{equation}
\phi \to \phi + n \label{eq:gaugeredphi}
\end{equation}
The full path integral is:
\begin{equation}
    Z = \int D\phi e^{i S_{k}[\phi]} \hspace{1 cm} \int D\phi = \prod_i \int_{-\frac{1}{2}}^{\frac{1}{2}} d\phi_i 
\end{equation}
where the integral measure is gauge-fixed under (\ref{eq:gaugeredphi}).

The most important aspect of the action (\ref{eq:3dAction}) for our present purposes is that it is a surface term which vanishes on a closed manifold. Hence we may obtain a theory solely on $\cM^2 = \partial \cN^3$:
\begin{equation}
    S_k = 2\pi i k \int_{\cM^2} \phi d\toZ{d\phi} = -2\pi i k \int \phi \rho_v \label{eq:2dAction}
\end{equation}
This is the model which we wish to present and, together with its gauged version we will see later, is the chiral lattice field theory. It is very simple. In indices on a square lattice, the action is:
\begin{multline}
    S_k[\phi] = -2\pi i k \sum_{i\in \cM^2} \phi_i \Big(\toZ{\phi_i - \phi_{i+x}} + \toZ{\phi_{i+x} - \phi_{i+x+y}} \\ - \toZ{\phi_{i+y} - \phi_{i+x+y}} - \toZ{\phi_i - \phi_{i+y}}\Big)
\end{multline}
where $i$ sums over the sites of the lattice. As before, under the redundancy (\ref{eq:gaugeredphi}), the action is invariant, since $\rho_v$ is invariant and $n\rho_v$ is integer-valued and so does not affect the exponential. 

\section{Anomalies}

This action has the unusual, anomalous $U(1)$ symmetry discussed in the previous chapter. Under $\phi\to\phi+ \theta$, with $\theta$ a global constant, the action transforms as:
\begin{equation}
    S_k[\phi] \to S_k[\phi] - 2\pi i k \theta \int_{\cM^2} \rho_v
\end{equation}
If $\cM^2$ is closed, then the total vortex number is zero, as $\int_{\cM^2}\rho_v = - \int_{\cM^2} d\toZ{d\phi} =0$ by summation by parts, and so the theory is $U(1)$ symmetric. If $\cM^2$ has a boundary $\cB^1 = \partial \cM^2$, then the action is not $U(1)$ invariant. The breaking of $U(1)$ symmetry in the presence of a boundary is one characterization of the $U(1)$ anomaly (we saw this behavior in the context of the ground state wavefunction in the previous chapter). It should be considered as a consequence of the familiar expression of the anomaly in a background gauge field (which we investigate next), as a background gauge field could impose an electric potential that would create a boundary. The resolution, of course, is to recall that the $1+1$d action (\ref{eq:2dAction}) should be considered as the boundary of the $2+1$d system (\ref{eq:3dAction}). Hence $\cB^1 = \partial \cM^2 = \partial^2 \cN^3 = \emptyset$, as boundaries do not have boundaries.

We can see the anomaly in a different light by coupling to a background gauge field. The resulting $2+1$d action was given in Chapter \ref{chap:U1SPT} and is:
\begin{multline}
    S_k[\phi;A ] = -2\pi i k \int_{\cN^3}\Big\{ (d\phi - A)(dA - \toZ{dA}) \\ - \toZ{dA} (d\phi - A)
    - d\Big[(d\phi - A)(d\phi - A - \toZ{d\phi - A})\Big]\Big\} \label{eq:gauged3dAction}
\end{multline}
As before, we quantize $A$ to have cycles in unity and this action has three gauge redundancies. Two are of the same ``rotor redundancy'' form:
\begin{align}
    \phi_i\to \phi_i+n_i \label{eq:gaugeredunphi2}\\
    A_{ij} \to A_{ij} + m_{ij } \label{eq:gaugeredunA2}
\end{align}
for $n_{i}, m_{ij} \in \mathbb Z$. The third is the typical gauge invariance:
\begin{equation}
    \phi \to \phi + \theta \hspace{1 cm}
    A \to A + d\theta \label{eq:gaugereduntheta}
\end{equation}
for $\theta$ an $\mathbb R/\mathbb Z$-valued field. We also assume $A$ to be weak, in the sense that $dA - \toZ{dA} \approx 0$. This implies that $d\toZ{dA} = 0$, i.e. that field configurations are free of monopoles.

Now we separate (\ref{eq:gauged3dAction}) into boundary terms and bulk terms. We can rewrite the action as:
\begin{multline}
    2\pi i k \int_{\cN^3}
    \Big\{
    -A(dA - \toZ{dA}) + \toZ{dA}{A} - d(A(A-\toZ{A}))
    \Big\}
    \\ + 2\pi i k \int_{\cM^2} \Big\{
    \phi(dA - \toZ{dA}) - \toZ{dA}\phi 
    \\- (d\phi - A)(d\phi - A -\toZ{d\phi - A}) + A(A -\toZ{A})
    \Big\}
    \label{eq:separeted3dgaugedAction}
\end{multline}
We have now split the action into `bulk terms' consisting only of $A$ and `boundary terms' which contain all the $\phi$. Each of them are separately invariant under $\phi \to \phi+n$, and we have added and subtracted terms to ensure that they are invariant under $A\to A+m$. They are not separately invariant under the gauge symmetry (\ref{eq:gaugereduntheta}).

The boundary integral is the proper gauged action for the edge mode, i.e. our $1+1$d chiral boson model. Specifically, it is:
\begin{multline}
    S = 2\pi i k \int_{\cM^2} \Big\{
    \phi(dA - \toZ{dA}) - \toZ{dA}\phi 
    \\- (d\phi - A)(d\phi - A -\toZ{d\phi - A}) + A(A -\toZ{A})
    \Big\} \label{eq:2dActionGauged}
\end{multline}
Or, writing it on a square lattice,
\begin{multline}
    S = 2\pi i k \sum_i \phi_i \Big(
    A_{i, i+x} + A_{i+x, i+x+y} - A_{i+x, i+x+y} -
    A_{i+y, i+x+y}\\ 
    - \toZ{A_{i, i+x} + A_{i+x, i+x+y} - A_{i+x, i+x+y} -
    A_{i+y, i+x+y}}\Big)
    \\ - \toZ{A_{i, i+x} + A_{i+x, i+x+y} - A_{i+x, i+x+y} -
    A_{i+y, i+x+y}} \phi_{i+x+y}
    \\
    - (\phi_{i} - \phi_{i+x} - A_{i, i+x})(\phi_{i+x} - \phi_{i+x+y} - A_{i+x, i+x+y} - \toZ{\phi_{i+x} - \phi_{i+x+y} - A_{i+x, i+x+y}})
    \\
    + (\phi_{i} - \phi_{i+y} - A_{i, i+x
    y})(\phi_{i+y} - \phi_{i+x+y} - A_{i+y, i+x+y} - \toZ{\phi_{i+y} - \phi_{i+x+y} - A_{i+y, i+x+y}})
    \\
    + A_{i, i+x}(A_{i+x, i+x+y} - \toZ{A_{i+x, i+x+y}}) - A_{i, i+y}(A_{i+y, i+x+y} - \toZ{A_{i+y, i+x+y}})
\end{multline}
Together with the ungauged $(A=0)$ model (\ref{eq:2dAction}), eq. (\ref{eq:2dActionGauged}) describes the chiral lattice field theory. One can check that the action is invariant under both of the symmetries (\ref{eq:gaugeredunphi2}) and (\ref{eq:gaugeredunA2}). However, it is not invariant under (\ref{eq:gaugereduntheta}), i.e. $\phi \to \phi + \theta$, $A\to A + d\theta$. The anomaly structure is in general complicated, but takes a simple form when we set $d\theta=0$. In that case, the action changes by a term:
\begin{equation}
    -2\pi (2k) i \theta \int_{M^2} \toZ{dA} \label{eq:anomaly}
\end{equation}
Now, $\int_{\cM^2}(dA - \toZ{dA})=-\int_{\cM^2}\toZ{dA}$ is the total flux of the gauge field over $\cM^2$, and so this is precisely $2\pi (2k) \int \theta F$, i.e. the anomaly required by the Hall conductance. In the context of the Laughlin thought experiment, we twist boundary conditions by $2\pi$, which sets $-\int \toZ{dA} = 1$. In that case, eq. (\ref{eq:anomaly}) tells us that charge $2k$ has appeared in the edge theory, i.e. has been added to the edge theory from the bulk.
We should recognize this as the expected Adler-Bell-Jackiw anomaly, i.e. as a lattice, discrete generalization of $4\pi i k \int \theta dA$. As is usual, this failure of gauge invariance is cancelled by an equal and opposite term from the bulk terms in eq. (\ref{eq:separeted3dgaugedAction}).

We have now examined the $1+1$d lattice model in detail and seen the $U(1)$ anomaly through both symmetry breaking in the presence of a boundary and through direct coupling to a background gauge field. Now we write down a continuum model for the edge theory and explain how it creates a chiral representation of $U(1)$ and a nonzero quantized Hall conductance. 

\section{Continuum Theory}

The ungauged action (\ref{eq:2dAction}) describes a bosonic field $\phi$ coupled to its vortices. To develop a continuum description, we describe the vortices using a hydrodynamic approach:
\begin{equation}
    \rho_v \sim \partial_{x}\partial_t \theta(x, t)
\end{equation}
We consider an action which ensures that that vortices shift $\phi$  by $2\pi$ i.e. that $[\phi(x), \partial_x\theta] = 2\pi i$:
\begin{equation}
    S \sim 2\pi i \int \phi \partial_x \partial_t \theta
\end{equation}
where $\phi$ has charge $1$ and, as discussed in the previous chapter, the vortex field $\theta$ has charge $2k$. We define the composite fields $\phi_R = \frac{1}{2}(\phi + \theta)$, $\phi_L = \frac{1}{2}(\phi - \theta)$ to get:
\begin{equation}
    2\pi i  \int(\phi_R\partial_x \partial_t \phi_R - \phi_L \partial_x \partial_t \phi_L)
\end{equation}
Thus the chiral model consists of a right-moving mode $\phi_R$ and a left moving mode $\phi_L$. As there are equal numbers of left and right moving modes, there is no gravitational anomaly or, equivalently, thermal Hall conductance.

On the other hand, there is a $U(1)$ anomaly which arises because $\phi_L$ and $\phi_R$ have differing charges. Denote the change of a field $\varphi$ by $C[\varphi]$, so that $C[\phi] = 1$. We have seen that vortices have charge $2k$ and so $C[\theta] = 2k$. Hence the anomaly has coefficient:
\begin{equation}
    C[\phi_R] - C[\phi_L] = C[\theta] = 2k
\end{equation}
This is consistent with the Hall conductance of the bulk system, which is $2k\frac{e^2}{h}$ \cite{DeMarco:2021erp}.

We have written down a boson theory with chiral $U(1)$ symmetry in $1+1$d by extracting the edge theory from an exactly solvable $2+1$d chiral model. We then demonstrated the $U(1)$ anomaly by both inspection of the ungauged theory and by explicitly coupling in a background gauge field and calculating the variation of the edge action. Finally, we wrote down a continuum theory for our model and showed that it carries chiral $U(1)$ charge as expected. They key to our model, and indeed the main lesson from this thesis, is that discontinuous actions are essential when considering exactly solvable fixed-point theories on lattices, and that with these we can build exactly solvable theories that we previously believed to be impossible.

\chapter{Outlook}

This thesis has been a deep exploration of chiral phases, and their attendant chiral field theories, on the lattice. The key to our work was the relationship between a renormalization-group class of Quantum Field Theories and phases of matter. In essence, we defined our field theories as the low-energy effective theories (on the edge or in the bulk) of phases of matter, and then used techniques from condensed matter physics to understand the field theories. Here we review our work, lay out immediate next steps, and speculate on the sources of future progress in the field. 

\section{Review}

We began with a numerical realization of the `Mirror Fermion' approach. There we went beyond previous Abelian models to create an $SU(2)$ chiral gauge theory. We saw the gap opened, and numerically demonstrated the preservation of symmetry, but were faced with significant technical challenges: we could not sufficiently address how to perform the integral over Higgs configurations, nor could we reduce the model in such a way that the Higgs field was unnecessary. We won the battle of opening a gap in the fermion spectrum, but the Higgs problem means that this victory was a Pyrrhic one.

Following the difficulties in defining $1+1$d chiral edge modes, we changed our approach in two key ways: we pivoted to the bulk theory in $2+1$ dimensions, and also restricted to Abelian Chern-Simons theories. This led to a significant success: we were able to define a lattice model for any Abelian topological order, the low-energy behavior of which we could reliably determine semi-classically. Moreover, this theory was one of local rotor degrees of freedom, a feature which no one had been able to reliably incorporate into Chern-Simons theories. The added appearance of exact $1$-symmetries then provided extremely strong evidence for the validity of the model.

The next step was the mathematically obvious one. Given a model of topological order in terms of a gauge field, we ungauged it to get an SPT model. However, the utility of the model went far beyond that. On the heels of a number of results arguing that no commuting projector model could host Hall conductance, our model provided exactly that. Moreover, we later connected it to the theory of the $XY$ transition between a superfluid and a topological Mott insulator, and understood both the transition and the Hall conductance as a condensation of charged vortices. 

Of course, the ungauged model had another critical application. Because it is a zero-correlation-length, fixed-point model, the edge theory could be exactly decoupled from the bulk. This gave us an exact, $1+1$d $U(1)$ chiral lattice field theory. Moreover, we could directly compute its anomaly and confirm that it had the expected behavior. This is remarkable: we retreated from gapless chiral edge theories, went to the bulk theory and, after several intermediate steps, found a perfect chiral edge theory handed back to us. Moreover, that formulation paves the way for far more complicated chiral field theories, which we elaborate on in the next section.

\section{Further Lattice Field Theories}

The discontinuous cocycle models we have discovered are an extraordinary tool with which we can create new exactly solvable models of SPT phases and their chiral gapless edge theories. An obvious $U(1)\times \T$ phase in $1+1$d is one coming from the action:
\begin{equation}
    S = i\pi k\int_{\cM^2} d\toZ{d\phi}
\end{equation}
On the other hand, $d+1$ dimensional, with even $d>0$, generalizations of our model are given by:
\begin{equation}
    S = 2\pi i k \int_{\cM^{d+1}}(d\phi - \toZ{d\phi})(d\toZ{d\phi})^{\frac{d}{2}}
\end{equation}
where the exponent is taken using the cup product. If, as in $2+1$ dimensions, these are SPT states for integral $k$, they would exhaust all the $U(1)$ SPTs predicted by group cohomology \cite{PhysRevB.87.155114}. This leads to a $4+1$d SPT action:
\begin{equation}
S = 2\pi i k \int_{\cM^{4+1}}d\phi d\toZ{d\phi} d\toZ{d\phi}
\end{equation}
which would host a $3+1$d $U(1)$ chiral edge theory on the boundary $\cB^{3+1} = \partial \cM^{4+1}$:
\begin{equation}
S = 2\pi i k \int_{\cB^{3+1}}\phi d\toZ{d\phi} d\toZ{d\phi}
\end{equation}
This theory thus defines chiral fermions with $U(1)$ symmetry propagating in spacetime just like ours. Many more possibilities arise with multiple $U(1)$ fields $\phi^I$, and it is probably that this formalism would capture all of them. We are also preparing a discontinuous model which captures all the time-reversal SPTs. 

One could also consider extending this formalism to SPTs with non-Abelian symmetry. The most obviously relevant theories are the $SU(N)$ SPTs in $D = 3 + 4n$ spacetime dimensions. 

In the particular case of $SU(2)$ SPTs, we can use the fact that $SU(2) \simeq S^3$ to establish how these theories should be constructed. Let $g_i$ be the zero-cochain lattice variable. In that case, any tetrahedron with elements $g_i$ of $S^3$ defined at its corners defines a tetrahedron on the surface of $S^3$. Let the three-cochain $\Omega[g]$ be the volume of that tetrahedron in $S^3$ (this is easier to thing about in one lower dimension, i.e. with elements of $S^2$ on the corner of triangles, and $\Omega$ a solid angle). At the same time, any four-simplex (i.e. one higher dimension than a tetrahedron) with elements of $S^3$ at its corners defines a wrapping number from its surface to $S^3$. Let $W[g]$ be the four-cochain which is that wrapping number. Now, these $SU(2)$ theories would have the form:
\begin{equation}
S \propto i k \int_{\cM^{3+4n}} \Omega \cup W^n
\end{equation}
where $k$ is the level and $n$ is adjusted for the appropriate dimension. Moreover, one can see that $\Omega$ must be exact up to an integral part, and $W$ is always exact, so we once again have a surface term formalism and commuting projector model. All of this follows as an immediate generalization of our work here; the difficulty lies in actually writing down formulae for $\Omega$ and $W$. 

While the discontinuous cochain formalism describes SPT phases and their chiral edges which suffer gauge anomalies, we are also interested in topological orders and their gravitational anomalies. The formalism we described in Chapter \ref{chap:KMatrix} provides a description of all Abelian topological orders which is semi-classically solvable. However, we would like to develop an exactly solvable formalism. Doing so is more difficult, and probably cannot be described quite the same sense as the SPT phases: we have seen that the surface term formalism leads to a commuting projector model. However, it is believed that commuting projector models cannot have thermal Hall conductance, which corresponds to gravitational anomalies, because thermal Hall conductance would lead to energy flow through the system. The change in energy at a site $i$ is given by:
\begin{equation}
\dot{E}_i = \partial_t \braket{\hat H_i} = -i \left[\hat H, \hat H_i\right] = -i\sum_{j} \left[\hat H_j, \hat H_i\right] = 0 
\end{equation}
Here $\hat H_i$ are the commuting projectors and $\hat H = \sum_i \hat H_i$ is the Hamiltonian. Hence commuting projector models cannot host gravitational anomalies and so we do not expect one for chiral topological orders. Of course, that has not stopped us from some trials towards exactly solved models of topologically ordered phases, and we describe one incomplete approach in Appendix \ref{app:ESCS}.

\section{What's Next}
In Chapter \ref{chap:CFTQM}, we briefly mentioned that the meeting of condensed matter physics and high-energy theory had triggered a golden age. This continues apace, and has been the source of enormous progress through the past decade. There are new materials, field theories, and phases of matter being discovered almost daily.

A new player has also arrived: the exploding fields of quantum computing and quantum information (QC/QI). New hardware, materials, and algorithms have unleashed an entire new field of science and an entire new industry. Progress in quantum algorithms is proceeding extremely fast, and the first demonstrations of quantum supremacy have been achieved in the past few years. Huge teams of quantum engineers are being hired, while a quantum computing startup recently received a $\$2$B valuation. The future is extraordinarily bright.

I will be working in quantum computing after graduation, so I am biased, but I do genuinely believe that what is next in physics will come from the intersection of QC/QI and other fields. Indeed, the entire group cohomology formalism arose when a QI student under Ike Chuang began working with Xiao-Gang. Now I make the reverse journey, as I begin a fellowship working with Ike and I look forward to what's next.

\begin{singlespace}
\bibliography{mainU1,main,publst,all,ES_CS}
\bibliographystyle{plain}
\end{singlespace}

\appendix
\chapter{Cochains and Cohomology}
\label{sec:CaC}

Let us first set some notation. We consider a three-dimensional simplicial
complex $M$, which we take to contain $0$-simplices (vertices), $1$-simplices
(links), $2$-simplices (faces), and $3$-simplices (faces). In this thesis, we
consider our spacetime lattice to be a simplicial complex with matter fields living on the vertices and gauge fields living on the
links. Moreover, we assume that these fields take values in an Abelian group. 

The matter fields, denoted $g_{i}$, form a map from the $0$-simplices of $M$ (or more formally, form a map from the space of $0$-chains of $M$) to the target space. This map is called a $0$-cochain, as it defines a linear map from the free abelian group on the $0$-simplices of $M$ to the target space. Similarly,  a gauge field $a_{ij}$ living on the links of the lattice defines a $1$-cochain. One can continue this, with $n$-cochains defining maps from the $n$-simplices to the target space. 

To see the explicit action of an $n$-cochain, let us label simplices by their vertices, so that an $n$-simplex is given by $[v_{0},..., v_{n}]$. Then an $n$-cochain $a$ assigns a target space element $a([v_{0}, ..., v_{n}])$  to any $n$-simplex. Furthermore, this map is multilinear over a formal sum of $n$-simplices with coefficients in $\mathbb{Z}$. For example, let $\sigma_{1} = [v_{0}, ..., v_{n}]$ and $\sigma_{2} = [w_{0}, ..., w_{n}]$. Then 
\begin{align}
a(2\sigma_{1} - \sigma_{2}) = 2a(\sigma_{1}) - a(\sigma_{2})
\end{align}

Given any $n$-cochain $a$, we can create an $n+1$ cochain $\dd a$ via:
\begin{align}
\dd a([v_{0}, ..., v_{n+1}]) = \sum_{i=0}^{n+1} (-1)^{i} a([v_{0}, ..., \hat{v_{i}}, ..., v_{n+1}])
\end{align}
where $\hat{v_{i}}$ means that we omit that index. One can then see that $\dd(\dd a)= 0$. This allows us to construct cohomology groups as the cohomology of the complex of $n$-chains:
\begin{align}
\leftarrow C^{n} \leftarrow C^{n-1} ...~C^{1}\leftarrow C^{0} \leftarrow 0
\end{align}
here the maps are just given by $d_{n}$, eg $d$ acting on elements of $C^{n}$, and the cohomology groups are just $\Im~d_{n}/\ker d_{n}$.

In the case that the target space is a ring, we have an additional structure called the cup product. The cup product takes an $m$-form $a_m$ and an $n$-form $b_n$ and returns an $n+m$ form $a\smile b$, defined by:
\begin{align}
&\ \ \ \ 
a_m\smile b_n([v_{0}, ..., v_{n+m}]) 
\nonumber\\
&= a_m([v_{0}, ..., v_{m}])b_n([v_{m}, ..., v_{m+n}])
\end{align}
Furthermore, when considered on cohomology classes, this cup product is graded
anticommutative. This means that there is a $n+m-1$ cochain $c_{m+n-1}$ such
that:
\begin{align}
a_m\smile b_n = (-1)^{nm}b_n\smile a_m + \dd c_{m+n-1}
\end{align}
To get an explicit expression for $c_{m+n-1}$, we need to introduce higher cup
product $a_m \hcup{k} b_n$ which gives rise to a $(m+n-k)$-cochain\cite{S4790}:
\begin{align}
\label{hcupdef}
&\ \ \ \
 a_m \hcup{k} b_n([0,1,\cdots,m+n-k])
\nonumber\\
&
 = 
\hskip -1em 
\sum_{0\leq i_0<\cdots< i_k \leq n+m-k} 
\hskip -3em  
(-)^p
a_m([0 \to i_0, i_1\to i_2, \cdots])\times
\nonumber\\
&
\ \ \ \ \ \ \ \ \ \
\ \ \ \ \ \ \ \ \ \
b_n([i_0\to i_1, i_2\to i_3, \cdots]),
\end{align} 
and $a_m \hcup{k} b_n =0$ for  $k<0$ or for $k>m \text{ or } n$.  Here $i\to j$
is the sequence $i,i+1,\cdots,j-1,j$, and $p$ is the number of permutations to
bring the sequence
\begin{align}
 0 \to i_0, i_1\to i_2, \cdots; i_0+1\to i_1-1, i_2+1\to i_3-1,\cdots
\end{align}
to the sequence
\begin{align}
 0 \to m+n-k.
\end{align}
For example
\begin{align}
&
 a_m \hcup1 b_n([0\to m+n-1]) 
 = \sum_{i=0}^{m-1} (-)^{(m-i)(n+1)}\times
\nonumber\\
&
a_m([0 \to i, i+n\to m+n-1])
b_n([i\to i+n]).
\end{align} 
We can see that $\hcup0 =\smile$.  
Unlike cup product at $k=0$, the higher cup product of two
cocycles may not be a cocycle. For cochains $a_m, b_n$, we have\cite{S4790}
\begin{align}
\label{cupkrel}
& \dd( a_m \hcup{k} b_n)=
\dd a_m \hcup{k} b_n +
(-)^{m} a_m \hcup{k} \dd b_n+
\\
& \ \ \
(-)^{m+n-k} a_m \hcup{k-1} b_n +
(-)^{mn+m+n} b_n \hcup{k-1} a_m 
\nonumber 
\end{align}
The above result also allows us to see that
the cup product interacts with $\dd$ in the familiar way:
\begin{align}
\dd(a_m\smile b_n) = (\dd a_m)\smile b_n + (-1)^{m}a_m\smile \dd b_n 
\end{align}
which we can interpret as the Leibniz rule. In the case that there is no boundary, we can interpret this as
yielding a form of integration by parts, so that $\dd a_m\smile b_n =
-(-1)^{m}a_m\smile \dd b_n $. We will abbreviate the cup product
with using `$\smile$,' so that $ab\equiv a\smile b$. 

\chapter{Physical Interpretation of Level Shift Symmetry}\label{app:LS}
In the main text, we saw that the fact that the cocycle was a surface term for our model implied that the bulk correlation functions of local operators are identical to the trivial case, and that this implied that the critical exponents for the SF-tMI transition were the same as for the SF-MI transition. Now we discuss why we might expect that on general grounds.

We wish to consider the SF-tMI transition, and we begin in the tMI phase. Rather than directly breaking the symmetry in the tMI, suppose that we stack the tMI with a trivial MI. We can then break the $U(1)$ symmetry in the trivial state, which involves a phase transition with the usual critical exponents. After the symmetry is broken in the trivial phase, we couple in the tMI. With the provision of symmetry breaking, the tMI can be trivialized, and also reduced to a trivial state. The critical exponents for this transition are thus the same as the usual SF-MI transition.

This argument shows that there must always be a SF-tMI critical point with the same exponents as the SF-MI critical point. What we have shown in this paper is that the XY SF-tMI transition, in the absence of a background gauge field, is precisely this transition.

A similar argument may be applied to any SPT, thus showing that there exists a symmetry-breaking transition out of the SPT that has the same exponents as the symmetry breaking transition of a symmetric trivial state. On the mathematical side, this reflects the fact that all group cocycles $\nu$ have the form $\nu = d\mu$, where $\mu$ may not be $G$-symmetric, i.e. that all SPTs are trivial in the absence of symmetry. 
\chapter{Cech-Deligne-Beilinson Approach to Chern-Simons Theory}\label{app:ESCS}

\section{Introduction} 
It is quite rare in theoretical physics for a system to be both exactly soluble and well defined. When we find a system that is, like the harmonic oscillator, we tend to make extensive use of it. Topological Field Theories (TFTs) are indeed exactly soluble, but their realization as a field integral is difficult to define, while the mathematical approaches to TQFTs tend to abandon the field integral entirely. In this Appendix, we propose a method to define TQFTs on a spacetime lattice that remains exactly soluble. We argue that we can reduce the field integral to a few real integrals and sums that can be evaluated using standard methods of calculus, and we calculate the partition function exactly in several example cases. While we largely focus here on Chern-Simons theories and $\Z_N$ gauge theories in three spacetime dimensions, the method we have developed can be extended to topological field theories in any dimension. 

The model we present is a lattice gauge theory. Gauge invariance is central to the calculations we present, and will allow us to calculate exact partition functions on various manifolds. In that sense, the integral measure---and hence any Hilbert space of a related Hamiltonian formulation---is not a product of local integrals. This stands in contrast to the models presented in the main body of this thesis. 

Throughout this Appendix, as in Chapters \ref{chap:KMatrix}-\ref{chap:Chiral}, we use a normalization so that the cycles of $U(1)$ variables are quantized to unity. This means that for a scalar field $\phi$, $\phi\sim\phi+1$, and that the na\"{i}ve level-$k$ Chern-Simons Lagrangian is $\pi k ada$. One can always return to the usual normalization by replacing $\phi\to \phi/2\pi, a \to a/2\pi$. 

Furthermore, we freely use notation from Differential Geometry and Algrebraic Topology \cite{BottTu, hatcher2002algebraic}. When working in the continuum, we represent differential forms as, e.g., $a \equiv a_{\mu}dx^{\mu}$. We will make extensive use of the wedge product without writing it, so that $a b \equiv a_{\mu}b_{\nu}dx^{\mu}dx^{\nu}$. When working on the lattice, we will denote lattice indices by latin letters $i,j,k,...$ and dual lattice indices greek letters $\alpha,\beta, \gamma,...$ Like in the continuum case, we will extensive use of the cup product without writing it, though we symmetrize it as we describe in Section \ref{sec:Sum}.

The best way to demonstrate our formalism is by example. In this section, we demonstrate the calculation of the partition function of Chern-Simons Theory in three spacetime dimensions on $T^{2}\times [0, 1]$. We assume that the lattice is cubical. The generalization to other lattices and spacetimes of general topology will be treated in the subsequent sections. 

\begin{figure}
\begin{center}
\includegraphics[width = \textwidth]{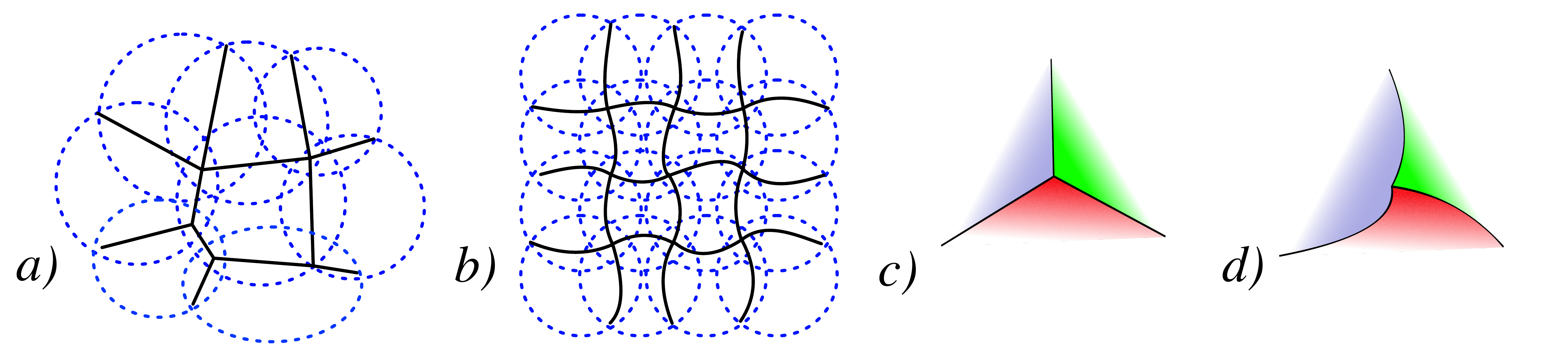}
\caption[Patchwise Gauge Fields]{\emph{(color online)} a) and b) A gauge field $A$ may only be expressed as a one-form $\aa$ in open sets $U_{\alpha}$, represented by the dashed lines. To define physical quantities, such as an action, we must divide the manifold into closed sets with overlaps of zero measure, within which we use a particular one-form $a^{1}, a^{2}, ...$. Here we denote such a choice by the solid black lines. We refer to this choice as a \emph{division}. On the lattice, the division is exactly the lattice cells. c) and d) Schematic depiction of two choices of division, with open sets not drawn. In order to ensure that the physical action is independent of choice of division, we have to add extra terms to the action along the black 1-cell and 0-cell boundaries. It is this amended action that we are able to put on the lattice.}\label{fig:OverviewPatches}
\end{center}
\end{figure}

To define gauge theories on a lattice, we rely on a simple general idea from continuum field theory. The basic principle that we review in Section \ref{sec:Classical} is that $U(1)$ gauge fields can only be defined as one-forms on contractible patches, while on overlaps they differ by a gauge transformation. To define physical quantities, such as an action or a Lagrangian, we cut the open patches into closed sets with overlaps of zero measure, as shown in Figure \ref{fig:OverviewPatches}. At these lower-dimensional overlaps, we must add extra terms to the Lagrangian to ensure that the Lagrangian is independent of how we cut the open patches into closed sets. Here we will simply quote the resulting Lagrangian, with a derivation to follow. 

We focus only on square and cubic lattices. We denote zero-cells of the dual lattice by $\Va^0$, one-cells of the dual lattice by $\Vab \equiv \Va^0 \cap \Vb^0$, two-cells of the dual lattice as $\Vabgs \equiv \Va^0 \cap \Vb^0 \cap V_{\gamma}^{2} \cap V_{\sigma}^{2}$, and so forth. We assume that $\Vab$ inherits its orientation from $\Va^0$, $\Vabgs$ inherits its orientation from $\Vab$, etc. 

The crux of our lattice formalism is each lattice cell $\Va^0$ carries its own gauge field $\aa_{ij}$ that lives on the links of the cell. When a link is shared by two lattice cells $\Va^0, V_\beta^0$, we do not require the gauge fields in the lattice cells to be equal, but allow them to differ by a gauge transformation:
\eq{
\aa_{ij}-\ab_{ij} = (d\pab)_{ij}\neq 0
}
A gauge field in our formulation is the collection $\aa$, together with the $\pab$ that glue cells together. One can do the same thing for a scalar field---on overlaps we require that a scalar field differ by an integer $\ta_i - \theta^{\beta}_i = m^{\alpha\beta}\in\Z$. Below we formalize these ideas for both a scalar field and a gauge field. 

\subsection{Definition of a Scalar Field on the Lattice}

Let us examine first how this works for a $U(1)$ scalar field $\Theta$. Recall that we enumerate lattice sites by $i,j,k...$ and dual lattice sites by $\alpha,\beta,\gamma...$. 

In our formalism, a scalar field $\Theta$ consists of:
\begin{itemize}
\item On each lattice cell (i.e. zero-cell of the dual lattice), a zero cochain $\ta_{i}$
\item One each overlap of lattice cells (i.e. one-cell of the dual lattice), a constant\footnote{To unify notation with the situation of a gauge field $A$ described below, one may consider a integer constant to be a ``$-1$'' form, as $\Z \to U(1)\to 0$ is exact.} $\mab \in \Z$ 
\end{itemize}
such that:
\begin{itemize}
\item On each one-cell of the dual lattice,
\eq{
\ta_{i} - \tb_{i} = \mab \label{eq:Sum:tm}
}
\item On each two-cell of the dual lattice
\eq{
m^{\alpha\beta} + m^{\beta\sigma}- m^{\alpha\gamma} - m^{\gamma\sigma} = 0\label{eq:Sum:Mconsist}
}
\end{itemize}
where $\alpha < \beta < \gamma < \sigma$. One can in fact derive (\ref{eq:Sum:Mconsist}) from (\ref{eq:Sum:tm}). Finally, a scalar field comes with the following redundancy:
\eq{
\label{Sum:ScalarRed}
\ta_{i} \to \ta_{i} + \ell^{\alpha} \\ \nonumber
\mab \to \mab + \ell^{\alpha} - \ell^{\beta}
}
In our formalism, a scalar field $\Theta$ consists of both the collection of zero-cochains $\ta_{i}$ and constants $\mab$. We will often write this as $\Theta = (\ta, \mab)$. 

Let us unpack the physical picture behind these definitions. Most of the lattices we will refer to in this paper---such as square lattices, cubic lattices, and so forth---carry a one-to-one correspondence of points to lattice cells. Denote the lattice cell (i.e. zero-cell of the dual lattice) corresponding to the $ith$ lattice point as $\alpha(i)$. Then one can fix the redundancy (\ref{Sum:ScalarRed}) by requiring that $\theta^{\alpha(i)}_{i}\in [0, 1)$. Thus at each lattice site, we have an angular variable $\theta^{\alpha(i)}_{i}$, but between lattice sites we also have an integer $m^{\alpha(i)\alpha(j)}$. The `continuum' picture to have in mind is that of of scalar field $\theta$ that has the value $\theta^{\alpha(i)}_{i}$ at $i$ and $\theta^{\alpha(j)}_{j}$ at $j$, but winds $m^{\alpha\beta}$ times between $i$ and $j$. For instance, the winding number about a nontrivial cycle is just the sum of the $\mab$ along that cycle.

In the conventional lattice formulation of a scalar field $\tilde\theta_i$, we lack the integers $\mab$, and in lieu of (\ref{Sum:ScalarRed}) we impose the redundancy $\tilde{\theta}_{i}\to \tilde{\theta}_{i} + \ell_{i}$, $\ell_i \in \Z$. As such, a typical `kinetic' term is $-\cos(\tilde\theta_i - \tilde\theta_j)$ and a typical `potential' term would be $-\cos(\tilde\theta_i)$. Instead, in our regularization, the difference of any two field values in the same patch $\alpha$ is well defined as a value in $\R$, and so a typical `kinetic term' might be 
\eq{
-\frac{1}{2}(\theta_{i}^{\alpha(i)}-\theta_{j}^{\alpha(i)})^{2}  
= -\frac{1}{2}(\theta_{i}^{\alpha(i)}-\theta_{j}^{\alpha(j)} + m^{\alpha(i)\alpha(j)})^{2}
}
On the other hand, a `potential' term with just a single $\theta^\alpha_i$ must still be periodic:
\eq{
-\cos(\theta^{\alpha(i)}_{i})
}
The inclusion of the integers $\mab$ are why this regularization differs radically from the usual, but they are also what will render our model exactly soluble.

\subsection{Definition of a Gauge Field on the Lattice}
A gauge field $A$ in our approach consists of:
\begin{itemize}
\item On each lattice cell (i.e. zero cell of the dual lattice), a one cochain $\aa_{ij}$
\item On each overlap (i.e. one-cell of the dual lattice), a scalar field $\Phi^{\alpha\beta}_{i} = (\pab_{i}, m^{\alpha\beta\gamma\sigma})$.
\end{itemize}
such that:
\begin{itemize}
\item on each one-cell of the dual lattice:
\eq{
\aa_{ij} - \ab_{ij} = (d\pab)_{ij}
}
\item on each two-cell of the dual lattice:
\eq{
\pab_{i} + \phi^{\beta\sigma}_{i} -\phi^{\alpha\gamma}_{i} -\phi^{\gamma\sigma} \equiv n^{\alpha\beta\gamma\sigma}\in \Z
}
\end{itemize}
Every lattice gauge field comes equipped with the gauge redundancy:
\eq{
\label{eq:Sum:ARed}
\aa_{ij}\to \aa_{ij} + (d\theta^{a})_{ij} \\ \nonumber
\pab_{i} \to \pab_{i} + \ta_{i} - \theta^{\beta}_{i}
}
where $\theta^{\alpha}_{i}$ is a scalar field independently defined on each lattice cell. 
 
As in the case of a scalar field, our treatment of a gauge field is radically different from the usual. For one, the flux through a plaquette is unbounded, whereas typical treatments of lattice field theory have flux that is only defined modulo $2\pi$. Again, this property is essential in making our approach well defined. 

Furthermore, one can show \cite{alvarez1985} that on any closed, oriented two-dimensional surface $S$, the sum of $da$ over the generator of $H_{2}(S)$ is given by the sum of the $n^{\alpha\beta\gamma\sigma}$ over $S$. In other words, flux is still quantized on closed surfaces. This will be a useful tool in the partition functions below.

\section{Lattice Lagrangians}
We use the Alvarez approach \cite{alvarez1985} to derive lattice Lagrangians that can evaluate the action for gauge fields $A$ that are only defined by one-forms $\aa$ locally. In our lattice formalism, we translate these Lagrangians to the lattice in the most na\"{i}ve possible way: substitute the continuum differential ``$d$'' for the lattice differential ``$d$'', replace the wedge product ``$\wedge$'' with a symmetrized cup product ``$\cup$,'' and exchange integration for evaluation against restrictions of the generator of the top cohomology of the manifold.

\begin{figure}
\begin{center}
\includegraphics[width = \textwidth]{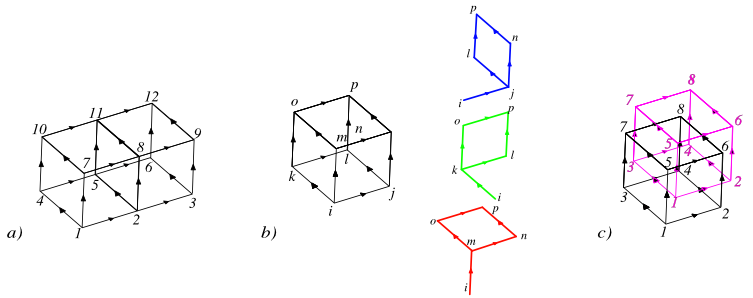}
\caption[Cubic Lattices, the Cup Product, and the Dual Lattice]{\emph{(color online)} a) Example of a cubic lattice on two cells. For more larger cubic lattices, we number lattice points first in the $x$ direction, then in the $y$ direction, and lastly in the $z$ direction, as we have done for this two cell lattice. b) Visual representation of the symmetrized cup product of a one-cochain and a two cochain. c) The interwoven branching structures of the lattice (black) and dual lattice (pink), where we use the self-duality of the cubic lattice to assign a branching structure to the dual lattice.}\label{fig:Sum:CS1}
\end{center}
\end{figure} 

We focus on the case of a cubic lattice on $S^{1}\times S^{1} \times I$ where $I = [0, 1]$. Labeling the three directions $x, y, $ and $z$, we assume that lattice sites have been numbered by proceeding first in the $x$ direction, then in the $y$ direction, and then in the $z$ direction. To each lattice cell, we assign the number of the site in the $(-\hat x, -\hat y, -\hat z )$ corner. See Figure \ref{fig:Sum:CS1}. In later sections, we can remove this awkward requirement by phrasing it as a certain branching structure condition, which will be essential to generalize approach to more complicated manifolds.

For boundary conditions, we take the gauge field to have zero flux on $T^{2}\times \{0\}$ and $T^{2}\times \{1\}$, as well as identical holonomy at both times. The zero flux condition is typical and represents a truncation of the Hilbert space to `low-energy states', while the requirement of identical holonomy reflects the fact that the partition function is a trace. 

We we also need a cyclic symmetrization of the usual cup product on this cubic lattice. Let $\omega_{ij}$ be a one-cochain and $\nu_{ijkl}$ a two cochain on our cubic lattice. Then we define the cup product as:
\eq{
(\omega\cup\nu)_{ijklmnop} = \omega_{ij}\nu_{ikmo} + \omega_{ik}\nu_{imjn} + \omega_{im}\nu_{ijkm}
}
This may be understood as a symmetrization of the usual cup product over cyclic permutations of indices; a visual demonstration is given in Figure \ref{fig:Sum:CS1}b. As before, we will always drop the cup ``$\cup$.''

\begin{figure}
\begin{center}
\includegraphics[width = .55 \textwidth]{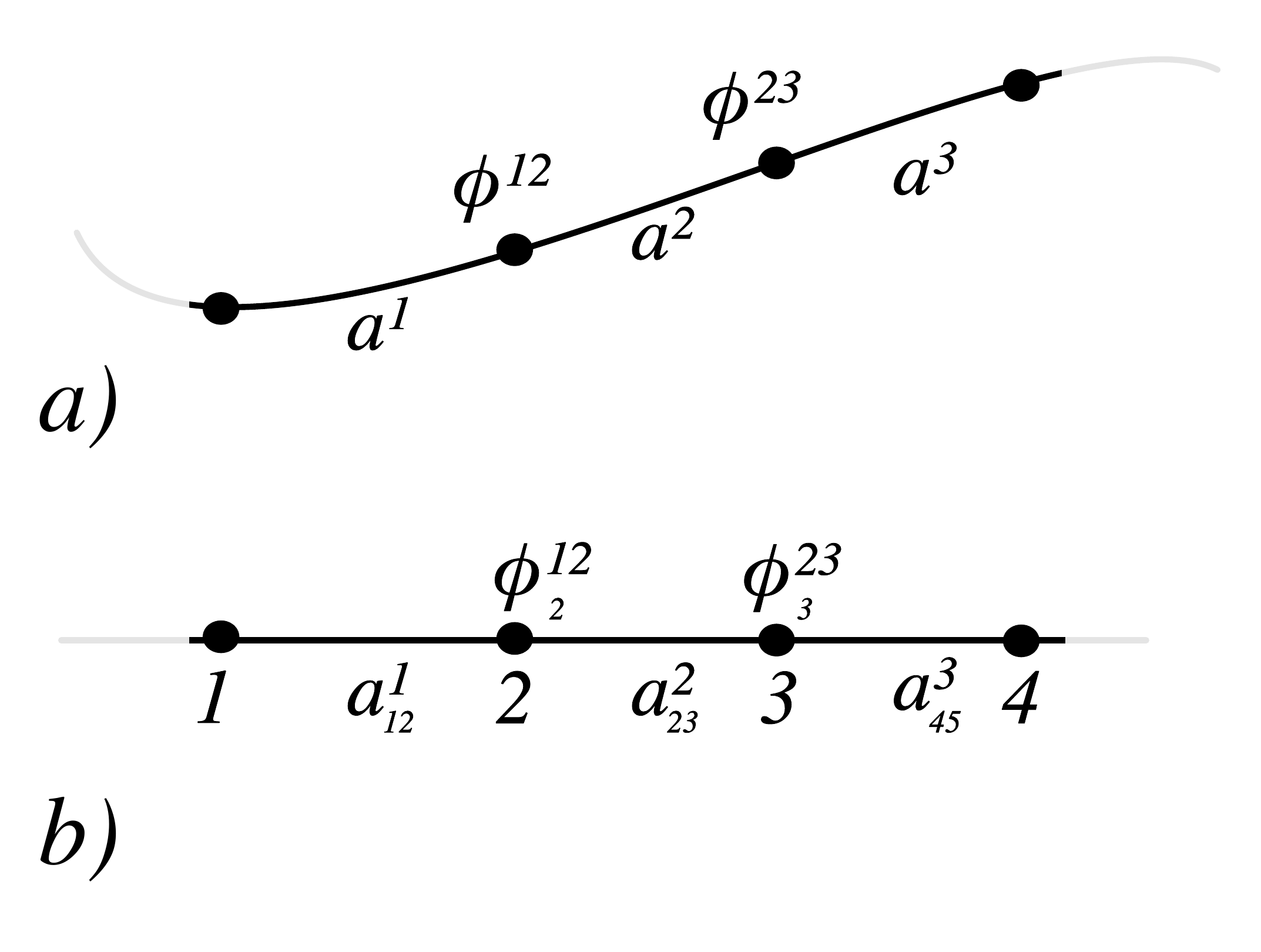}
\caption[Wilson Lines in the CDB Formalism]{a) Schematic depiction of the Wilson line when $A$ is only represented by one-forms $\aa$ locally. The corresponding picture on the lattice, where $A$ is represented by one-cochains $\aa_{ij}$ in each lattice cell.}
\label{fig:Sum:WilsonLine}
\end{center}
\end{figure}

We will also need the expression for the holonomy. Consider the situation in Figure \ref{fig:Sum:WilsonLine}. In each closed interval, we use a particular one form $aa$, however where two closed sections intersect between regions $\alpha$ and $\beta$ we must add a term $-\pab$, where $\aa-\ab = \pab$. So the continuum expression for a Wilson line is

The action we obtain from the Alvarez procedure is:
\eq{ S = \pi k \left\{
\sum_{\alpha}a^{\alpha}da^{\alpha} 
- \sum_{V_{\alpha\beta}}\phi^{\alpha\beta} da^{\beta}
+ \sum_{V_{\alpha\beta\gamma\sigma}}n^{\alpha\beta\gamma\sigma}a^{\sigma} 
\right.
\\
\nonumber
\left.
-\sum_{V_{\alpha\beta\gamma\sigma\delta\lambda\rho\zeta}}(n^{\alpha\beta\delta\lambda}\phi^{\lambda\zeta} - n^{\alpha\beta\gamma\sigma}\phi^{\sigma\zeta} - n^{\alpha\gamma\lambda\rho}\phi^{\rho\zeta})
\right\}\label{eq:Sum:CSLag}
}
where we sum over all lattice cubes $V_{\alpha}$, faces $V_{\alpha\beta}$, edges $V_{\alpha\beta\gamma\sigma}$ and points $V_{\alpha\beta\gamma\sigma\delta\lambda\rho\zeta}$. Each term is evaluated on the appropriate restriction of the top cohomology of $T^{2} \times I$. For example, consider the cube in Figure \ref{fig:Sum:CS1}c above. The terms in the action for this cube are:
\eq{
a^{1}_{12}(da^{1})_{1,4,7, 10} + a^{1}_{14}(da^{1})_{1728} + a^{1}_{17}(da^{1})_{1245} 
-\phi^{12}_{1} (d\ab)_{1357} - \phi^{1 3}_{2} (da^{3})_{2637} - \phi^{15}(da^{5})_{4567}\\
\nonumber
+ n^{1256}a^{6}_{57} + n^{1357}a^{7}_{67} + n^{1234}a^{4}_{37}
-(n^{1256}\phi^{68}_{7}
 +n^{1234}\phi^{48}_{7}
  + n^{1367}\phi^{78}_{7})
}
where $d\aa_{ijkl} = \aa_{ij} + \aa_{jl} - \aa_{ik} - \aa_{kl}$. One can then proceed over all cubes of an arbitrarily large cubic lattice, assigning terms as above. 

At this point one can see already because $\pab$ carries an integer redundancy $\phi\to\phi+\ell_{i}$, the action is only well defined if $k\in 2\Z$, reflecting the fact that our theory describes bosonic Chern Simons theory\footnote{Indeed, the Alvarez approach \cite{alvarez1985} was developed specifically to explain level quantization without resorting to the homology of the underlying manifold.}. For convenience, we relax this restriction to $k\in Z$below.

\subsection{Evaluation of Partition Functions}\label{sec:Sum:Eval}

We believe that our theory describes a renormalization group fixed point, in that one can evaluate the integrals over certain lattice links and end up with the same action on a reduced lattice. With that in mind, here we evaluate the partition function on the simplest possible lattice: a single lattice cell covering $M = S^{1}\times S^{1}\times I$, `glued' to itself around the nontrivial loops of $M$. 

\begin{figure}
\begin{center}
\includegraphics[width = .9\textwidth]{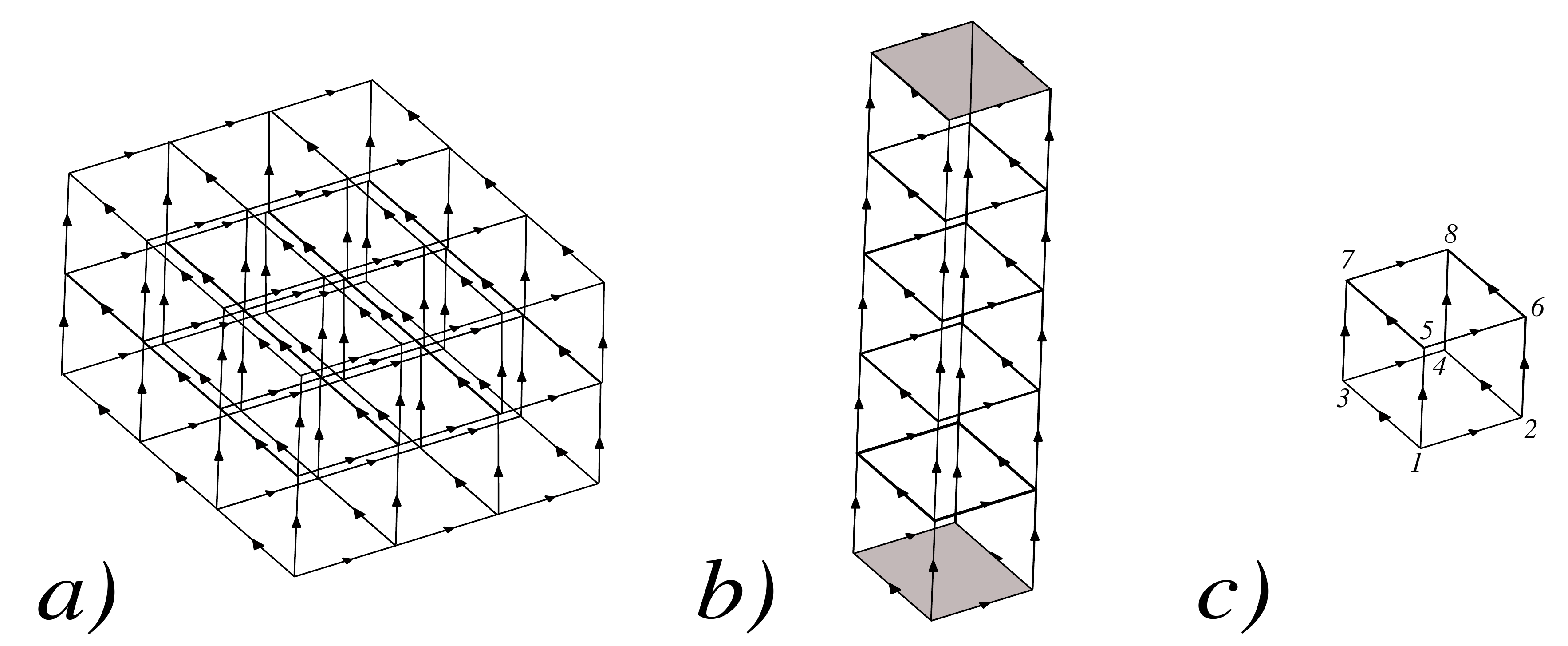}
\caption[Lattice Pruning]{
Pruning of the lattice for Chern-Simons Theory.
a) We begin with a large cubic lattice on $T^{2} \times I$.
b) In Section (), we show that by integrating out lattice variables, we can obtain the same theory on a reduced lattice. We do so gradually until it is a single ``stack'' of non-simply-connected cubes along $I$. In this frame, we denote the flat-connection boundary conditions by shading.
c) After integrating out yet more lattice points, we can reduce the lattice until it is a single cubic cell, non-simply connected to itself. In this diagram, $1\sim 2 \sim 3\sim 4$ are a single actual point, though we denote them differently to make it easier to notate links. Similarly, $5\sim 6\sim 7\sim 8$ are a single point. We have dropped the shading that was present in (b).}\label{fig:Sum:CS_SingleCell}
\end{center}
\end{figure}

The single lattice is shown in Figure \label{fig:Sum:CS_SingleCell}. Since there is only one cell, we drop the cell index. 

\begin{figure}
\begin{center}
\includegraphics[width = .7\textwidth]{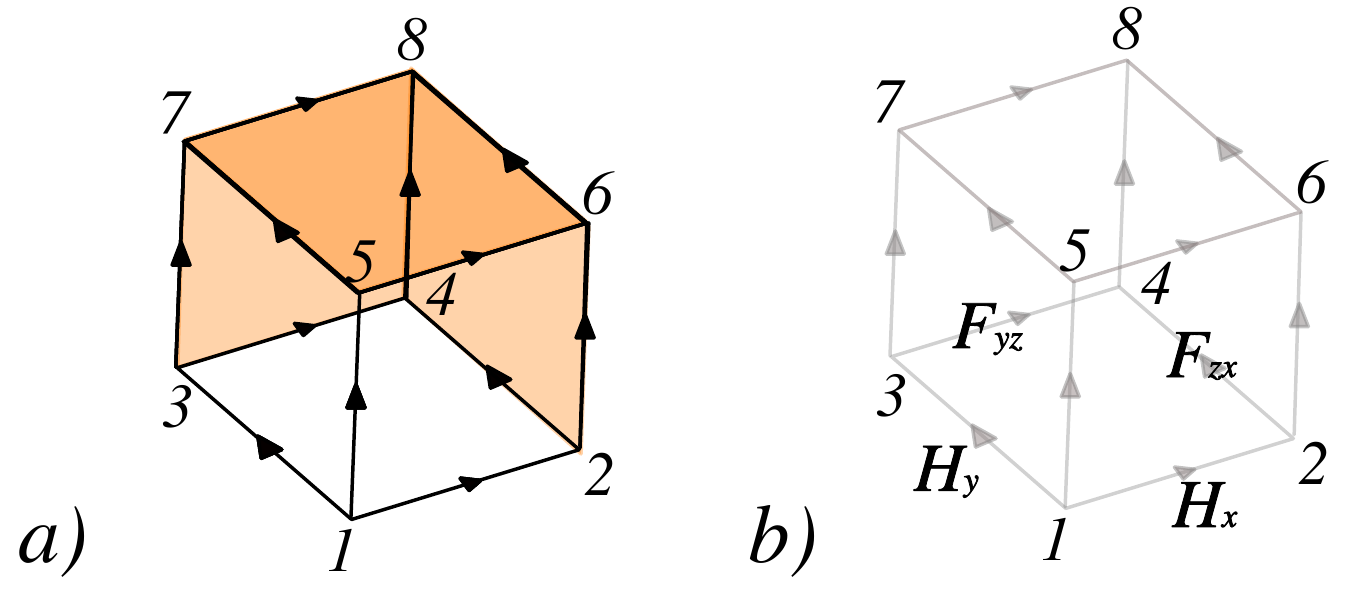}
\caption[The Single Cell Lattice]{Detailed view of the single-cell lattice. a) We do not require that the lattice variables are periodic, but instead allow them to differ by a gauge transformation $d\phi$, where $\phi$ is defined on the orange faces. b) Taking the temporal gauge with all links in the vertical direction zero, we can rewrite the Lagrangian in terms of the holonomy in the $x$ and $y$ directions, $H_x$ and $H_y$ respectively, as well as the flux on the $xz$ and $yz$ faces.}
\label{fig:Sum:SingleCell_Detail}
\end{center}
\end{figure}

Instead of periodic boundary conditions, we allow the links to differ by a gauge transformation. Specifically, we assume that there is a field $\phi$ defined on faces $5678, 2468,$ and $3478$ (see Figure \ref{fig:Sum:SingleCell_Detail}) such that:
\begin{align}
a_{34} - a_{12} = (d\phi)_{34} \hspace{1cm} a_{78} - a_{56} = (d\phi)_{56} \\
a_{37} - a_{15} = (d\phi)_{48} = a_{15} - a_{26} \\
a_{24} - a_{13} = (d\phi)_{24} \hspace{1cm} a_{68} - a_{57} = (d\phi)_{57}
\end{align}
Note that, because $1\sim2\sim3\sim4$ represent the same point as do $5\sim6\sim7\sim8$, $(d\phi)_{12}, (d\phi)_{13}, (d\phi)_{56}, (d\phi)_{57}\in Z$. On the other hand, $(d\phi)_{15} \in \R$, and because the link $15$ is contractible we can take the temporal gauge with $a_{14} = 0$.

Now we can write down the Lagrangian on our lattice. Directly translating eq. (\ref{eq:Sum:CSLag}), we obtain:
\eq{
S = \pi k \Big[ a_{13}(da)_{1256} + a_{12}(da)_{1357} - \phi_{1}(da)_{1256} - \phi_{1} (da)_{1357} 
\\ \nonumber
n_{1256}a_{13} + n_{1357}a_{12} - n_{1256} \phi_{1} - n_{1357}\phi_{1}\Big]
}

Let us rename these quantities in a more transparent way. We rename the holonomies as:
\eq{
H_x \equiv \Hol(A, 12) \hspace{2 cm} H_y = \Hol(A, 13)
}
and the fluxes as:
\eq{
F_{zx} = (da)_{1256} \hspace{2 cm} F_{yz} = (da)_{1357}
}
The action becomes:
\eq{
S = 2\pi k\left[H_y F_{zx} + H_x F_{yz}\right]
}
which has the virtue of being written in completely gauge-invariant, non-redundant quantities. 

Now we can evaluate the partition function. When doing so, we can sum over the `$x$' variables first or the `$y$' variables first. Whichever variable we sum over first plays the role of the canonical momentum from canonical quantization, while the second play the role the canonical position. Let us sum over the `$x$' variables first, $H_x, F_{zx}$. The partition function is:
\eq{
Z = 
\sum_{F_{yz}\in\Z}
\int\limits_0^1 dH_y 
\sum_{F_{zx}\in\Z}
\int\limits_0^1 dH_x
\exp\left\{2\pi i k \left[H_y F_{zx} + H_x F_{yz}\right]\right\}
}
We have reduced the quantum mechanical problem to two real integrals over $[0, 1)$, and two sums over $\Z$. Evaluating from right to left, we have:
\begin{enumerate}
\item Integrate $H_x$ from $0$ to $1$. Recalling that $F_{yz}\in \Z$ we have,
\eq{
\int\limits_0^1 dHx e^{2\pi i k H_x F_{yz}} = 
\begin{cases}
1 & F_{yz} = 0 \\ 
0 & F_{yz} \neq 0
\end{cases}
}
Hence the integral over $H_x$ sets up a Kronecker $\delta$-function $\delta_{F_{yz}, 0}$.
\item Sum $F_{zx}$ over $\Z$. This sets up a Dirac comb:
\eq{
\sum_{F_{zx}\in\Z}e^{2\pi i k H_y F_{zx}} = \sum_{\ell\in\Z}\delta(\ell - k H_y) = \frac{1}{k}\sum_{\ell\in\Z}\delta\left( H_y - \frac{\ell}{k}\right)
}
These Dirac $\delta$-functions force the $y$-holonomy $H_y$ to be in $\Z_k$.
\end{enumerate}

Let us unpack the physical picture here. The Kronecker $\delta$-function sets the flux $F_{yz}$ to be zero. This is analogous to setting $H_y$ to be time-independent. In the Hamiltonian picture, the holonomy around one nontrivial loop is the canonical position and uniquely labels ground states. Similiarly, the Dirac $\delta$-function constrains that same holonomy to be in $\Z_k$, reflecting the $k$ ground states on a torus. One may exchange the roles of $H_x$ and $H_y$ by exchanging the order of integrating over variables, namely by integrating over $H_y$ and summing over $F_{yz}$ and then integrating over $H_x$ and summing over $F_{xz}$.

Performing the remaining integrations, we have 
\eq{
Z = 
\sum_{F_{yz}\in\Z}
\int\limits_0^1 dH_y 
~\left(\delta_{F_{yz}, 0}\right)
\sum_{\ell= 0}^{k-1}\frac{1}{k}\delta\left(H_y - \frac{\ell}{k}\right) 
= 1
}
The $k$ contributions to the partition function from the $k$ ground states on a torus are multiplied by the $\frac{1}{k}$, reflecting the fact that our formalism calculates a normalized partition function; future work will generalize this calculation to the ground state degeneracy and braiding of Wilson loops.

\chapter{Commentary}

I have not asked him, but I believe that Xiao-Gang and I began to work with each other because we both believe that the universe must fundamentally be computable. To that end, I took the least computable ideas I could imagine: chiral quantum field theories, and sought to render them well defined on the lattice so that we can simulate them on a computer. Of course, the great joy of reality becomes apparent when, instead of somehow taming the wildness of Quantum Field Theory, defining them on the lattice only reveals an even more extraordinary nature. 

Not all of my drive comes from abstract considerations. I hope deeply that this work proves useful to other scientists over the long term. Most of my work in the latter days of my PhD has been seeking to communicate these ideas to those from the Lattice Gauge Theory and Quantum Information fields, in the hopes that it solve several related problems for them not only in principle, but in practice. In my experience, progress in abstract considerations is best measured by its contribution to hard, down-to-earth problems. 

Many physicists would ask why, given my obsession with computability, I did not work in quantum information. On the one hand, I will spend my coming post-doctoral fellowship, and likely the rest of my career, building quantum computers, and I could not be more excited for that. But I do not regret working in condensed matter quantum field theory at all. There is something extraordinary about quantum mechanics in spacetime, where dimensionality plays a central, even defining role. I did not want to study states in an abstract Hilbert space, but rather correlation functions stretched out across space and time. And, most of all, I wanted to work with Xiao-Gang. 

Feynman first said that we do not understand a theory unless we can explain it to freshmen. About halfway through my PhD, I first heard the ``Wen Principle:'' that we do not understand a theory unless we can program it into a computer\footnote{This is very similar to Wilson's principle, named for the founder of modern renormalization group theory Kenneth Wilson. From his Nobel lecture \cite{Wilson}:

\begin{displayquote}
``In thinking and trying out ideas about `what is a field theory' I found it very helpful to demand that a correctly formulated field theory should be soluble by computer, the same way an ordinary differential equation can be solved on a computer, namely with arbitrary accuracy in return for sufficient computing power.''
\end{displayquote}

Yet there is a key difference between the viewpoints of Wen and Wilson. Wilson saw the lattice as a useful way to define a quantum field theory, while field theory itself was closer to the fundamental theory of the universe. Wen sees the lattice as a candidate for a fundamental theory of our universe, and quantum field theories as useful ways to reason about many-body systems. I am clearly sympathetic to Wen's view.
}. (Surely this is more flexible than Feynman's rule.) The Wen principle has drastic implication for the nature of our world: If you believe that the universe can be understood, or even, as I do, that it is our destiny to understand it, then the universe must be computable. All of the quantum field theories that control the behavior of the physical world must be computable. That is what my PhD work, and this thesis, have been about. 

The connection between understanding and computation goes much farther. In the ``it-from-bit'' and ``it-from-qubit'' formulations, we often speak of the universe \emph{as a} computation. However, it is not that reality is modeled after computation, but the opposite: all of our computation, and even all of our cognition, is modeled after the nature of the universe. Equations encode words, words communicate ideas, and ideas emerge from the world around us. I think often of C.S. Lewis' belief that there is an order to things, and the thing in our experience which that order most resembles is a mind. The Wen principle then follows naturally: computers model minds, and minds model reality.

Again from Feynman: ``what I cannot create I do not understand.'' Perhaps to understand \emph{is} to create \cite{2004qftm.book.....W}. Or, from Wen's principle, to compute. Because I suspect that, at the center of things, at the beginning of it all, lays the idea, the order. The \emph{logos}.

\end{document}